\documentclass[aps,twocolumn,nofootinbib,showpacs,prd,aps,10pt]{revtex4-1}
\usepackage[dvips]{graphicx}
\usepackage[english]{babel}
\usepackage[T1]{fontenc}
\usepackage{mathrsfs}
\usepackage[tbtags]{amsmath}
\usepackage{amssymb}
\usepackage{amsxtra}
\usepackage{amsopn}
\usepackage{latexsym}
\usepackage[mathcal]{eucal}
\usepackage{mathtools}
\usepackage{float}

\newcommand{\BE}{\begin{equation}}
\newcommand{\EE}{\end{equation}}
\newcommand{\BA}{\begin{align}}
\newcommand{\EA}{\end{align}}
\newcommand{\Tr}{\mathrm Tr}
\newcommand{\nn}{\nonumber}
\newcommand{\bk}{ {\bf k} }
\newcommand{\bp}{ {\bf p} }
\newcommand{\kkk}{ \frac{{\rm d}^4k}{(2\pi)^4}}

\newcommand{\ppp}{ \frac{{\rm d}^4p}{(2\pi)^4}}

\newcommand{\Rerm}{\mathop{\rm Re}}

\renewcommand{\Im}{\mathop{\rm Im}}

\newcommand{\omkm}{\epsilon_{\bk,m}}
\newcommand{\omka}{\epsilon_{\bk,\alpha}}
\newcommand{\omkb}{\epsilon_{\bk,\beta}}

\newcommand{\ompkb}{\epsilon_{\bp-\bk,\beta}}

\newcommand{\opa}{\epsilon_{y+x,\alpha}}
\newcommand{\oma}{\epsilon_{y-x,\alpha}}
\newcommand{\opb}{\epsilon_{y+x,\beta}}
\newcommand{\omb}{\epsilon_{y-x,\beta}}

\newcommand{\omxm}{\epsilon_{x,m}}
\newcommand{\omxa}{\epsilon_{x,\alpha}}
\newcommand{\omxb}{\epsilon_{x,\beta}}

\begin{document}

\title{Thermal extension of the screened massive expansion in the Landau gauge}

\author{Fabio Siringo}
\email{fabio.siringo@ct.infn.it}
\author{Giorgio Comitini}
\email{giorgio.comitini@dfa.unict.it}

\affiliation{Dipartimento di Fisica e Astronomia 
{\it E. Majorana} dell'Universit\`a di Catania,\\ 
INFN Sezione di Catania,
Via S.Sofia 64, I-95123 Catania, Italy}

\date{\today}

\begin{abstract}

The massive screened expansion for pure  SU(3) Yang-Mills  theory
is extended to finite temperature in the Landau gauge. 
All thermal integrals are evaluated analytically up to an external
one-dimensional integration, yielding 
explicit integral representations of analytic functions which can
be continued to the whole complex plane. The gluon propagator is first explored in the
Euclidean space by making use of parameters obtained from first principles, which were already found to
accurately reproduce the lattice data at zero temperature. Within such a scheme, the agreement with the
lattice at $T\neq 0$ turns out to be only qualitative. The description improves provided that the parameters
are tuned in a temperature-dependent way by a fit to the data, carried out separately for each component of the propagator; in particular, the transverse
component closely follows the lattice data, while the agreement of the longitudinal component with the data is poor at small momenta and moderately high temperatures. The dispersion relations of the quasi-gluon are then extracted from 
the pole trajectory in the complex plane using the fitted parameters. 
A crossover is found for the mass, suppressed by temperature like an order parameter in the confined phase, while increasing like an ordinary thermal mass in the
deconfined phase.
\end{abstract}




\maketitle

\section{introduction}
In the last decades, considerable efforts have been devoted to the study of the complex behavior 
of quarks and gluons under the extreme conditions which are reached in heavy-ion collisions.
In principle, the dynamical and thermal properties of a quark-gluon plasma should descend from
the relatively simple Lagrangian of the SU(3) gauge theory which describes QCD. However, things are not so easy because 
the standard perturbative approach breaks down in the strong-coupling IR limit and is also plagued by 
further resummation problems at any finite temperature. As a matter of fact, we still miss a full
theoretical treatment of the problem.

Even the pure gauge theory, without quarks, is not fully understood, despite its
relevance for describing the quark-gluon plasma. Many important advances have been made by
the numerical simulation of the pure Yang-Mills (YM) Lagrangian on a lattice, providing insights 
into the gluon dynamics and the phase diagram.
Among them, the confirmation of a dynamically generated 
gluon mass~\cite{duarte,cucchieri08,cucchieri08b,bogolubsky,dudal,binosi12,oliveira12,burgio15},
as predicted by Cornwall in 1982~\cite{cornwall82},
and the occurrence of a phase transition, 
with the gluons that become confining below a critical temperature~\cite{lucini,silva,aouane}.

It would be a desirable progress if the dynamical and transport parameters,  
like masses, widths, dispersion relations, transport coefficients, etc., which are currently 
regarded as phenomenological parameters~\cite{werner2016,castorina2012,alba2012,greco2011},
could be directly evaluated from first principles. That program might be accomplished in part if the elementary correlators 
and their analytic properties were known in the Minkowski space. 
Unfortunately, all lattice calculations and most numerical works provide information in
the Euclidean space and the analytic continuation is a difficult ill-defined problem for the numerical data~\cite{dudal14}.

In the last years, a very predictive analytical method has been developed~\cite{ptqcd,ptqcd2,analyt,xigauge}
by a mere change of the expansion point of ordinary
perturbation theory (PT) for the exact gauge-fixed Becchi-Rouet-Stora-Tyutin (BRST) invariant YM Lagrangian, 
yielding a screened massive expansion which is safe in the IR while 
recovering the correct results of ordinary PT in the UV. At one-loop and zero temperature, 
the screened expansion provides analytical results which are in
excellent agreement with the lattice and can be easily continued to Minkowski space~\cite{xigauge,scaling,ghost,beta,beta2}.
Thus the method provides a way to extract dynamical details like masses and damping rates from first principles.

In this paper, the formalism is extended to a finite temperature $T\not=0$, with the aim to provide a complementary tool for the study
of the gluon plasma from first principles. As discussed in \cite{damp,varT}, 
the screened expansion can be extended to finite temperature, providing a quasi-particle picture 
for the gluon which is damped, with a very short finite lifetime, and canceled from the asymptotic states.
Here, we give a full account of the details of the calculation and report a comprehensive set of results for the
gluon sector, including propagators, analytic properties, poles, masses, widths and dispersion relations.
We discuss different optimization strategies and, by a comparison with the available lattice data,
we explore how robust the screened expansion is when it is extended to finite temperature.
 
While the existence of a screening mass mitigates the effects of the hard thermal loops, several problems arise
at a finite temperature, ranging from the temperature dependence of the optimal mass scale, to the analytic continuation
of the numerical integrals. Actually, even if a formal extension to finite temperature is straightforward and based on
standard thermal Feynman graphs, the ambition to extract analytical results requires a quite tedious and lengthy analytical
calculation of the integrals and, even so, a final one-dimensional numerical integration cannot be avoided. 
Nonetheless, the resulting numerical integrals are shown to define analytic functions which can be evaluated in the complex plane. 
Then, the poles of the gluon propagator and the resulting dispersion relations can be easily extracted numerically.

Overall, despite the expected difficulties, the one-loop screened expansion seems to be reliable at low temperature, with correct
predictions which become less quantitative at high temperature, especially for the longitudinal sector, when compared with the lattice data.

At $T=0$, the one-loop approximation is quite sensitive to the renormalization scheme and to the subtraction point, but it can be shown
to be basically {\it tangent} to the exact result, which is approached for a special choice of the ratio between the gluon mass parameter $m$ and
the renormalization scale $\mu$. 
Here, $m$ is just a mass parameter which defines the shift of the expansion point~\cite{ptqcd,ptqcd2,beta,beta2}, 
not to be confused with
the physical mass of the gluon. It seems that, for that special ratio $\mu/m$, the higher order terms become negligible, yielding very accurate
analytical expressions for the propagators. While that special ratio is scheme-dependent, it can be determined from first principles
by monitoring some identities which must be fulfilled by the exact propagators, 
like the Nielsen identities, which express the gauge-invariance of the poles~\cite{xigauge}.
We must mention that, once the ratio is optimized in the complex {\it Minkowski} space, where the poles are defined, the propagators are found in excellent agreement 
with the lattice data in the {\it Euclidean} space. Thus, the optimized analytical expression is not just a good interpolation formula, 
but a very good approximation for the whole analytic function which is defined in the complex plane. Moreover, at the optimal ratio $\mu/m$ there
is only one energy scale left in the calculation, say the mass parameter $m$, so that its actual value becomes irrelevant, since it can be
used as energy units and is eventually determined by a comparison with the phenomenology. For instance, sharing the same units of the 
lattice data, a value $m=0.656$ GeV was established in previous works~\cite{xigauge,beta}.

At a finite temperature $T\not=0$, there is a third energy scale and the optimal parameters $m$, $\mu$ become two independent functions
of temperature, $m(T)$, $\mu(T)$, since their optimal ratio is expected to depend on $T$. In principle, one could proceed as for $T=0$ and
fix the optimal ratio by monitoring the gauge-invariance of the poles. However, that would at least require a knowledge of the thermal propagators in a generic covariant gauge,
while the present formalism has been developed only in the Landau gauge. Moreover, no lattice data are available for a comparison in a generic gauge
and finite $T$. This is not a theoretical limitation by itself, but leads to a weakening of the control of the accuracy.

That of the gauge-invariance of the poles actually is an additional problem one encounters when extending the theory to finite $T$~\cite{kajantie,heinz,hansson,carrington}. Even though the poles of the propagator are constrained to be non-perturbatively gauge-independent by e.g. the Nielsen identities~\cite{kobes}, in the thermal formalism different powers of the coupling constant coexist at the same loop order when hard-thermal-loop effects are taken into account, so that consistent resummation schemes are needed in order to obtain truly gauge-invariant results for the poles' position. To first order in the coupling, this can be shown to only affect the imaginary part of the dispersion relations, i.e. the gluon's damping rate. In this work no attempt has been made to implement such resummation schemes or to keep under control the accuracy of the approximation with respect to the issue of gauge-invariance. Whereas at low, non-zero temperatures the screening provided by the gluon's mass may somewhat suppress the effects of the required resummed terms, at higher temperatures the latter are expected to become non-negligible, causing our predictions for the gluon damping rate to become less and less reliable as the temperature is increased.

In the Landau gauge, we explored two complementary strategies and checked that the qualitative description which emerges is robust enough
and does not depend on the optimization choice. The first, simpler, strategy consists in using the same $m$ and $\mu$ parameters that work at $T=0$. 
That choice was already made in Ref.~~~\cite{damp} (albeit with different values for the parameters) and makes sense at low temperature where we expect that $m(T)\approx m(0)$ and $\mu(T)\approx \mu(0)$.
With this choice, we find the correct qualitative behavior without any adjustment of parameters. 
In particular, the longitudinal propagator shows a non-monotonic behavior with a crossover at $T/m(0)\approx 0.15$. 
However, the agreement with the lattice data is not quantitative, and the predicted transition temperature is too small ($T\approx 100$ MeV),
thus indicating that we are already outside the safe low-temperature range. Nonetheless, the disagreement can be absorbed in part by a 
temperature-dependent optimization of the expansion.

Thus, as a second strategy, we relax the constraints of $m$ and $\mu$ being equal to their $T=0$ values and regard $m(T)$ and $\mu(T)$ as independent
unknown functions. Reversing the argument that led to their optimization at $T=0$, we tune the unknown functions in the Euclidean space by
looking for the best agreement with the lattice data. Then, {\it assuming} that the higher-order terms are smaller when the agreement is
better, the optimized propagators are continued to Minkowski space where the pole location gives information on the dispersion relations of the quasi-gluons
at finite temperature. We anticipate that, from a strictly quantitative point of view, the agreement with the lattice is not comparable
with the excellent result which was reached at $T=0$. Moreover, while the transverse propagator is generally well described, the longitudinal
projection becomes very poor deep in the IR for moderately high temperatures. Since most of the deviation occurs below 500-700 MeV, we expect that
the predictions for the pole position at high momenta might not be affected too much. We stress that there are no data available 
in the Minkowski space for a comparison, thus evidencing the power of the method for exploring the analytic properties of the propagators.

Irrespective of the optimization criterion, we confirm the finding of Ref.~~\cite{damp} and the quasi-gluon scenario which was described by
Stingl~\cite{stingl}, with a gluon which has a very short finite lifetime and can only exist as a short-lived 
intermediate state at the origin of a gluon-jet event.

This paper is organized as follows. In Sec. II we review the set-up and main features of the screened massive expansion and its extension to finite temperatures. In Sec. III we present our results for the Landau gauge gluon propagator at $T\neq 0$ and vanishing Matsubara frequency, $\omega=0$. In Sec. IV we derive the dispersion relations for the quasi-gluons at finite temperatures. In Sec. V we discuss our results and present our conclusions. In the Appendix we explicitly compute the gluon polarization and ghost self-energy at finite temperatures using the screened massive expansion.

\section{The screened expansion and its extension to finite temperature}

In a linear covariant $\xi$-gauge, the gauge-fixed BRST invariant Lagrangian of pure Yang-Mills SU(N) theory 
is
\BE
{\cal L}={\cal L}_{YM}+{\cal L}_{fix}+{\cal L}_{FP},
\label{Ltot}
\EE
where 
\begin{align}
{\cal L}_{YM}&=-\frac{1}{2} \Tr\left(  \hat F_{\mu\nu}\hat F^{\mu\nu}\right),\nn\\
{\cal L}_{fix}&=-\frac{1}{\xi} \Tr\left[(\partial_\mu \hat A^\mu)(\partial_\nu \hat A^\nu)\right],
\label{LYMfix}
\end{align}
and ${\cal L}_{FP}$ is the ghost term
arising from the Faddeev-Popov (FP) determinant.
The tensor operator is defined as
\BE
\hat F_{\mu\nu}=\partial_\mu \hat A_\nu-\partial_\nu \hat A_\mu
-i g \left[\hat A_\mu, \hat A_\nu\right],
\label{F}
\EE
where the gauge field operators satisfy the $SU(N)$ algebra
\begin{align}
\hat A^\mu&=\sum_{a} \hat X_a A_a^\mu,\nn\\
\left[ \hat X_a, \hat X_b\right]&= i f_{abc} \hat X_c,\quad
f_{abc} f_{dbc}= N\delta_{ad}.
\end{align}

In the standard PT formalism, the total action is split as $S_{tot}=S_0+S_I$, where the quadratic part can
be written as
\begin{align}
S_0&=\frac{1}{2}\int A_{a\mu}(x)\delta_{ab} {\Delta_0^{-1}}^{\mu\nu}(x,y) A_{b\nu}(y) {\rm d}^4 x\,{\rm d}^4 y \nn \\
&+\int c^\star_a(x) \delta_{ab}{{\cal G}_0^{-1}}(x,y) c_b (y) {\rm d}^4 x\, {\rm d}^4 y,
\label{S0}
\end{align}
while the interaction contains three vertices
\BE
S_I=\int{\rm d}^4x \left[ {\cal L}_{gh} + {\cal L}_3 +   {\cal L}_4\right],
\label{SI}
\EE
\begin{align}
{\cal L}_{3g}&=-g  f_{abc} (\partial_\mu A_{a\nu}) A_b^\mu A_c^\nu,\nn\\
{\cal L}_{4g}&=-\frac{1}{4}g^2 f_{abc} f_{ade} A_{b\mu} A_{c\nu} A_d^\mu A_e^\nu,\nn\\
{\cal L}_{ccg}&=-g f_{abc} (\partial_\mu c^\star_a)c_b A_c^\mu.
\label{Lint}
\end{align}
In Eq.~(\ref{S0}), the standard free-particle propagators for
gluons and ghosts, $\Delta_0$ and ${\cal G}_0$ respectively, are defined by their Fourier transforms
\begin{align}
{\Delta_0}^{\mu\nu} (p)&=\Delta_0(p)\left[t^{\mu\nu}(p)  
+\xi \ell^{\mu\nu}(p) \right],\nn\\
\Delta_0(p)&=\frac{1}{-p^2}, \qquad {{\cal G}_0} (p)=\frac{1}{p^2},
\label{D0}
\end{align}
where the transverse and longitudinal projectors are used
\BE
t_{\mu\nu} (p)=g_{\mu\nu}  - \frac{p_\mu p_\nu}{p^2},\quad
\ell_{\mu\nu} (p)=\frac{p_\mu p_\nu}{p^2}.
\label{tl}
\EE
Later, we will take the limit $\xi\to 0$ and use the Landau gauge which is a Renormalization Group (RG) fixed point and
is the most studied gauge on the lattice.
In the above equations, the fields and the coupling must be regarded as renormalized objects and the inclusion of
the usual set of counterterms is understood in the total Lagrangian.

\begin{figure}[b] \label{fig:graphs}
\centering
\includegraphics[width=0.25\textwidth,angle=-90]{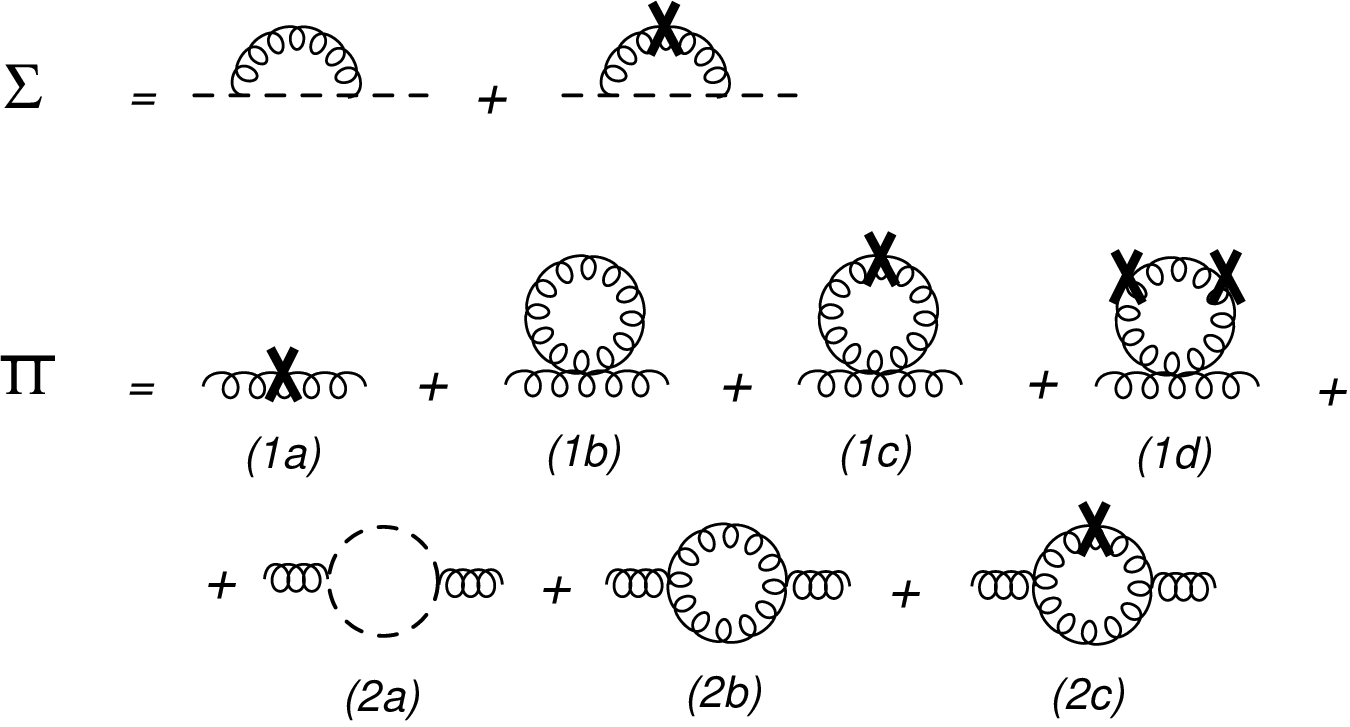}
\caption{Two-point graphs with no more than three vertices and no more than one loop. 
The cross is the transverse mass counterterm of Eq.~(\ref{dG2}) and is regarded as
a two-point vertex.
In the Appendix, a detailed description of the calculation at finite $T$ is given for all
the polarization graphs in the figure.}
\end{figure}

The massive screened version of PT was developed in Refs.~\cite{ptqcd,ptqcd2,analyt}. At $T=0$ and in a generic
covariant gauge, the method is very accurate and predictive if the expansion is optimized 
by the constraints of BRST symmetry~\cite{xigauge,beta,beta2}.
The expansion arises by a mere change of the expansion point of ordinary PT.
Following Refs.~\cite{ptqcd2,xigauge}, the new {\it massive} expansion is recovered by just adding 
a transverse mass term to the quadratic part of the action and subtracting it again from the interaction, 
leaving the total action unchanged. In more detail, we add and subtract the action term
\BE
\delta S= \frac{1}{2}\int A_{a\mu}(x)\>\delta_{ab}\> \delta\Gamma^{\mu\nu}(x,y)\>
A_{b\nu}(y) {\rm d}^4\, x{\rm d}^4y,
\label{dS1}
\EE
where the vertex function $\delta\Gamma$ is a shift of the inverse propagator,
\BE
\delta \Gamma^{\mu\nu}(x,y)=
\left[{\Delta_m^{-1}}^{\mu\nu}(x,y)- {\Delta_0^{-1}}^{\mu\nu}(x,y)\right],
\label{dG}
\EE
and ${\Delta_m}^{\mu\nu}$ is a new {\it massive} free-particle propagator,
\begin{align}
{\Delta_m^{-1}}^{\mu\nu} (p)&=
(-p^2+m^2)\,t^{\mu\nu}(p)  
+\frac{-p^2}{\xi}\ell^{\mu\nu}(p).
\label{Deltam}
\end{align}
Adding that term is equivalent to substituting the new massive propagator ${\Delta_m}^{\mu\nu}$ for the 
old massless one ${\Delta_0}^{\mu\nu}$ in the quadratic part. Thus, the new expansion point is a massive
free-particle propagator for the gluon, which is much closer to the exact propagator in the IR. 
The mass-shift
parameter $m$ is irrelevant in the UV, but acts as a natural cutoff which screens the theory in the IR. 

Of course, in order to leave the total action unaffected by the change, the same term is subtracted from the interaction,
providing a new interaction vertex $-\delta\Gamma$, a two-point vertex which can be regarded as a new counterterm.
Dropping all color indices in the diagonal matrices and
inserting Eq.~(\ref{D0}) and (\ref{Deltam}) in Eq.~(\ref{dG}), the vertex is just the transverse mass shift
of the quadratic part,
\BE
-\delta \Gamma^{\mu\nu} (p)=-m^2 t^{\mu\nu}(p),
\label{dG2}
\EE
and must be added to the standard set of vertices arising from Eq.~(\ref{Lint}). The new vertex 
is now part of the interaction, even if it does not depend on the coupling. Thus,
the expansion has the nature of a $\delta$-expansion, since different powers of the coupling
coexist at each order in powers of the total interaction.

The proper gluon polarization and ghost self energy can be evaluated, order by order, by the modified PT.
In all Feynman graphs, 
any internal gluon line is a massive free-particle propagator ${\Delta_m}^{\mu\nu}$ and the
new insertions  of the (transverse) two-point vertex $\delta \Gamma^{\mu\nu}$ are denoted by a cross, as
shown in Fig.~~1. For further details we refer to Refs.~\cite{ptqcd,ptqcd2,xigauge}. 

Since the total gauge-fixed FP Lagrangian is not modified and because of BRST invariance,
the longitudinal polarization is known exactly and is zero. 
At $T=0$, the exact polarization and the dressed gluon propagator are defined by a single function,
\BE
\Pi^{\mu\nu}(p)=\Pi(p)\, t^{\mu\nu}(p),
\label{pol}
\EE
so that, in the Landau gauge, the exact gluon propagator is transverse, 
\BE
\Delta_{\mu\nu}(p)=\Delta (p)\,t_{\mu\nu}(p),
\EE
and defined by the scalar function $\Delta(p)$. This feature is lost at any finite temperature $T>0$,
since Lorentz-invariance is broken, and two scalar functions are required instead. In that perspective, it is convenient to
maintain the Lorentz structure explicit and to switch to the Euclidean 
formalism. Then, denoting with $p^2$ the Euclidean squared momentum, the exact (dressed) gluon and ghost propagators can be written as
\begin{align}
{{\Delta}^{-1}}_{\mu\nu} (p)&=(p^2+m^2)t_{\mu\nu}(p)+\frac{p^2}{\xi}\ell_{\mu\nu}(p)-\Pi_{\mu\nu}(p),\nn\\
{\cal G}^{-1}(p)&=-p^2-\Sigma (p),
\label{dressprop}
\end{align}
where $t_{\mu\nu}$ and $\ell_{\mu\nu}$ are the Euclidean projectors of Eq.~(\ref{tlE}).
The proper gluon polarization $\Pi_{\mu\nu}$ and the ghost self-energy $\Sigma$ are the
sum of all one-particle-irreducible (1PI) graphs in the screened expansion, including all counterterms.
In Fig.~~1, the two-point 1PI graphs are shown up to one-loop and third order in the delta-expansion.
In the exact self energies, we can single out the tree-level terms and write
\begin{align}
\Pi_{\mu\nu}(p)&=m^2t_{\mu\nu}(p)-p^2t_{\mu\nu}(p)\delta Z_A+\Pi^{loop}_{\mu\nu}(p),\nn\\
\Sigma (p)&=p^2\delta Z_c+\Sigma^{loop}(p),
\label{selfs}
\end{align}
where the first term $m^2t_{\mu\nu}(p)$ is the tree graph (1a) in Fig.~~1 
and arises from the insertion of the new two-point 
vertex $-\delta \Gamma_{\mu\nu}$ of Eq.~(\ref{dG2}). We observe that this first tree term cancels the mass shift
of the gluon propagator in Eq.~(\ref{dressprop}). Indeed, the physical mass of the gluon arises from
the loops and is not merely given by the mass-shift parameter $m^2$. 
The other tree-level terms, $-p^2\,t_{\mu\nu}\,\delta Z_A$, $p^2\,\delta Z_c$, are not shown in Fig.~~1 and are the usual
field-strength renormalization counterterms. Their UV diverging parts are not affected by the mass parameter
and are the same of standard PT~\cite{ptqcd,ptqcd2}.
The proper functions, $\Pi^{loop}_{\mu\nu}$, $\Sigma^{loop}$, are given by the sum of all 1PI graphs containing loops.
The finite parts of $\delta Z_A$, $\delta Z_c$ are arbitrary and depend on the scheme and 
on the renormalization scale $\mu$~~\cite{beta,beta2}.
The diverging parts of $\delta Z_A$, $\delta Z_c$ cancel the UV divergences of the functions
$\Pi^{loop}_{\mu\nu}/p^2$ and 
$\Sigma_{loop}/p^2$ which become finite dimensionless functions of the variable $p_\mu/m$. They are
defined up to a constant which depends on the dimensionless renormalization scale parameter $t=\mu^2/m^2$. 
Thus, at $T=0$, there are two energy scales in the calculation, $m$ and $\mu$.
For instance, in a momentum subtraction scheme (MOM) and in the Landau gauge, 
the one-loop dressed propagators can be written as
\begin{align}
{\Delta}(p)^{-1}&=p^2-Ng^2\left[\Pi^{(1)}(p) -\Pi^{(1)}(\mu)\right],    \nn\\
{\cal G}(p)^{-1}&=-p^2-Ng^2\left[\Sigma^{(1)}(p)-\Sigma^{(1)}(\mu)\right],
\label{dressprop1}
\end{align}
having made explicit the dependence on $N$ and $g^2$ as factors in the one-loop functions
$\Pi^{(1)}$, $\Sigma^{(1)}$, according to the notation of  Appendix A, where all details of the calculation are reported.
In Eq.~(\ref{dressprop1}), an explicit choice has been made for the finite parts of the renormalization constants
$\delta Z_A$, $\delta Z_c$. Of course, that choice depends on the scheme and on the renormalization scale $\mu$.
A more general way to get rid of all the scheme-dependent parameters, including the renormalized coupling $g^2$, 
was discussed in  previous papers on the screened expansion~\cite{ptqcd,ptqcd2,xigauge,beta}, where two dimensionless
one-loop functions were defined (see Appendix B.1 for their explicit expressions),
\begin{align}
\pi_1(p^2/m^2)&=-\left(\frac{16\pi^2}{3}\right)\frac{\Pi^{(1)}(p)}{p^2},   \nn\\
\sigma_1(p^2/m^2)&=\left(\frac{16\pi^2}{3}\right)\frac{\Sigma^{(1)}(p)}{p^2}, 
\label{pisigma}  
\end{align}
so that the one-loop propagators in Eq.~(\ref{dressprop1}) can be recast as functions of the dimensionless
variable $s=p^2/m^2$,
\begin{align}
p^2\,{\Delta}(p)&=\frac{z_\pi}{\pi_1(s)+\pi_0},    \nn\\
p^2\,{\cal G}(p)&=-\frac{z_\sigma}{\sigma_1(s)+\sigma_0}, 
\label{dressprop2}
\end{align}
where  $z_\pi$ and $z_\sigma$ are irrelevant normalization constants while all the scheme-dependent parameters are
embedded in the two constants $\pi_0$ and $\sigma_0$. With some abuse of language, we will refer to them as
{\it renormalization} constants. Eq.~(\ref{dressprop2}) is quite general since it does not require any specific 
renormalization scheme to be defined. Of course, our ignorance about those constants reflects a well known weakness
of the one-loop approximation which depends on the details of  the renormalization scheme and on the actual value
of the renormalization scale $\mu$. In this sense, we still have two scales, $m$ and $\mu$, and the arbitrary choice
of their ratio $t=\mu^2/m^2$ somehow determines the actual value of the renormalization constants $\pi_0$ and $\sigma_0$.

A nice feature of the one-loop result is its apparent {\it tangency} to the exact result which is approached for  special 
values of the renormalization constants. Those values are equivalent to a choice of the best renormalization scale 
$\mu$, where the approximation is more effective. It is just an example of the 
optimized perturbation theory by variation of the renormalization scheme~\cite{stevensonRS1,stevensonRS2}.
There might be a special scale $\mu$ where the expansion converges more quickly and the higher order terms are minimal.
Thus, from first principles, we could determine the optimal constants by monitoring some identities which must 
be satisfied by the exact propagators. For instance, in Ref.~~\cite{xigauge}, 
the Nielsen identities~\cite{nielsen1,nielsen2} were used, which are
a direct consequence of BRST symmetry. From the identities, one can prove the gauge-parameter-independence of the poles 
and residues of the exact gluon propagator~\cite{xigauge}. Then, we might expect that the renormalization constants 
are optimal when the poles have a minimal sensitivity  to the gauge parameter.
It is remarkable that the optimized one-loop  propagators turn out to be in excellent agreement with the lattice data in
the IR. Notably, while the comparison with the data requires an analytic continuation to the Euclidean space, 
the poles are found in the complex plane. Thus, the one-loop propagators in Eq.~(\ref{dressprop2}) are not just one of the
many interpolation formula for the data, but they provide a very accurate analytic function in the whole complex plane.
The existence of complex poles is one of the most important predictions of the screened expansion. 
While a thermal mass and a finite damping rate are expected by PT at high temperature,  the existence of finite
intrinsic values at $T=0$ can be regarded as a proof of confinement
as first discussed by Stingl~\cite{stingl}. The quasi-gluon has a finite lifetime and can only exist as a short-lived 
intermediate state. However, at finite temperature, the quasi-gluons play an important role for determining
the thermal properties
of the hot plasma. Thus, a finite temperature extension of the screened expansion is required for a full study of the
dispersion relations which emerge from the pole location.

At a finite temperature $T>0$, Eqs.~(\ref{dressprop}),(\ref{selfs}) are still valid, but the one-loop graphs
in Fig.~~1 acquire a finite thermal part which must be added to the vacuum (diverging) contribution at $T=0$.
The thermal parts are finite and no further renormalization is required. We only have to add the thermal parts
to the self-energies in Eq.~(\ref{selfs}). 

We write the Euclidean four-vector as
$p^\mu=({\bf p}, \omega)$
where $\omega=p_4=-ip_0$, while the
Lorentz four-vector was $(p_0, {\bf p})$.
In the finite-temperature formalism, $\omega=\omega_n=2\pi n T$ and the
Euclidean integral is replaced by a sum over $n$ and by a three-dimensional
integration,
\BE
\int \ppp \to T\sum_n\int\frac{{\rm d}^3{\bf p}} {(2\pi)^3}.
\EE
Since Lorentz invariance is obviously broken, we introduce a transverse projector $P^T_{\mu\nu}$, 
orthogonal to the fourth Euclidean direction, and its longitudinal complement $P^L_{\mu\nu}$, as defined in Eq.~(\ref{TL}),
so that the gluon polarization and propagator in Eqs.~(\ref{dressprop}),(\ref{selfs}) 
can be written in the Landau gauge, $\xi=0$, as
\begin{align}
\Pi_{\mu\nu}(p,T)&= \Pi_L(p,T)\>P^L_{\mu\nu}(p)+\Pi_T(p,T)\>P^T_{\mu\nu}(p),\nn\\
\Delta_{\mu\nu}(p,T)&=\Delta_L(p,T)\>P^L_{\mu\nu}(p)+\Delta_T(p,T)\>P^T_{\mu\nu}(p),
\end{align}
where the  projected one-loop dressed functions are
\begin{align}
\Delta_T(p,T)^{-1}&=p^2+p^2\delta Z_A-Ng^2\Pi^{(1)}_T(p,T),\nn\\
\Delta_L(p,T)^{-1}&=p^2+p^2\delta Z_A-Ng^2\Pi^{(1)}_L(p,T).
\end{align}
and $\Pi^{(1)}_{L,T}$ are the one-loop projected polarizations, evaluated by projection of the one-loop graphs in
Fig~1, omitting the tree graphs. As discussed in Appendix B, each graph contributing to $\Pi^{(1)}_{L,T}$ 
can be split as 
\BE
\Pi^{(1)}_{L,T}(p,T)=\left[\Pi^{(1)}_{L,T}\right]_{Th}+\left[\Pi^{(1)}_{L,T}\right]_{V},
\EE
where the vacuum part $\left[\Pi^{(1)}_{L,T}\right]_{V}=\Pi^{(1)}_{L,T}(p,0)$ is the same graph evaluated at $T=0$
and does not depend on $T$, while the thermal part, $\left[\Pi^{(1)}_{L,T}\right]_{Th}$, vanishes at $T=0$. Thus,
we can generalize Eqs.~(\ref{pisigma}),(\ref{dressprop2})
and define dimensionless functions 
\begin{align}
\left[\pi_{L,T}(p,T)\right]_V&=-\left(\frac{16\pi^2}{3}\right)
\frac{\left[\Pi^{(1)}_{L,T}(p,T)\right]_V}{p^2} =\pi_1(s),  \nn\\
\left[\pi_{L,T}(p,T)\right]_{Th}&=-\left(\frac{16\pi^2}{3}\right)
\frac{\left[\Pi^{(1)}_{L,T}(p,T)\right]_{Th}}{p^2}, 
\end{align}
so that the projections of the one-loop propagator can be recast as 
\BE
p^2\,{\Delta}_{L,T}(p,T)=\frac{z_\pi}{\pi_1(s)+\pi_0+\left[\pi_{L,T}(p,T)\right]_{Th}}.
\label{dresspropLT}
\EE
In this form Eq.~(\ref{dresspropLT}) is quite general since it does not require any specific 
renormalization scheme to be defined. All the scheme-dependent parameters are
embedded in the  {\it renormalization} constant $\pi_0$.

It is not obvious that the same scale $\mu$ and constant $\pi_0$ which were optimal at $T=0$ are 
still optimal at finite $T$. Indeed, they might depend on $T$ and even take a different value 
for the different projections. Moreover, the mass parameter $m$, which was the only energy scale
left after optimization at $T=0$,  might take a value $m(T)$ which depends on $T$. Thus we have
three energy scales: the optimal $\mu(T)$, the mass parameter $m(T)$ and $T$ itself. In other words, 
according to Eq.~(\ref{dresspropLT}), at any $T$ and in units of $m(0)$ we have two free parameters,
the ratio $m(T)/m(0)$ and the optimal renormalization constant $\pi_0(T)$. 
Having the role of variational parameters, to be optimized, their best 
values might be different for the two projections.

While at $T=0$ the optimal constant $\pi_0$ was determined from first principles~\cite{xigauge}, by requiring
a minimal sensitivity of the poles to any change of the gauge parameter, here we have the less ambitious aim
of exploring {\it if} a set of optimal parameters does exist such that the screened expansion is able to describe the
lattice data with reasonable accuracy. Thus, we work in the Landau gauge and, for each value of $T>0$, 
we fix the parameters by a fit of the available lattice data in the Euclidean space.

At low temperature, as we said, we also explored the alternative of maintaining
the parameters fixed at their optimal value for $T=0$, in order to give a general description at finite $T$ from first
principles, without any input from the lattice and from the known phenomenology. Of course, this approach can only be
reliable if $T$ is very low and the thermal effects are small. However, even extrapolating at higher temperatures, the
qualitative predictions turn out to be in agreement with the data. Thus, the screened expansion is able to capture the main
features of gluon thermodynamics at finite temperature. This is a very important aspect, since our final aim will be to extract
some dynamical properties of the quasi-gluons, like the dispersion relations, which cannot be measured on the lattice.
Moreover, even qualitative properties, like the existence of complex poles, are of central interest for understanding
the behavior of the gluon plasma at high temperature and its phase transition.

In order to fulfill that program, once optimized by one of the two alternatives discussed above, the gluon propagator
must be continued to the complex plane. This is a straightforward step if the one-loop graphs are expressed as analytic
functions of the Euclidean momentum. A very detailed but tedious analytical evaluation of the integrals 
is reported in the Appendix. Most of the integrals were encountered in a study of the Curci-Ferrari model~\cite{serreau}.
We basically use the same method for decomposing the integrals. However, in the screened expansion there are also
some different graphs, namely the crossed graphs in Fig.~~1, with one insertion of the mass counterterm. 
Their explicit expressions are obtained by a derivative in the Appendix. 

Unfortunately, at finite $T$, not all the multidimensional integrals can be evaluated analytically
and an external one-dimensional numerical integration cannot be avoided for almost all the one-loop graphs.
Thus, as shown in the Appendix, all the graphs can be written as analytic functions which are defined by
integral representations. The remaining integration can be carried out numerically for any complex value of the
external momentum, provided that no singularity is encountered along the integration path.
Actually, in general, the analytic continuation of integral functions is not trivial.
As discussed in Ref.~~\cite{blaizot2005}, we must check that the external integration on the real axis does not cross any
singular point of the logarithmic functions. Otherwise, a modified path must be chosen before the analytic continuation
can be undertaken. As shown in Ref.~~\cite{damp}, by inspection of the explicit expressions, the existence of singular
points on the integration path can be ruled out in the present case. For instance, 
denoting with $\Omega=p_0$ and $p^\mu=(\Omega, {\bf p})$ the external momentum in Minkowski space, the analytic
continuation of the thermal integral $I^{\alpha\beta}(y,-i\Omega)$ is defined by 
the integral representation of Eq.~(\ref{Iabth}), where $y$ is the external three vector modulus, 
$y=\vert {\bf p}\vert$.
We can continue the external energy $\Omega$ to the complex plane if there are no singular points on the positive
real axis of the integration variable. However,
some branch cuts might be present, originating at the
singular branch point of the logarithmic function in Eq.~(\ref{logalpha}) which reads 
\BE
L_\beta (z_\alpha;y,q)=\log\left[\frac{z_\alpha^2+\epsilon^2_{y+q,\beta}} {z_\alpha^2+\epsilon^2_{y-q,\beta}} \right],
\EE
where the complex
variable $z_\alpha$ is defined as
$z_\alpha=i\Omega\pm i\sqrt{q^2+\alpha^2}$ and $\epsilon^2_{y\pm q,\beta}=(y\pm q)^2+\beta^2$.
Here $\alpha$ and $\beta$ are masses equal to $0$ or $m$ and $q$ is the integration variable.
Assuming the existence of a
branch point at $q=q_0$ on the real axis, the latter must satisfy 
\BE
\pm 2q_0y=\alpha^2-\beta^2-y^2+\Omega^2\pm2\Omega\sqrt{q_0^2+\alpha^2},
\label{zero}
\EE
where the $\pm$ signs are independent of each other. Taking a complex energy $\Omega=\Rerm\Omega+i\Im\Omega$ with $\Im\Omega>0$, 
the imaginary part of Eq.~(\ref{zero})
gives 
\BE \Rerm\Omega=\mp \sqrt{q_0^2+\alpha^2},
\EE
and substituting back in the real part we obtain 
\BE
\epsilon^2_{y \pm q_0,\beta}+(\Im\Omega)^2=0,
\EE
which is never satisfied unless $\Im\Omega=\beta=0$. Thus, if $\Omega$ is not real, the branch point $q_0$ cannot be real and
the integral over $q$, on the real axis, defines an analytic function of $\Omega$.
The same argument holds for the other thermal integrals in Appendix B.
Thus, we can safely continue the numerical integrals from 
the Euclidean space ($\Rerm\Omega=0$, $\Im\Omega>0$) to the whole upper  half-plane. 
Moreover, in the large wavelength limit $y\to 0$, there are no branch points at all because the logarithmic
function can be written as $L_\beta (z_\alpha;y,q)\approx \log\left[1+{\cal O} (y)\right]$ 
and the argument of the log does not vanish if $y$ is small enough.

Having ruled out the existence of singularities along the integration path,
the poles of the gluon propagator and the dispersion relations can be easily extracted numerically in the complex
plane by the integral representation of the thermal integrals which are derived in Appendix B.

\section{The gluon propagator at finite T}

The longitudinal and transverse projections of the polarization graphs entering in Eq.~(\ref{dresspropLT})
are decomposed as the sum of more basic Euclidean integrals in Appendix A, for all the one-loop graphs of Fig.~~1.
The explicit thermal parts of those integrals are presented in Appendix B by integral representations. For any given value
of the external three-momentum $y=\sqrt{{\bf p}^2}$ and Euclidean frequency $\omega=p_4=2\pi n T$, 
the one-dimensional integrals are evaluated numerically by a simple integration on the real axis and the result is
inserted in Eq.~(\ref{dresspropLT}).
We will first explore the projected propagators for $\pi_0$ and $m$ fixed at their zero-temperature values 
which were determined from first principles in Ref.~~\cite{xigauge}. 
Then, we will show how their values can be optimized by a comparison with the available lattice data.

\subsection{Expansion optimized at $\boldsymbol{T=0}$}

In the low-temperature limit, we assume that the optimal renormalization constant $\pi_0(T)$ and mass parameter $m(T)$
can be replaced by their zero-temperature values $\pi_0=-0.876$ and $m(0)=m_0=656$~MeV, as determined in Ref.~~\cite{xigauge} 
by requiring a minimal sensitivity of the pole structure to the gauge parameter. Strictly speaking, in the Landau gauge,
that condition fixes $\pi_0$, while $m_0$ is the only energy scale left and is fixed in order to match the energy units
of the lattice data.

\begin{figure}[b] \label{fig:longprop}
  \centering
  \includegraphics[width=0.32\textwidth,angle=-90]{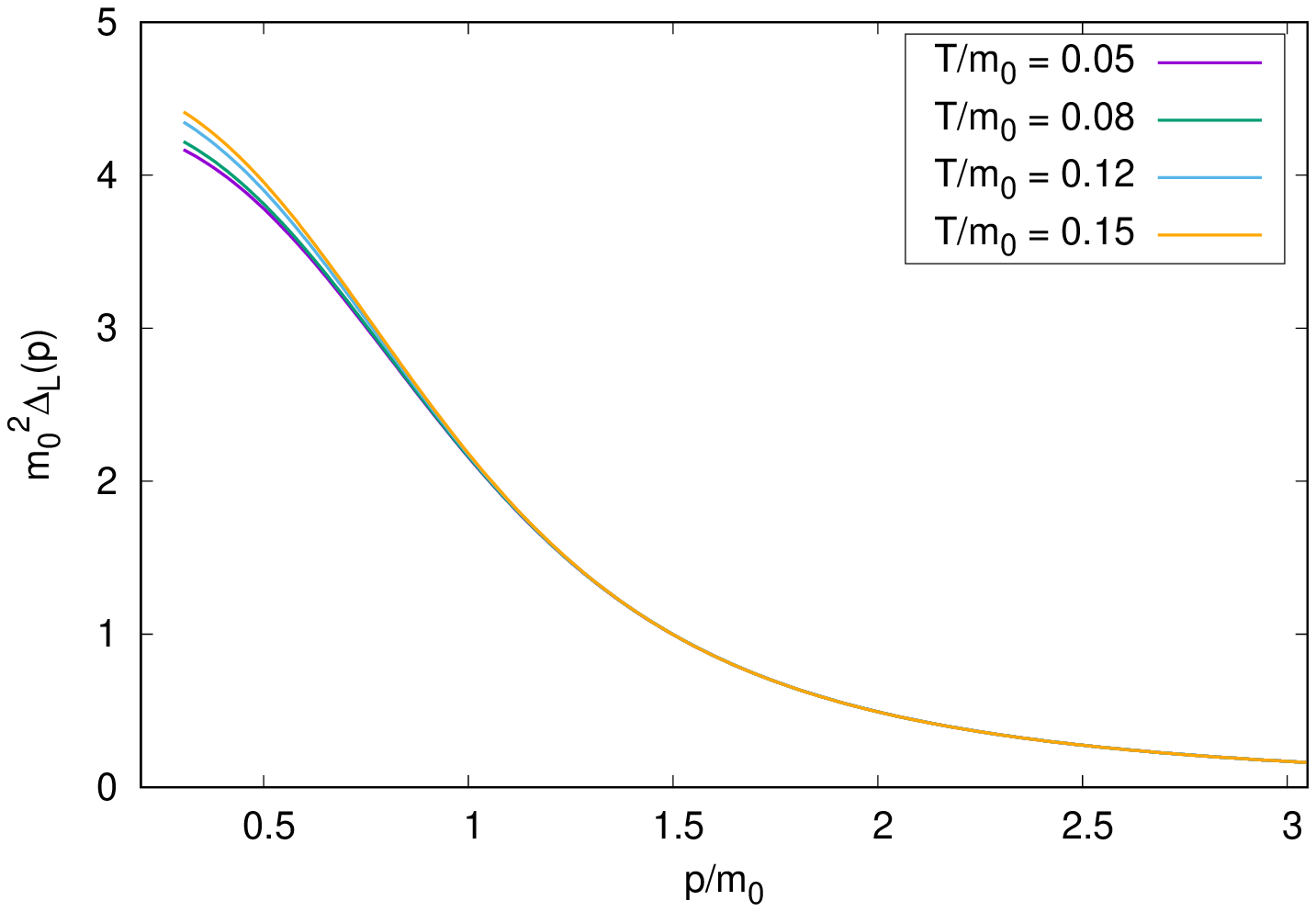}
  \
  \includegraphics[width=0.32\textwidth,angle=-90]{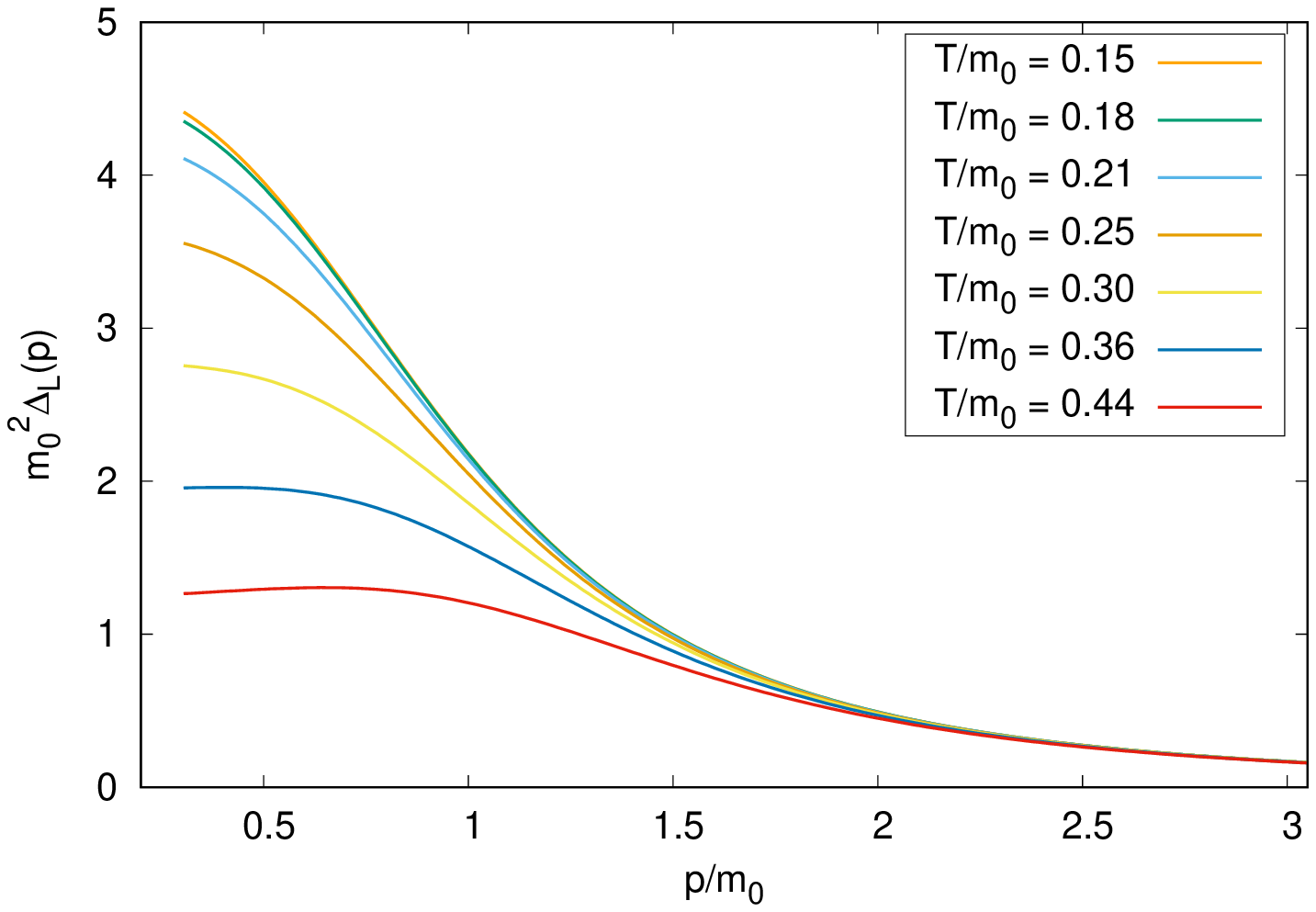}
\caption{Longitudinal propagator $\Delta_L$ in units of $m_0=m(0)$ at $\omega=0$ for the low temperature range $T/m_0<0.15$ (top) and the high temperature range $T/m_0>0.15$ (bottom). 
The renormalization constant and the mass parameter are fixed at their optimal $T=0$ values, $\pi_0(T)=\pi_{0}(0)=-0.876$
and $m(T)=m_0=656$~MeV. All the curves are multiplicatively renormalized at $\mu_{0}/m_{0}=6.098$ ($\mu_{0}=4$ GeV in physical units).}
\end{figure}

\begin{figure}[t] \label{fig:transprop}
  \centering
  \includegraphics[width=0.32\textwidth,angle=-90]{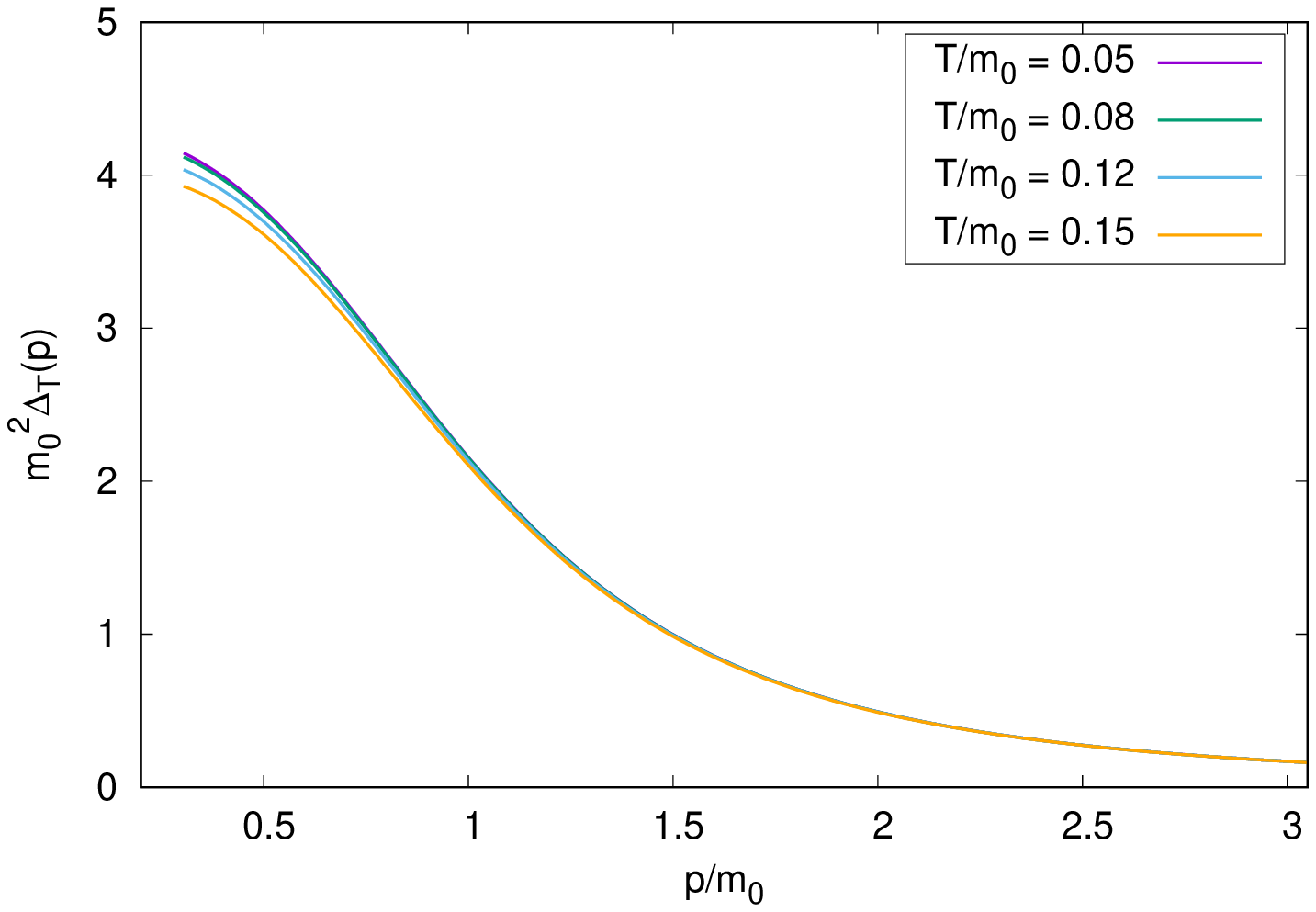}\quad\includegraphics[width=0.32\textwidth,angle=-90]{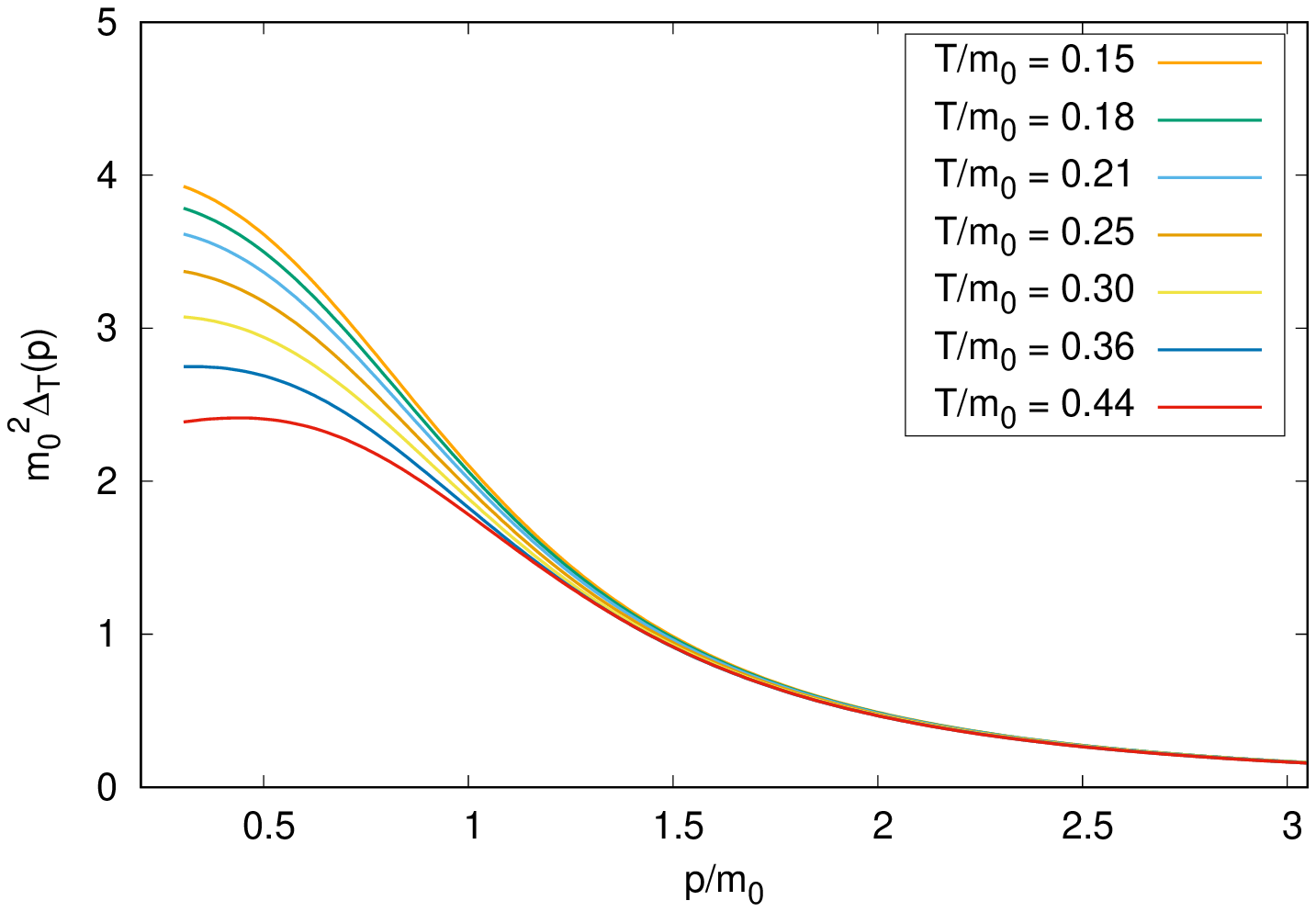}
\caption{Transverse propagator $\Delta_T$, with the same notation and parameters of Fig.~~2.}
\end{figure}

Let us first explore the behavior of the gluon propagators as a function of $T$ in the limit $\omega\to 0$,
where $p^2={\bf p}^2$, which is the most studied case on the lattice~\cite{silva,aouane}. The longitudinal and transverse 
propagators are shown in units of $m_0$ in Fig.~~2 and Fig.~~3, respectively. The former were multiplicatively renormalized by requiring that
\BE
\Delta_{L,T}(p,T)\Big|_{\omega=0,|{\bf p}|=\mu_{0}}=\frac{1}{\mu^{2}_{0}}
\EE
with $\mu_{0}/m_{0}=6.098$ (corresponding to $\mu_{0}=4$ GeV for $m_{0}=656$ MeV). We observe that, because of the chosen optimization, in the limit $T\to 0$ the longitudinal and transverse propagators coincide and reproduce the lattice
data extremely well~\cite{ptqcd,ptqcd2,xigauge,scaling,beta,beta2}, so that the low-temperature limit can be
regarded as exact. For reference, in Tab. I we report the physical equivalent of the adimensional temperatures $T/m_{0}$ used for the plots.
\begin{table}[h]
\def\arraystretch{1.3}
\begin{tabular}{|c|c|c|c|c|}
\hline
$T/m_{0}$&0.05&0.08&0.12&0.15\\
\hline
$T\ \text{(MeV)}$&32.80&52.48&78.72&98.40\\
\hline
\end{tabular}
\begin{tabular}{|c|c|c|c|c|c|c|}
\hline
$T/m_{0}$&0.18&0.21&0.25&0.30&0.36&0.44\\
\hline
$T\ \text{(MeV)}$&118.08&137.76&164.00&196.80&236.16&288.64\\
\hline
\end{tabular}
\caption{Dimensionful values of the adimensional temperatures $T/m_{0}$ plotted in Figs.~2 and 3, given $m_{0}=656$ MeV.}
\end{table}

We observe a crossover, in Fig.~~2, with the longitudinal propagator which increases in the IR for increasing $T$ below
$T_c\approx 0.15\cdot m_0$, but sharply decreases above $T_c$. This non-monotonic behavior is a well known feature
which has been reported by several lattice calculations~\cite{silva,aouane}.  
The transverse propagator in Fig.~~3, on the other hand, has a monotonic behavior, decreasing for increasing $T$, again in qualitative
agreement with the known predictions of the lattice. Actually, we cannot expect a quantitative agreement at $T\approx T_c$
or larger values, because we are extrapolating the optimization condition which was valid at $T=0$. Thus, the correct
qualitative behavior of the propagators at high temperature is an encouraging result. A crude estimate of $T_c$ 
is found by using the zero-temperature value $m_0=656$~MeV for restoring the energy units, yielding 
at the crossover $T_c\approx 100$~MeV. This value is quite smaller than the known transition temperature
$T_c\approx 270$~MeV which is measured on the lattice~\cite{lucini,silva,aouane}. The difference might well be the consequence
of a sub-optimal choice of the renormalization constant, but it could also arise from a change of the mass parameter
with temperature or from the more general failure of PT at high temperature. Thus, it becomes relevant to explore
whether a more quantitative agreement might be obtained by a tuning of the free parameters.

\subsection{Optimization by a fit of data at finite $\boldsymbol{T}$}

As the temperature increases, our previous assumption, $m(T)=m(0)$, $\pi_{0}(T)=\pi_{0}(0)$, becomes less valid. In what follows, we turn to fixing the optimal value of the parameters at $T\neq 0$ by a fit of the lattice data of Ref.~~~\cite{silva}. Since at non-zero temperatures the projections $\Delta_{L}(p,T)$ and $\Delta_{T}(p,T)$ have different behaviors with respect to a change in $T$, we may expect that the optimal values of the parameters will differ depending on which of the two components of the lattice propagator is used for the fit. This is indeed what we found. Of course, since in the subtracted Lagrangian of the present formalism the gluon mass parameter $m^{2}(T)$ is multiplied by the full four-dimensional transverse projector $t_{\mu\nu}(p)$, choosing different mass parameters/scales for the two components of the propagators is not allowed from first principles. This issue will be addressed at the end of this section.\\

\begin{figure}[t] \label{fig:longpropfit}
  \centering
  \includegraphics[width=0.32\textwidth,angle=-90]{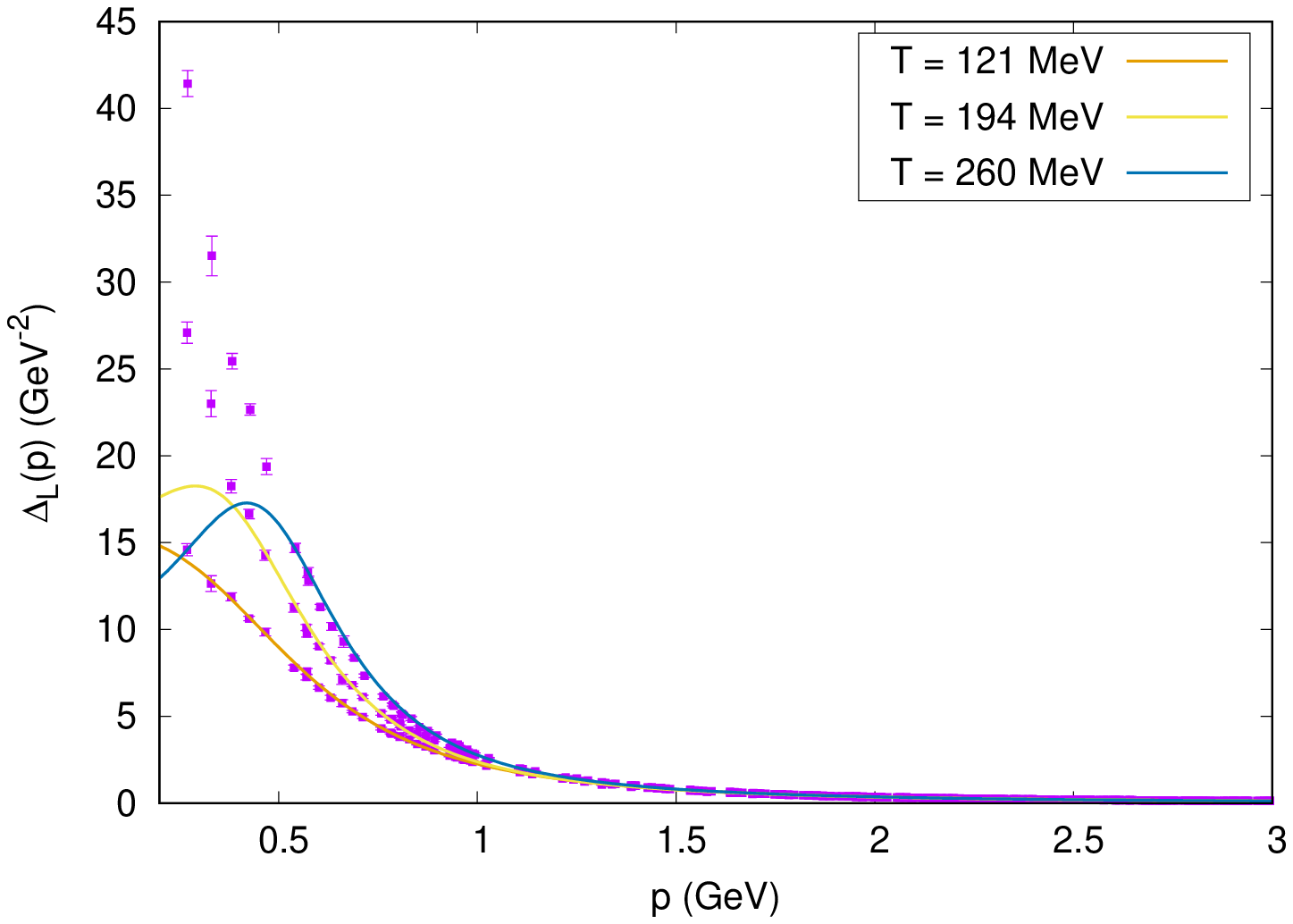}\quad\includegraphics[width=0.32\textwidth,angle=-90]{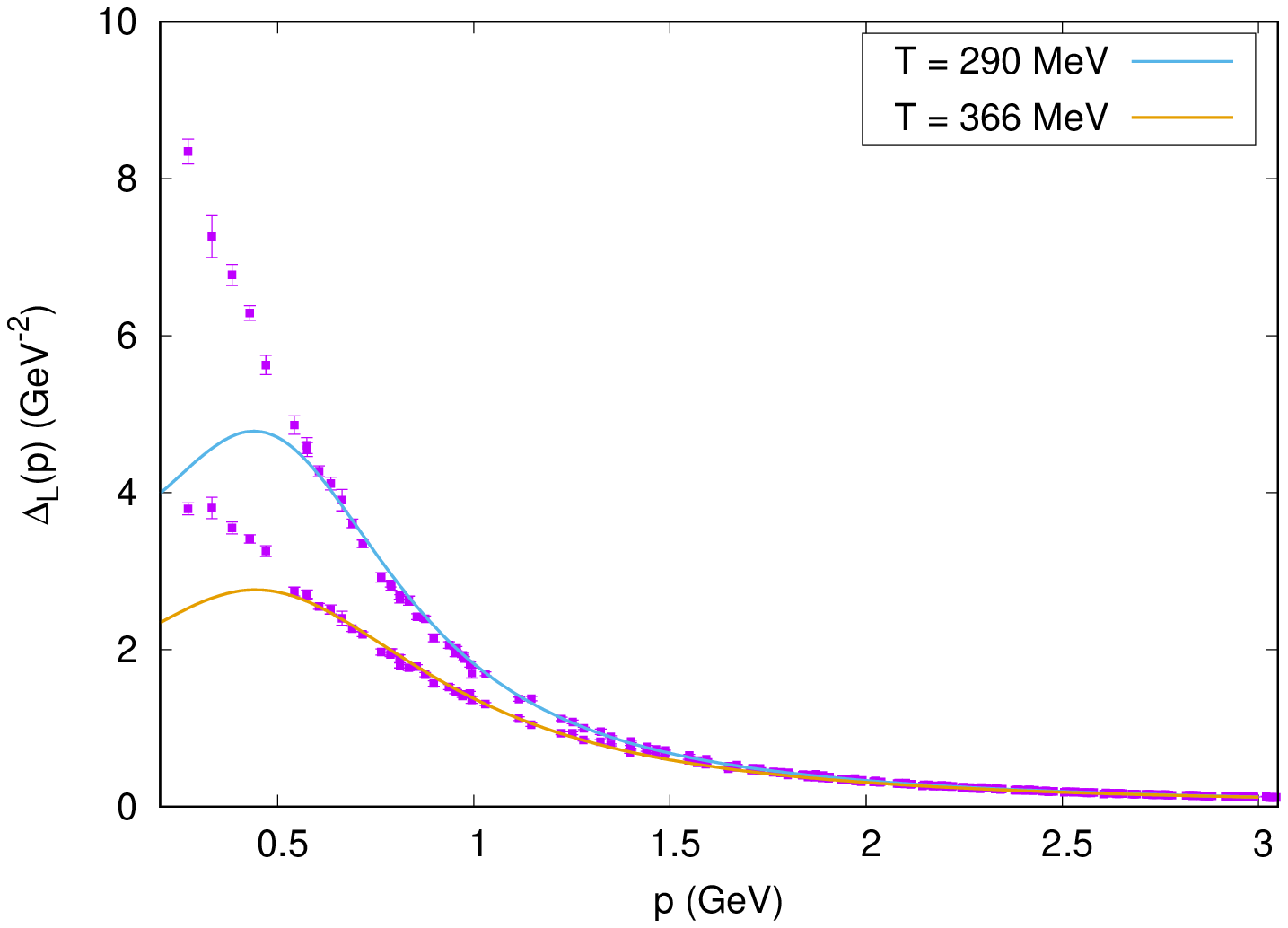}
\caption{Longitudinal propagator $\Delta_L$ at $\omega=0$ below (top) and above (bottom) the critical temperature $T_{c}\approx 270$ MeV. The curves are obtained using the parameters given in Tab.~II. The lattice data were taken from Ref.~~~\cite{silva}.}
\end{figure}
\begin{figure}[t] \label{fig:transpropfit}
  \centering
  \includegraphics[width=0.32\textwidth,angle=-90]{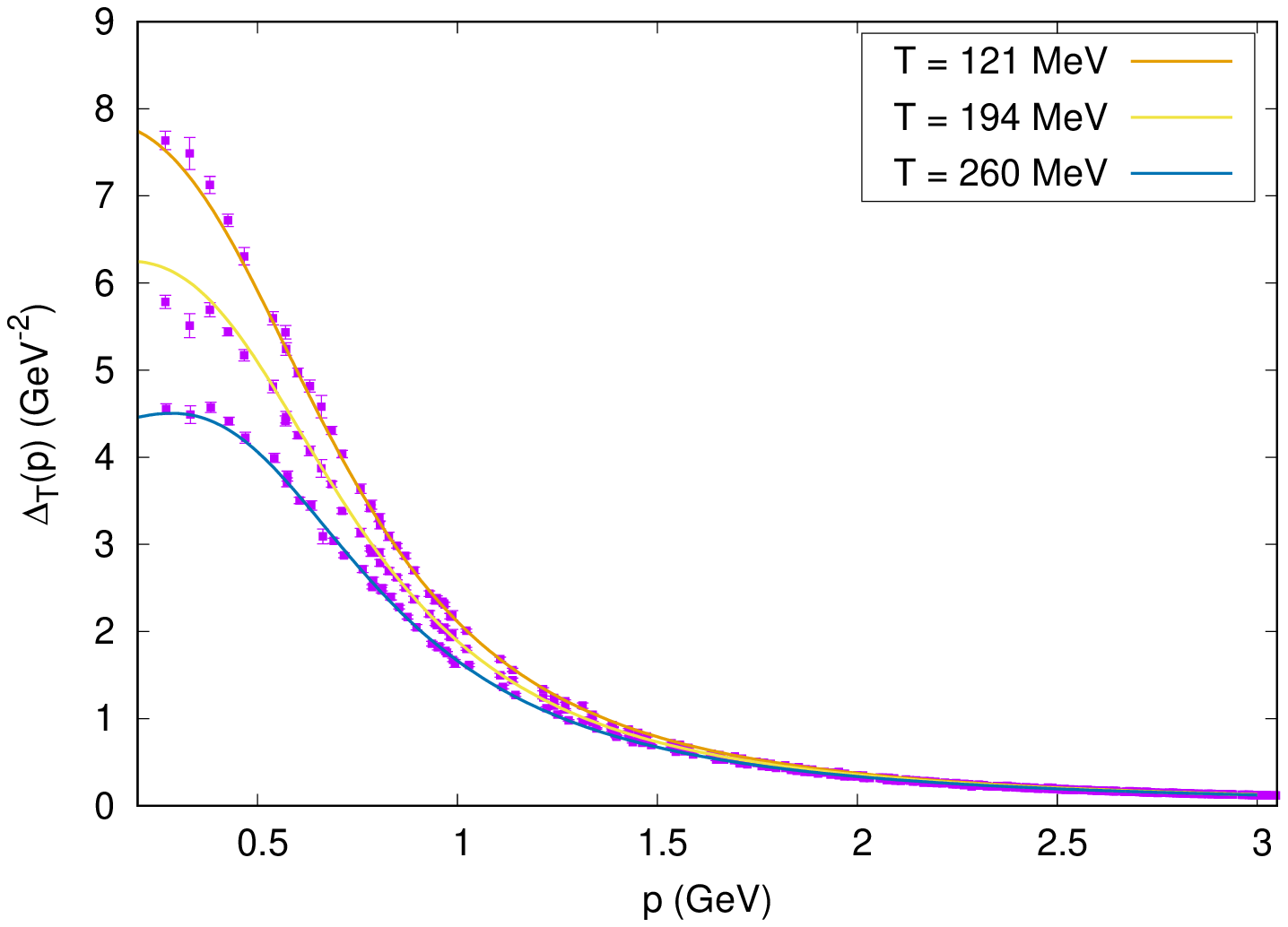}\quad\includegraphics[width=0.32\textwidth,angle=-90]{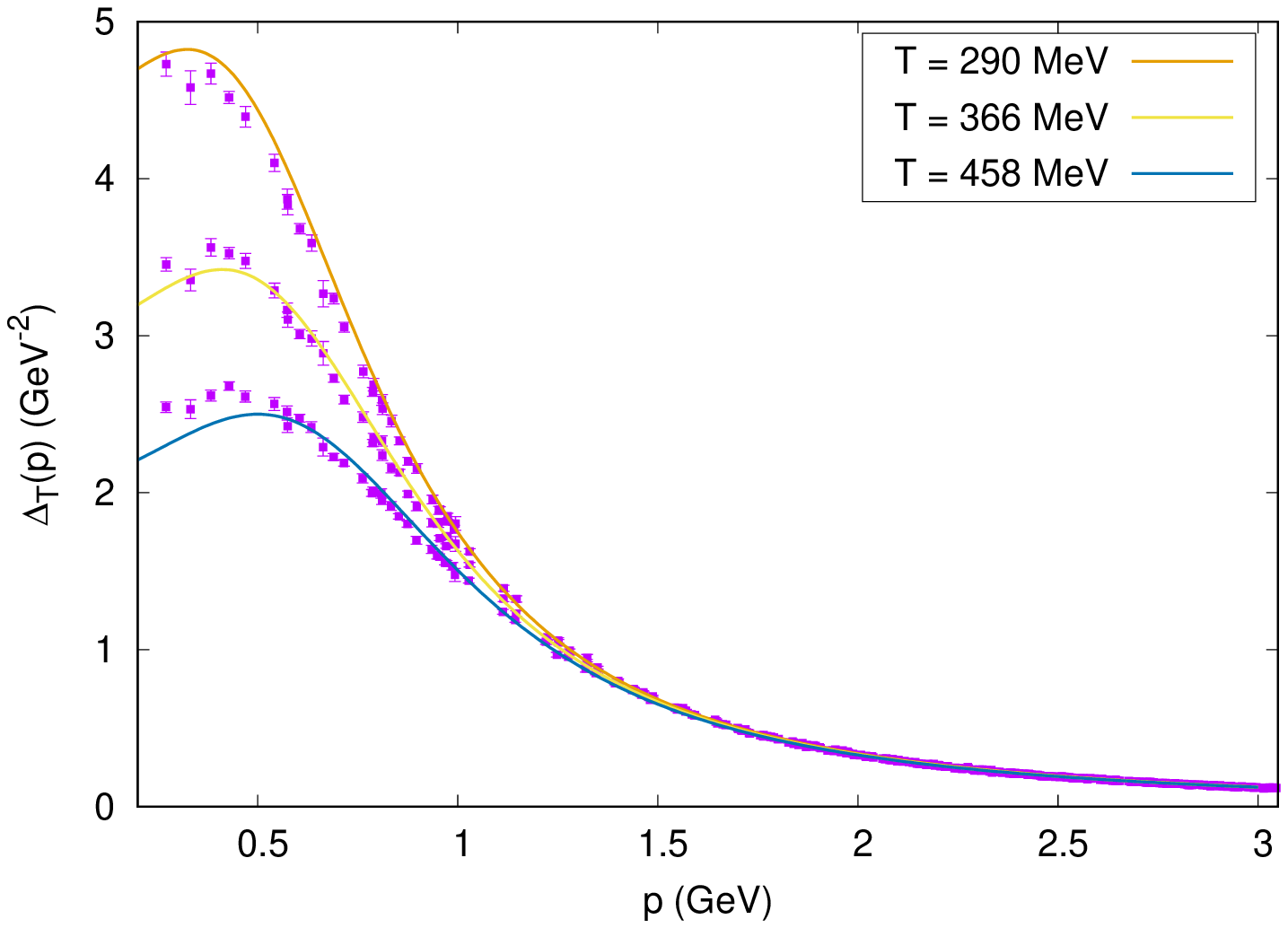}
\caption{Transverse propagator $\Delta_T$ at $\omega=0$ below (top) and above (bottom) the critical temperature $T_{c}\approx 270$ MeV. The curves are obtained using the parameters given in Tab.~II. The lattice data were taken from Ref.~~~\cite{silva}.}
\end{figure}
\begin{figure}[t] \label{fig:longpropfit458}
  \centering
  \includegraphics[width=0.32\textwidth,angle=-90]{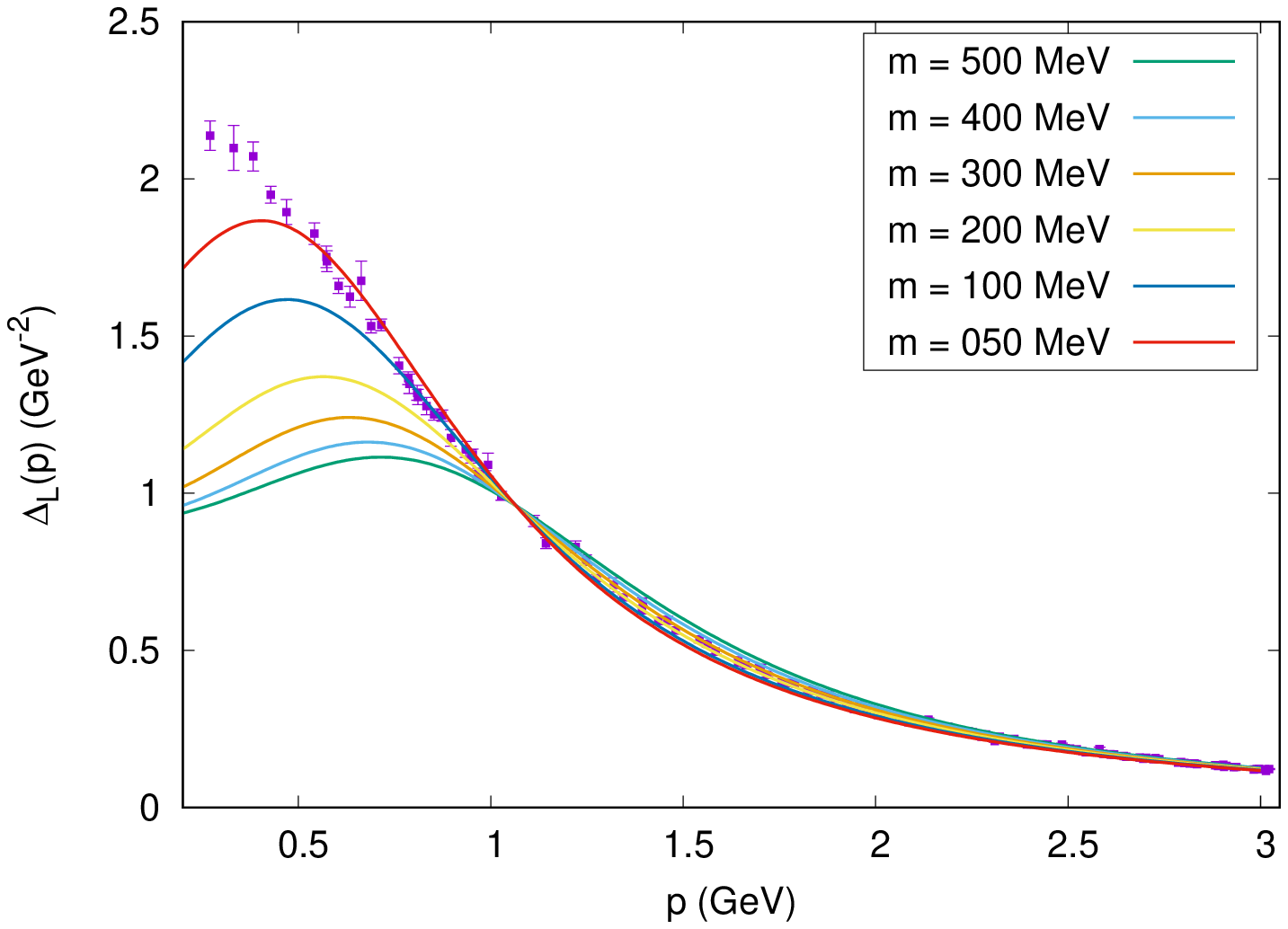}
\caption{Longitudinal propagator $\Delta_L$ for $\omega=0$, $T= 458$ MeV and different values of the gluon mass parameter. The lattice data were taken from Ref.~~~\cite{silva}.}
\end{figure}

In Figs.~4 and 5 we show, respectively, the longitudinal and transverse components of the gluon propagator at $\omega=0$ (multiplicatively renormalized at $\mu_{0}=4$~GeV), as functions of the three-dimensional momentum $|{\bf p}|=\sqrt{{\bf p}^{2}}$, with $m(T)$ and $\pi_{0}(T)$ as reported in Tab.~II. Such values where obtained by a separate fit of the two components to the lattice data of Ref.~~~\cite{silva}; the mass parameters should be understood to have an uncertainty of about $\pm 50$ MeV.

\begin{table}[h]
\def\arraystretch{1.3}
\begin{tabular}{|c|c|c|}
\hline
$T$ (MeV)&$m(T)$ (MeV) (long., trans.)&$\pi_{0}(T)$ (long., trans.)\\
\hline
121&550, 656&$-0.89$, $-0.84$\\
194&425, 550&$-1.10$, $-0.70$\\
260&425, 450&$-1.42$, $-0.42$\\
290&275, 450&$-0.97$, $-0.48$\\
366&150, 450&$-0.60$, $-0.20$\\
458&$\ \,$//, 450&$\ \ \ \ \,$//, $\, +0.21$\\
\hline
\end{tabular}
\caption{Parameters for the curves in Figs.~4 and 5, obtained by a separate fit of the lattice data for the longitudinal and transverse gluon propagator of Ref.~~~\cite{silva}.}
\end{table}
As we can see, once the parameters are tuned to fit the data, the screened expansion is able to reproduce the lattice propagators quite accurately down to momenta of approximately $0.5$ GeV. Moreover, the longitudinal propagator still shows the characteristic non-monotonic behavior with respect to a change in the temperature, increasing at fixed momentum below $T=T_{c}\approx 270$ MeV and decreasing above $T=T_{c}$.

Below $|{\bf p}|\approx0.5$ GeV, the transverse propagator is still in good agreement with the data, while the longitudinal one shows significant deviations, especially at high temperatures. In particular, from a qualitative standpoint, the longitudinal propagator shows an infrared turnover as a function of momentum which has no counterpart in the lattice data. From a numerical point of view, the difficulty in obtaining a good match with the data is exemplified in Fig.~6, where we display the longitudinal propagator for $T=458$ MeV and different values of the mass parameter\footnote{For each value of the mass parameter, the renormalization constant $\pi_{0}(T)$ was optimized so as to obtain the best fit with the data at large momenta.}. When tuning the mass parameter $m(T)$, there is a tension between the low- and intermediate-momentum behavior of the propagator: at lower values of $m$, the propagator is enhanced (resp. suppressed) below (resp. above) $|{\bf p}|\approx1$ GeV, so that achieving a good match at low momenta results in a loss of accuracy at intermediate momenta. This behavior is actually shared by both the components of the propagator and at every $T\neq 0$, albeit being less significant for the transverse component and at low temperatures. In particular, already at $T=458$~MeV the optimal longitudinal values of the mass parameter and of the renormalization constant strongly depend on the choice of a lower cutoff momentum for the fit to the lattice data; for this reason, we do not report them.

As anticipated earlier, the optimal mass parameters (and renormalization constants) needed to reproduce the lattice data differ for the two components of the propagator. In Fig.~~7 we plot the parameters of Tab.~II as functions of the temperature. With the exception of the point $T=260$ MeV, which is very close to the critical temperature $T_{c}\approx 270$ MeV, the optimal mass parameter $m(T)$ is a non-increasing function of the temperature for both the projections. When fitted from the transverse propagator, $m(T)$ shows plateaux both at small and at large temperatures, decreasing from $m(T)= m(0)=656$~MeV to $m(T)\approx 450$ MeV. As for the longitudinal propagator, except for $T=260$ MeV, $m(T)$ is approximately linear, with a behavior which is well-described by the equation
\BE
m(T)\approx 656\ \text{MeV}-1.307\ T\qquad\text{(long.)}.
\EE
At $T=260$ MeV $\approx T_{c}$, the optimal value of $m(T)$ is nearly equal for both the projections, namely $m(T)=425-450$ MeV. As for the renormalization constant, except for the point at $T=290$ MeV $\approx T_{c}$, the optimal $\pi_{0}(T)$ increases with the temperature when fitted from the transverse propagator. When optimized by the longitudinal propagator, on the other hand, it shows a non-monotonic behavior, decreasing below $T_{c}$ and increasing again above $T_{c}$.

\begin{figure}[t] \label{fig:mf0}
  \centering
  \includegraphics[width=0.32\textwidth,angle=-90]{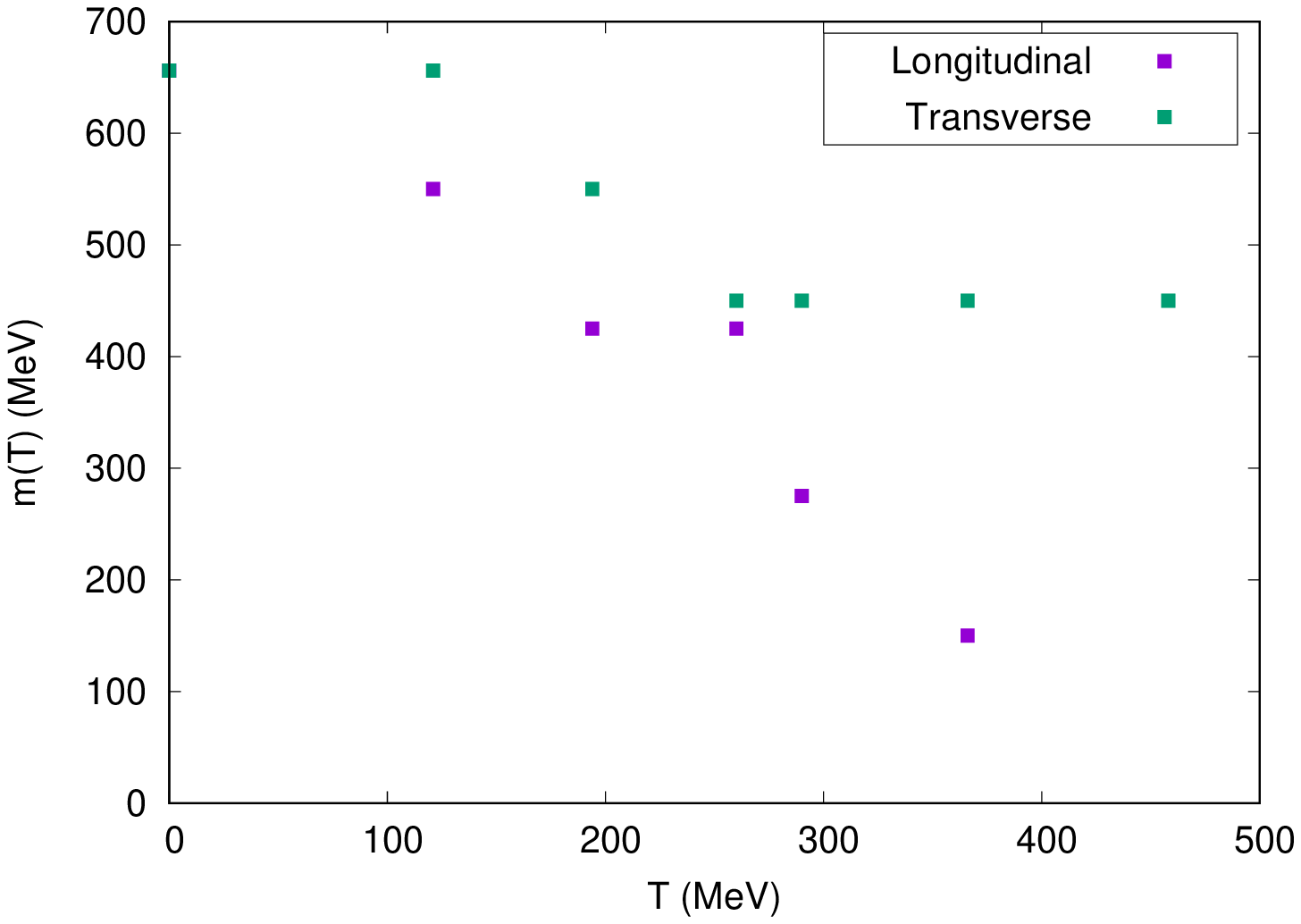}\quad\includegraphics[width=0.32\textwidth,angle=-90]{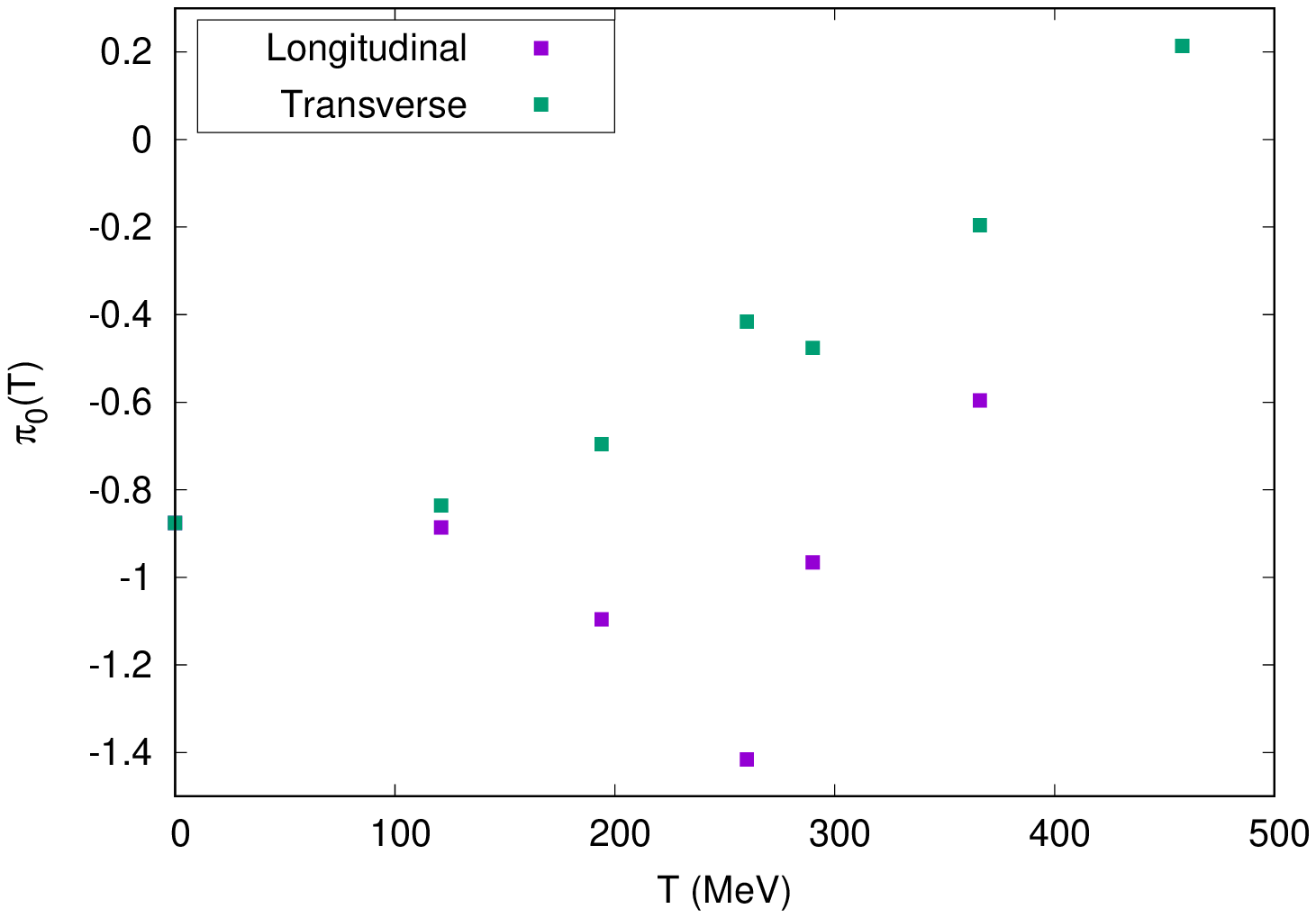}
\caption{Mass parameters (top) and renormalization constants (bottom) of Tab.~II, as extracted from the lattice data of Ref.~~~\cite{silva}.}
\end{figure}
\

The large differences in the optimal values of $m(T)$ and $\pi_{0}(T)$ obtained for the two projections make it clear that, in the present formalism, it is not possible to quantitatively recover both the longitudinal and the transverse component of the gluon propagator by a unique choice of parameters. Thus at $T\neq 0$ the screened expansion appears to be suboptimal as a ``variational'' ansatz. At least in part, this could be expected on the basis of what is known about the high-temperature, low-momentum behavior of the Yang-Mills propagators: at large temperatures and low momenta, the gluons' thermal mass is best described by a momentum- and direction-dependent Hard Thermal Loop (HTL) term in the Lagrangian, given by~\cite{braaten90b}
\BE
\Delta \mathcal{L}_{\text{HTL}}=-\frac{1}{2}\,m_{el}^{2}(T)\ \text{Tr}\left\{F_{\mu\nu}\int\frac{d\Omega}{4\pi}\frac{\hat{y}^{\nu}\hat{y}^{\lambda}}{(\hat{y}\cdot D)^{2}}\ F_{\lambda}^{\ \mu}\right\},
\EE
where $m_{el}^{2}(T)=g^{2}NT^{2}/3$, $\hat{y}$ is a light-like four-vector and the integration is over the directions of $\hat{y}$. To first order in the coupling, $\Delta\mathcal{L}_{\text{HTL}}$ generates two different thermal masses for the three-dimensional projections $\Delta_{L}(p,T)$ and $\Delta_{T}(p,T)$ of the gluon propagator. By not taking into account this difference, the screened expansion lends itself to a breakdown at large temperatures, which can be partially avoided if the mass parameter and renormalization constant are tuned to separately fit the two projections.

The simplest way of solving this issue in the context of the screened expansion, i.e. without resorting to a HTL resummation, would be to change the expansion point of perturbation theory in such a way that the two three-dimensional projections of the \textit{zero-order} gluon propagator, $\Delta_{m}^{T}$ and $\Delta_{m}^{L}$, have different masses \textit{ab initio}. This can be achieved by redefining the kernel $\delta\Gamma_{\mu\nu}(p;T)=m^{2}(T)\,t_{\mu\nu}(p)$ of the shift of the action $\delta S$ as
\BE
\delta\Gamma_{\mu\nu}(p;T)\to m_{T}^{2}(T)\,P^{T}_{\mu\nu}(p)+m_{L}^{2}(T)\,P^{L}_{\mu\nu}(p),
\EE
where $m_{T}(T)$ and $m_{L}(T)$ are independent mass-parameter functions for the two projections. With such a prescription, in a general covariant gauge the zero-order Euclidean gluon propagator $\Delta_{m}^{\mu\nu}(p;T)$ would read
\begin{align}\label{newshift}
\nn\Delta_{m}(p;T)_{\mu\nu}&\to \Delta_{m}^{T}(p;T)\ P^{T}_{\mu\nu}(p)+\Delta_{m}^{L}(p;T)\ P^{L}_{\mu\nu}(p)+\\
&\quad+\frac{\xi}{p^{2}}\ \ell_{\mu\nu}(p),
\end{align}
where
\BE
\Delta_{m}^{T,L}(p;T)=\frac{1}{p^{2}+m_{T,L}^{2}(T)}
\EE
are the sought-after zero-order propagators. Setting-up the perturbation theory with independent mass functions for the two projections would give us the freedom to optimize the former separately from first principles, according to the behavior of the respective dressed propagators. Implementing the shift in Eq.~~\eqref{newshift}, however, is a non-trivial task: having different longitudinal and transverse masses running in the loops breaks the Lorentz-invariance even of the simplest vacuum integrals and, more generally, requires a complete recalculation of the gluon polarization.

\section{Dispersion relations at finite T}

Being in possession of analytical expressions (modulo a one-dimensional integration at finite $T$) for the Euclidean gluon propagator allows us to analytically continue the latter to the whole complex plane so as to study its singularities. As is well known, the location of the poles of the propagator gives us information on the dispersion relations of the gluonic quasi-particles: the energy $\varepsilon_{T,L}({\bf p},T)$ and damping rate $\gamma_{T,L}({\bf p},T)$ of the quasi-particles, as functions of the three-dimensional momentum ${\bf p}$ and of the temperature $T$, are obtained by solving the equation
\BE
\Delta_{T,L}^{-1}(-i\omega_{T,L}({\bf p},T),{\bf p},T)=0,
\EE
where $\omega=\varepsilon-i\gamma$ (modulo a factor of $i$) extends the real and discrete Matsubara frequencies $\omega_{n}=2\pi n T$ to the complex plane and the subscripts $T,L$ refer to the components of the propagator. At non-zero temperatures and momenta, the poles of the two components are expected to be found at different locations, yielding two separate branches of the dispersion relations.

The limit $T\to 0$ of the dispersion relations was already studied in the framework of the screened massive expansion in Refs.~~\cite{analyt,scaling,xigauge}. In~\cite{xigauge} we found that the zero-temperature gluon propagator (whose longitudinal and transverse three-dimensional components are constrained to be equal by Lorentz simmetry) has two complex-conjugate poles at $-p^{2}=m_{\text{pole}}^{2},\,(m_{\text{pole}}^{2})^{*}$, where, setting $m_{0}=656$ MeV by 
sharing the same units of the lattice,
\BE
m_{R}^{2}=0.197\ \text{GeV}^{2}\ ,\qquad m_{I}^{2}=0.436\ \text{GeV}^{2},
\EE
with $m_{\text{pole}}^{2}=m_{R}^{2}+i\,m_{I}^{2}$. In terms of $\varepsilon_{\text{vac}}({\bf p})=\lim_{T\to 0}\varepsilon_{T,L}({\bf p},T)$ and $\gamma_{\text{vac}}({\bf p})=\lim_{T\to 0}\gamma_{T,L}({\bf p},T)$ -- and singling out one of the poles --, this translates into the dispersion relations
\begin{widetext}
\begin{align}\label{vacuumdisp}
\nn\varepsilon_{\text{vac}}({\bf p})&=\left[\frac{1}{2}\ \sqrt{({\bf p}^{2}+m_{R}^{2})^{2}+(m_{I}^{2})^{2}}+\frac{1}{2}\,({\bf p}^{2}+m_{R}^{2})\right]^{1/2}\ ,\\
\gamma_{\text{vac}}({\bf p})&=\left[\frac{1}{2}\ \sqrt{({\bf p}^{2}+m_{R}^{2})^{2}+(m_{I}^{2})^{2}}-\frac{1}{2}\,({\bf p}^{2}+m_{R}^{2})\right]^{1/2}\ .
\end{align}
\end{widetext}
Clearly, $m_{R}^{2}=(\varepsilon_{\text{vac}}^{2}-\gamma_{\text{vac}}^{2})|_{{\bf p}=0}$ and $m_{I}^{2}=2\,\varepsilon_{\text{vac}}\,\gamma_{\text{vac}}|_{{\bf p}=0}$, where
\BE
\varepsilon_{\text{vac}}({\bf 0})=581\ \text{MeV}\ ,\qquad\gamma_{\text{vac}}({\bf 0})=375\ \text{MeV}\ .
\EE
At the other end of the spectrum, as $|{\bf p}|\to \infty$, the gluon's vacuum dispersion relations reduce to those of a massless particle, $\varepsilon_{\text{vac}}({\bf p})\to |{\bf p}|$, $\gamma_{\text{vac}}({\bf p})\to 0$.

Under the assumption that the optimal masses $m(T)$ and renormalization constants $\pi_{0}(T)$ reported in the previous section only depend on the temperature, and not on the Matsubara frequency $\omega_{n}$, the finite-$T$ dispersion relations of the gluon quasi-particles can be easily extracted from the screened expansion's gluon propagator, making use of said parameters (cf. Tab.~II). We remark that, since at low momenta the longitudinal projection was not found to be in good agreement with the lattice data for any value of the parameters, the longitudinal dispersion relations are expected to be reliable only at sufficiently high momenta (say above $|{\bf p}|\approx0.5-0.7$ GeV).

In Figs.~8 and 9 we plot the energy $\varepsilon_{T,L}({\bf p},T)$ and the damping rate $\gamma_{T,L}({\bf p},T)$ of the transverse and longitudinal gluons at fixed $T$, as functions of the momentum $|{\bf p}|$. As we can see, below the critical temperature $T_{c}\approx 270$~MeV both the transverse energy and the transverse damping rate (Fig.~~8) are suppressed with respect to their zero-temperature (vacuum) limit, with the effect being more pronounced for $\varepsilon_{T}$ than for $\gamma_{T}$. Above $T_{c}$ this behavior is reversed; the transverse energy starts to approach again its vacuum limit, while the damping rate grows larger than it. The longitudinal branch (Fig.~~9) shows a more significant suppression in both the energy and the damping rate below $T_{c}$, with $\gamma_{L}$ becoming quite small at high momenta around the critical temperature. At higher temperatures both $\varepsilon_{L}$ and $\gamma_{L}$ start to approach back their vacuum limit.\footnote{Here we are disregarding the low-momentum behavior of the longitudinal dispersion relations due to their lack of reliability, as previously discussed.}\\

\begin{figure*} \label{figdispersion1}
  \centering
  \includegraphics[width=0.30\textwidth,angle=-90]{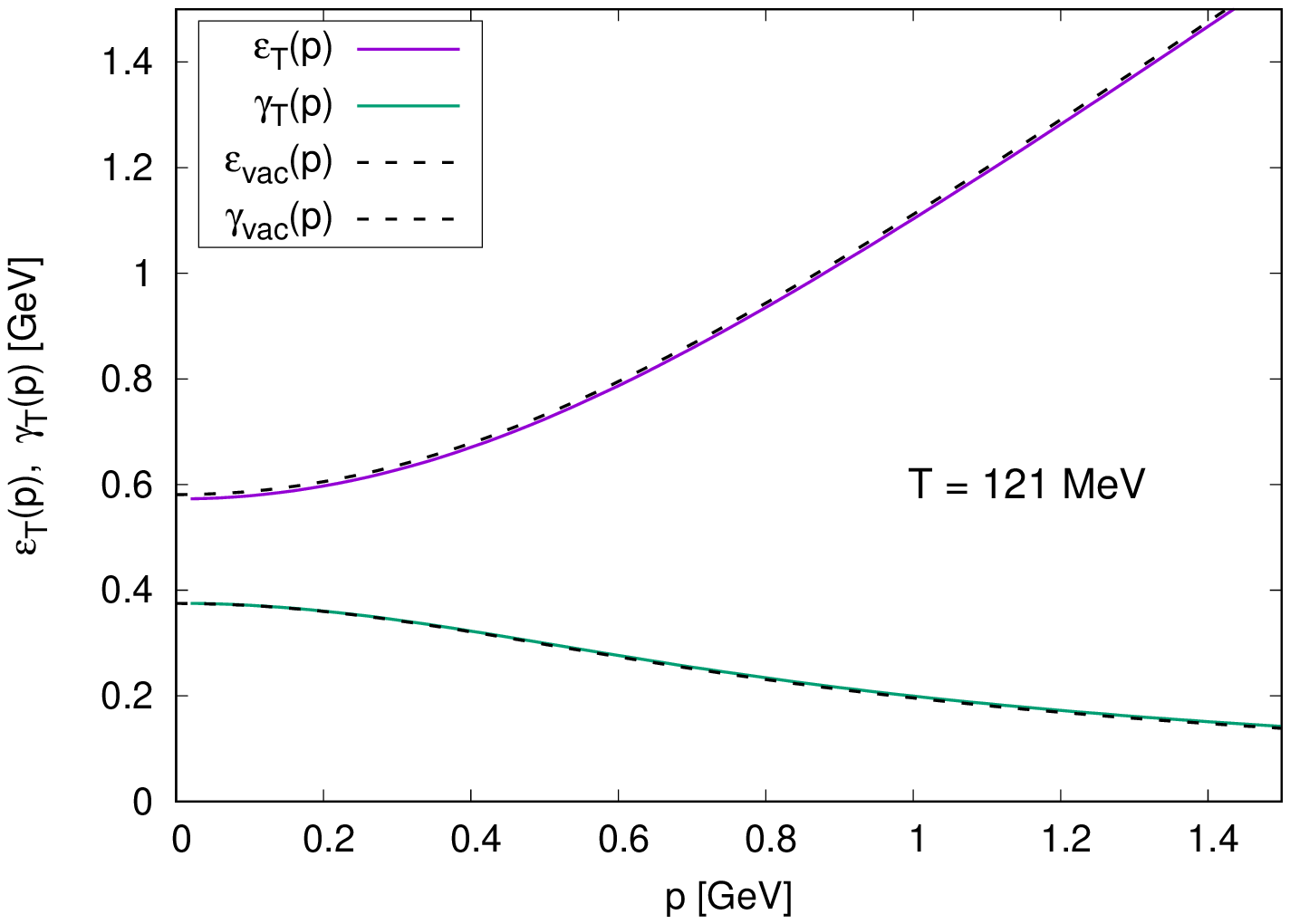}\qquad\ \includegraphics[width=0.30\textwidth,angle=-90]{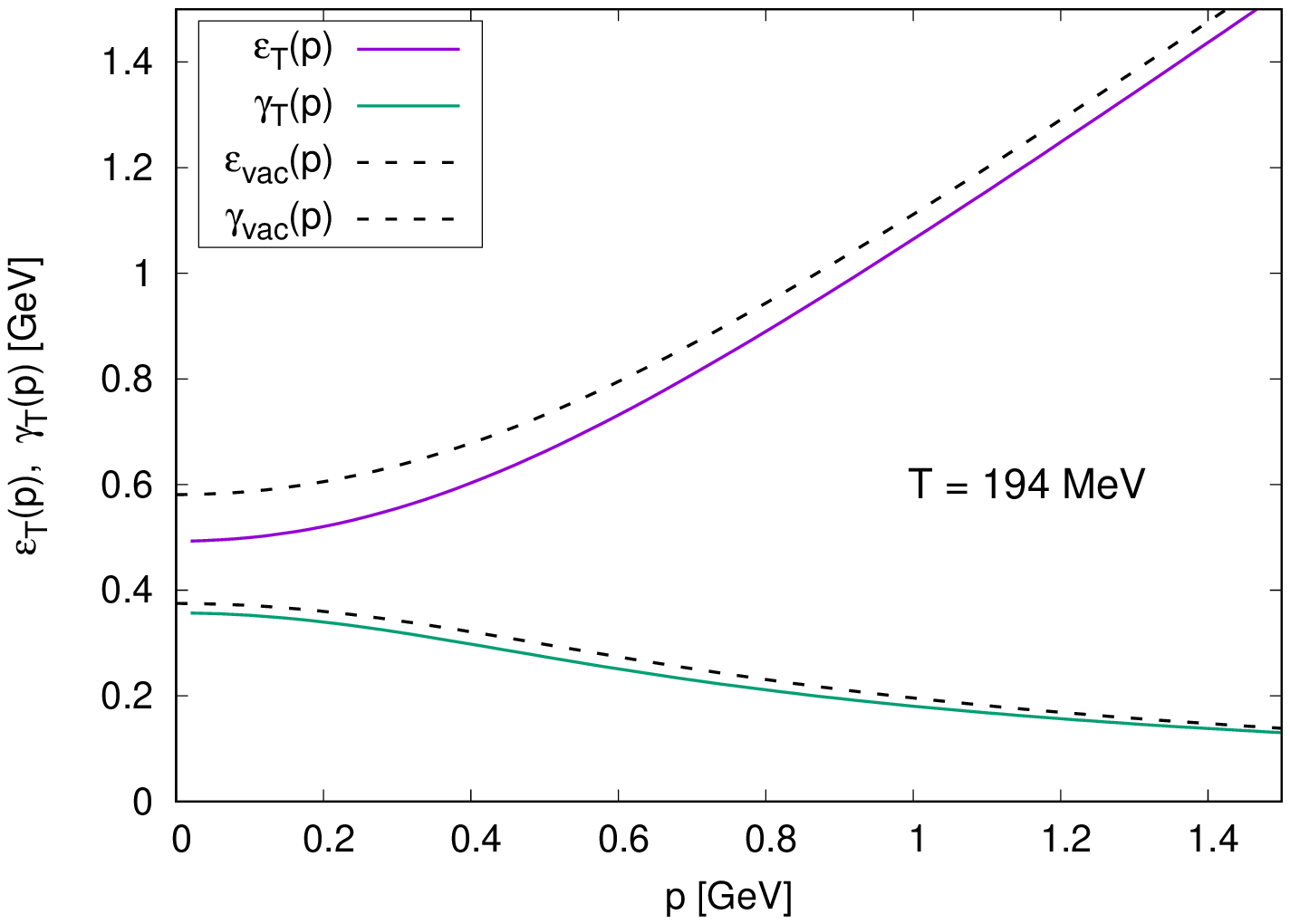}\\
  \includegraphics[width=0.30\textwidth,angle=-90]{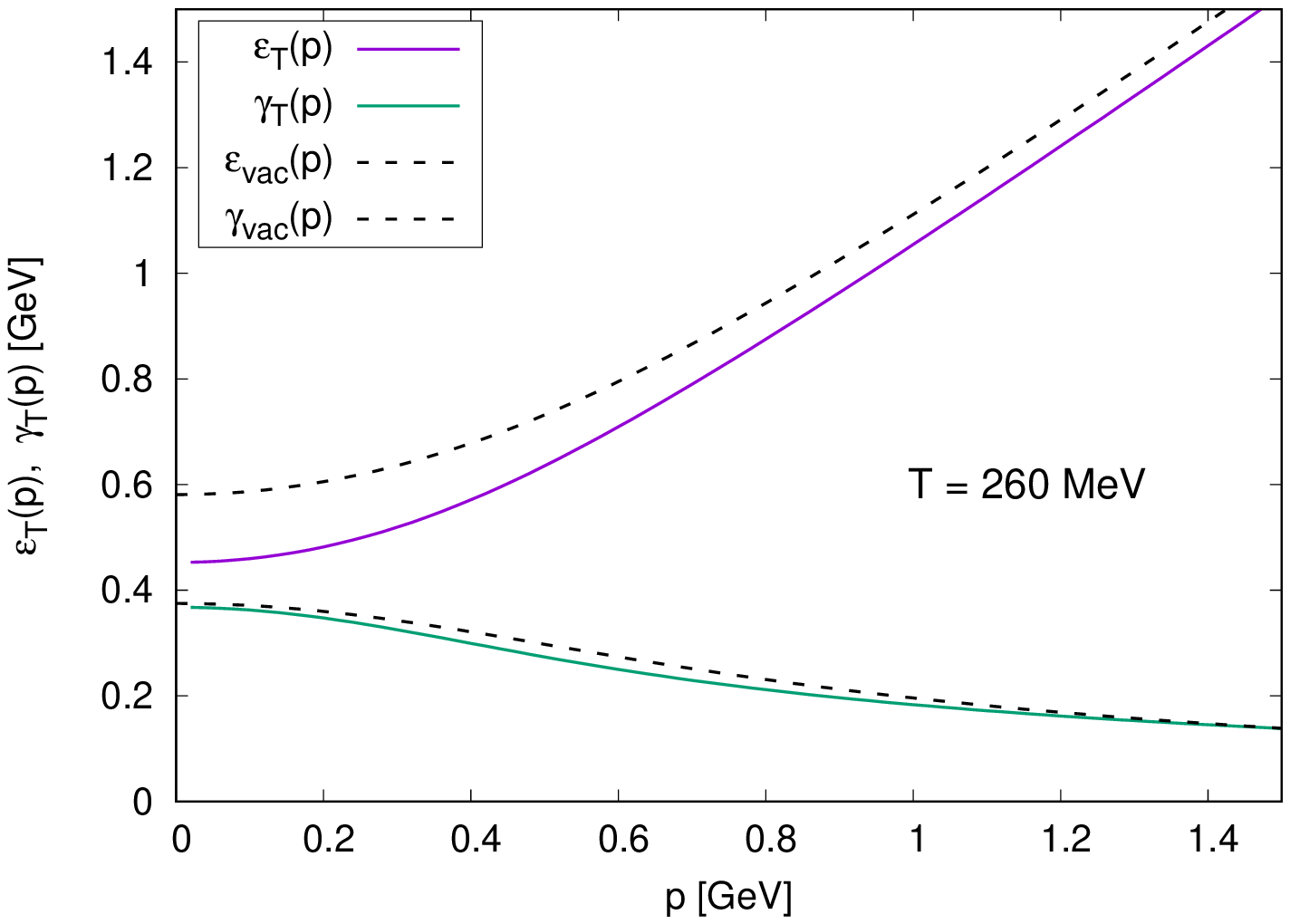}\qquad\ \includegraphics[width=0.30\textwidth,angle=-90]{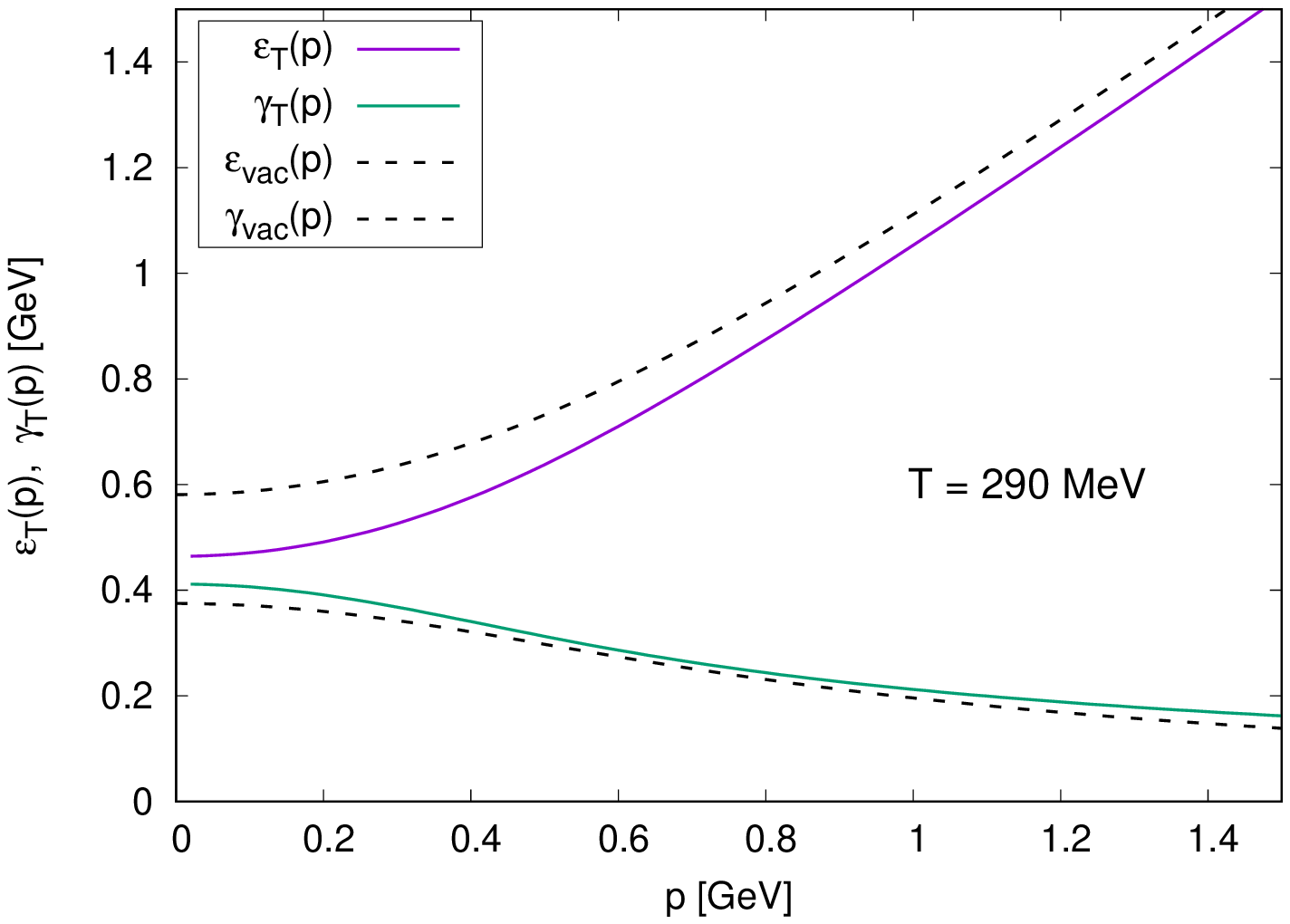}\\
  \includegraphics[width=0.30\textwidth,angle=-90]{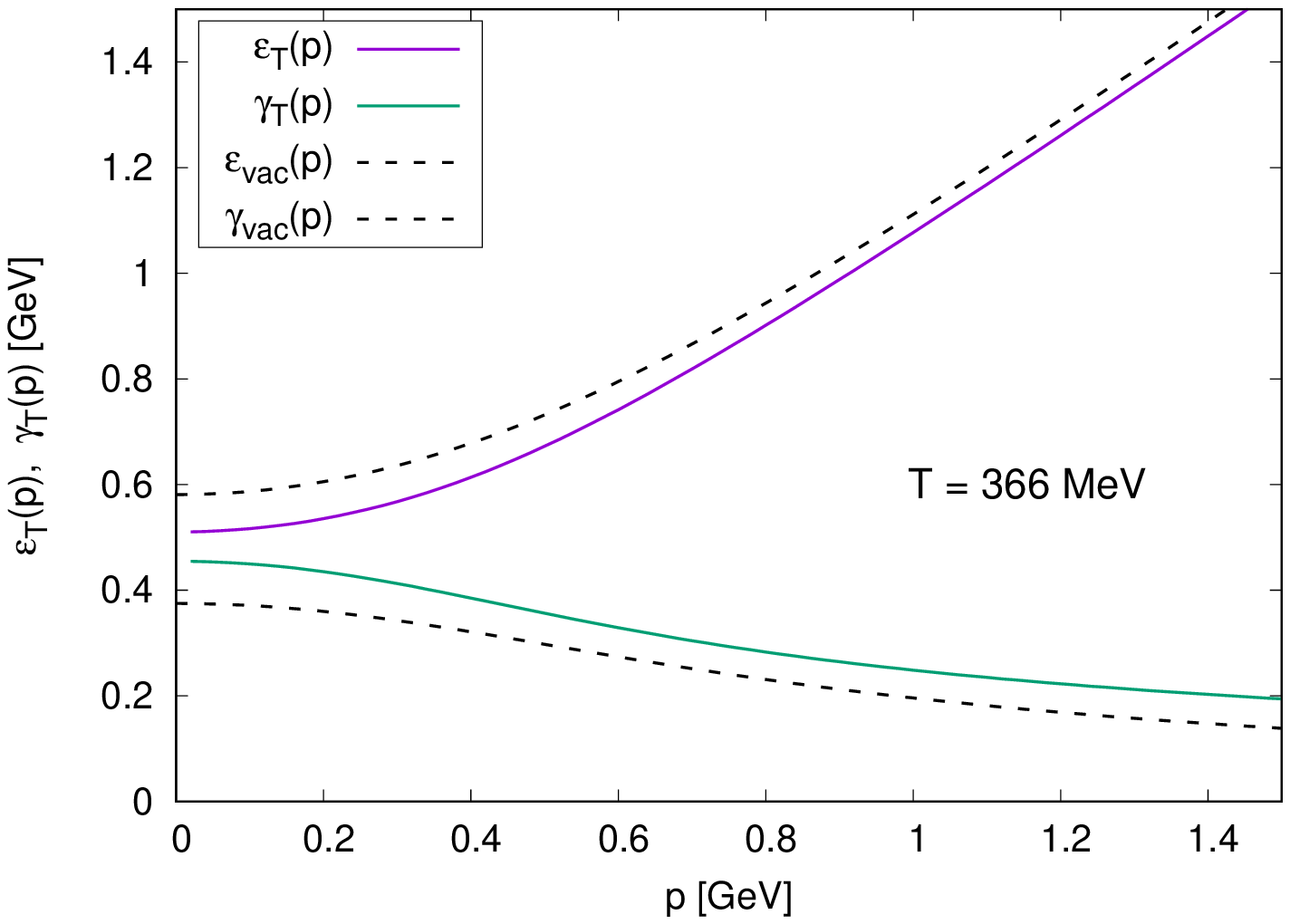}\qquad\ \includegraphics[width=0.30\textwidth,angle=-90]{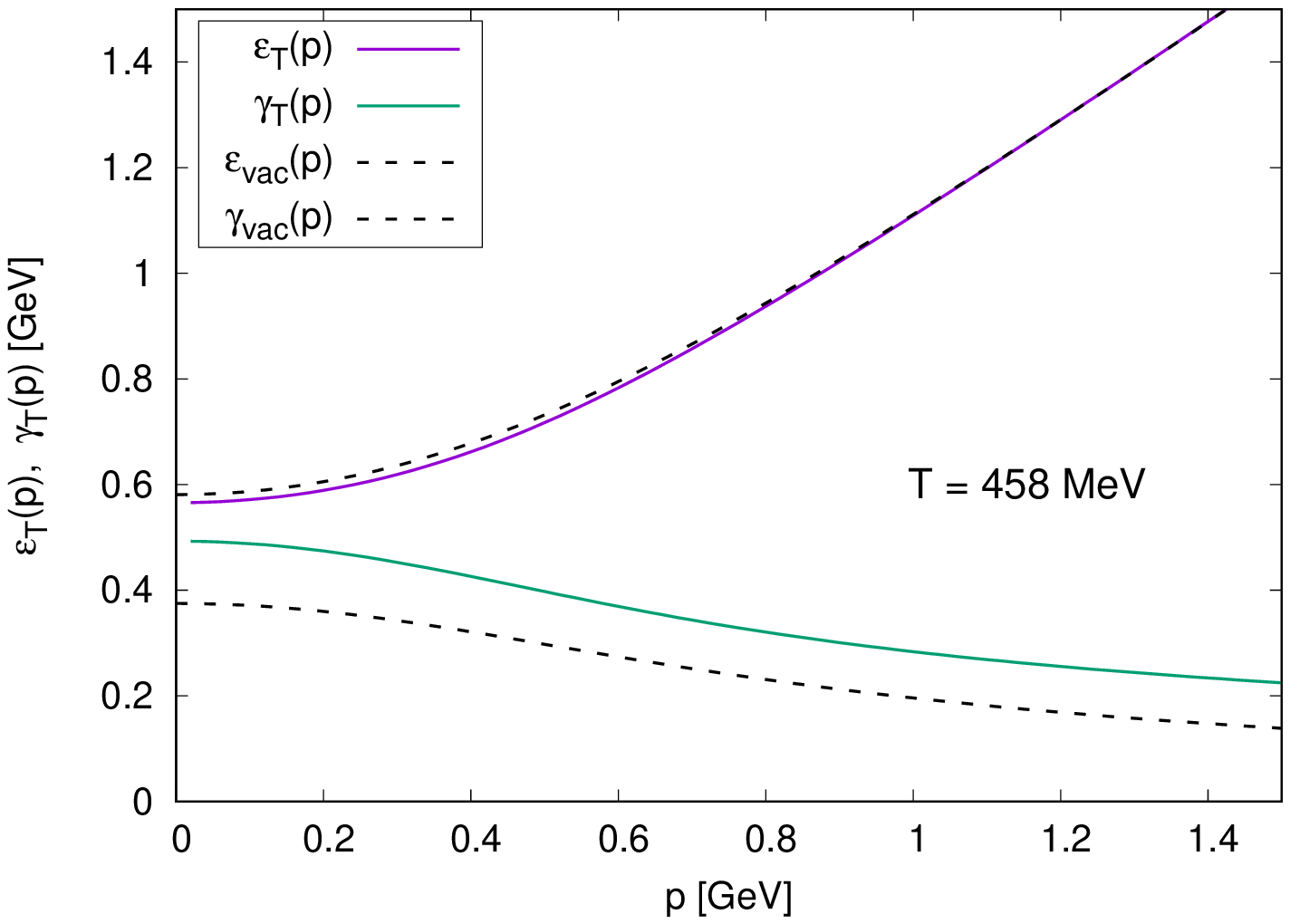}
\caption{Transverse dispersion relations for the gluon quasi-particles. The broken lines are the vacuum dispersion relations, common to both projections and given by Eq.~~\eqref{vacuumdisp}. The gluon mass parameters $m(T)$ and renormalization constants $\pi_{0}(T)$ used for the plots are reported in Tab.~II.}
\end{figure*}
\begin{figure*} \label{figdispersion2}
  \centering
    \includegraphics[width=0.30\textwidth,angle=-90]{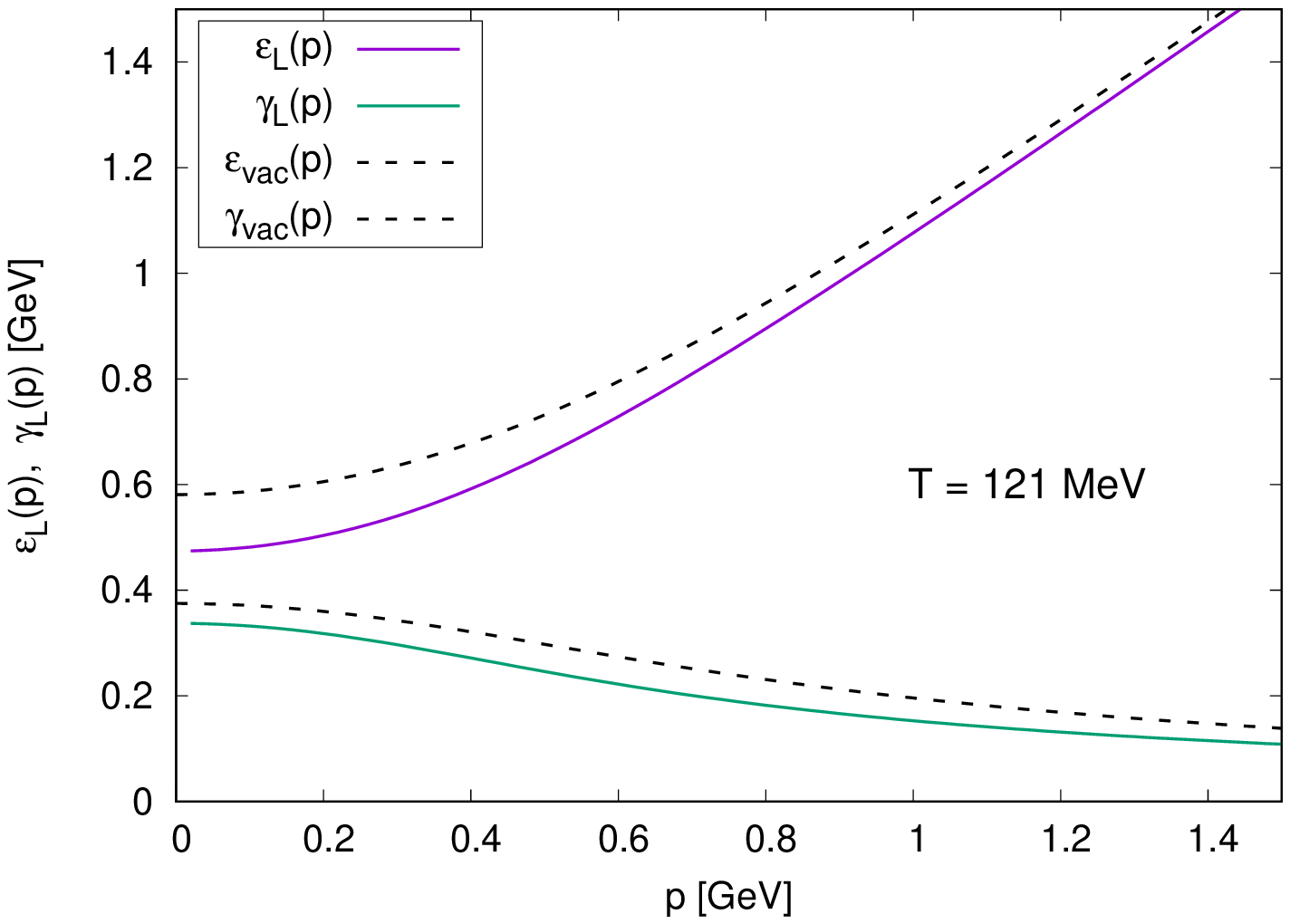}\qquad\ \includegraphics[width=0.30\textwidth,angle=-90]{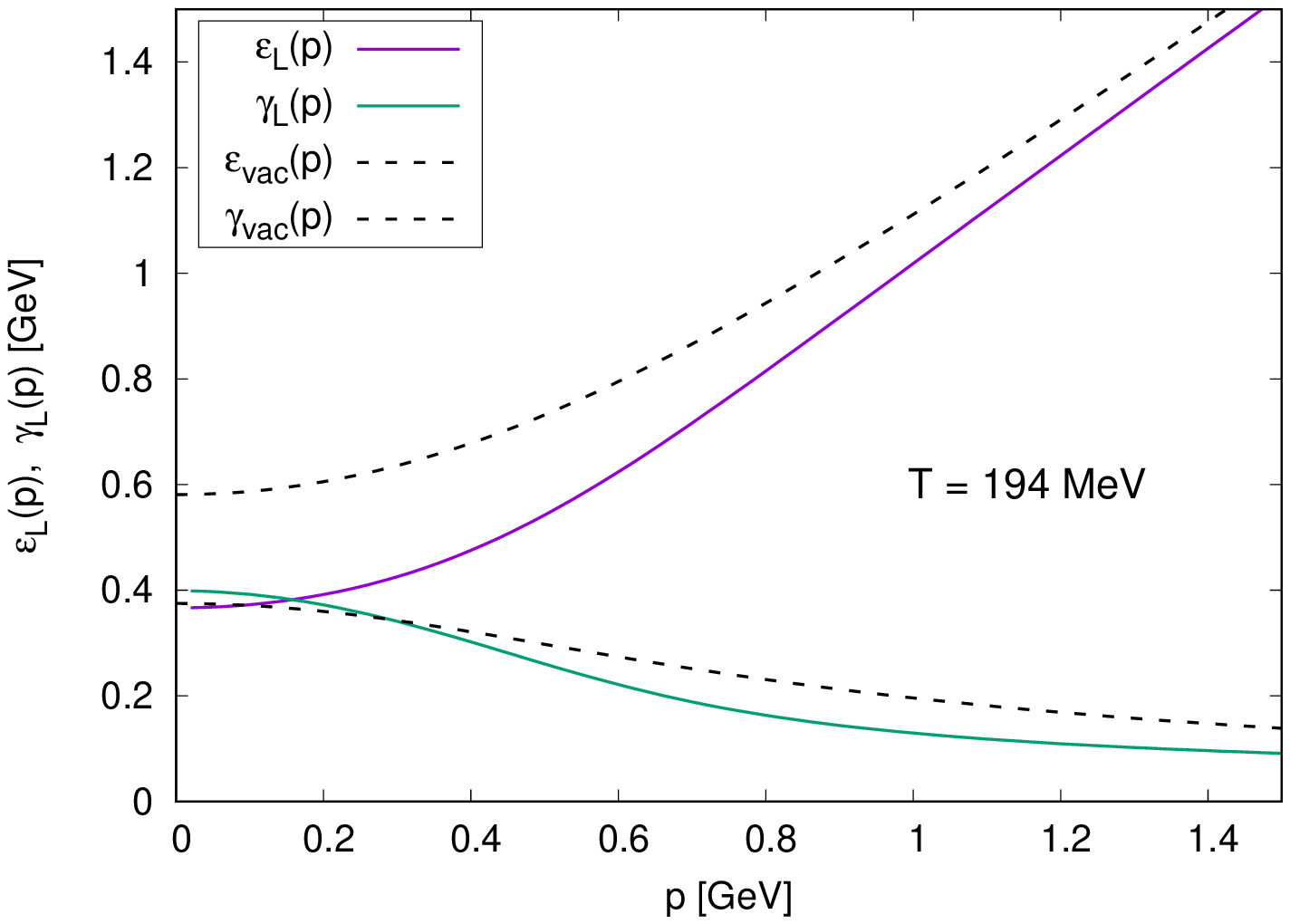}\\
  \includegraphics[width=0.30\textwidth,angle=-90]{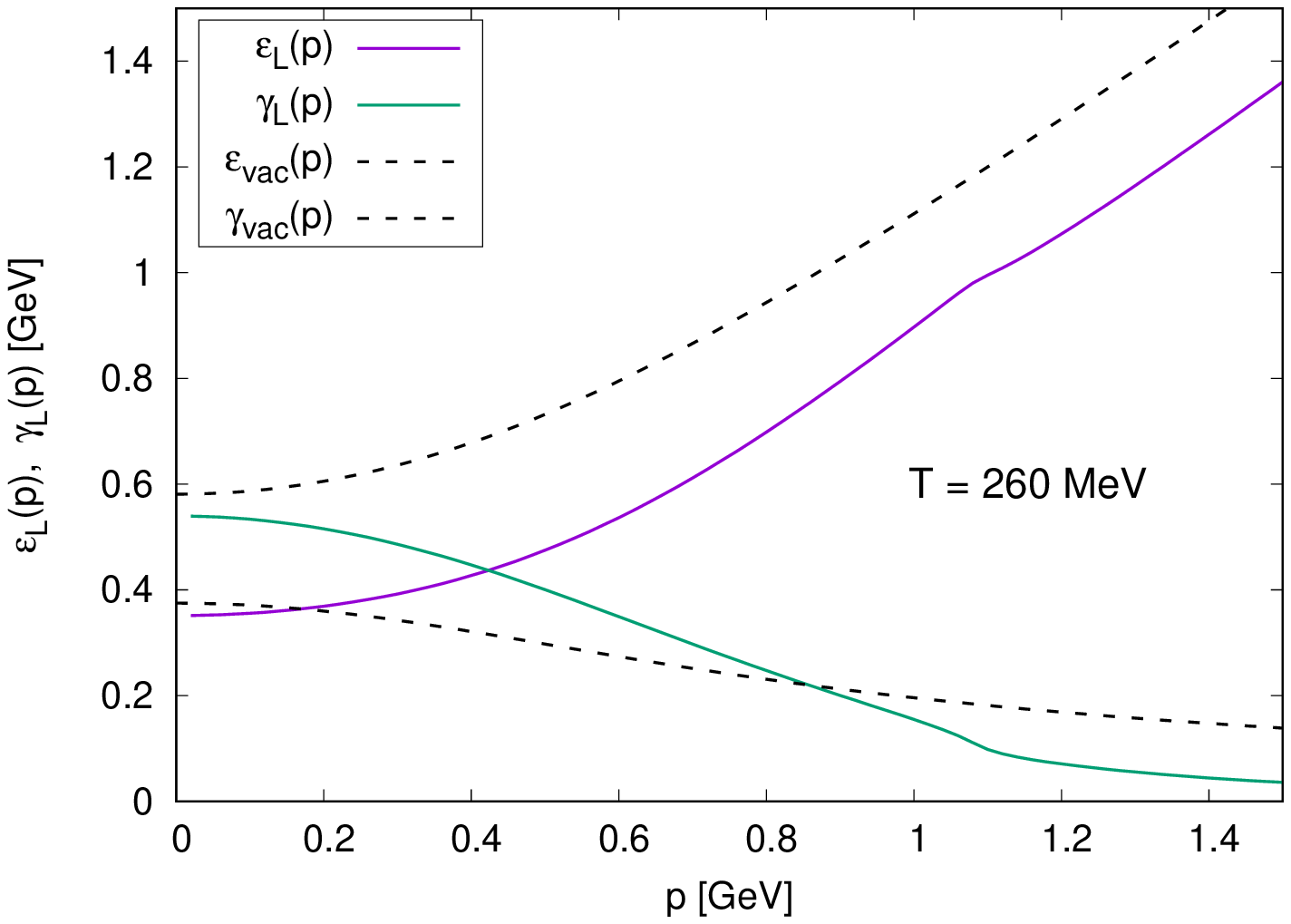}\qquad\ \includegraphics[width=0.30\textwidth,angle=-90]{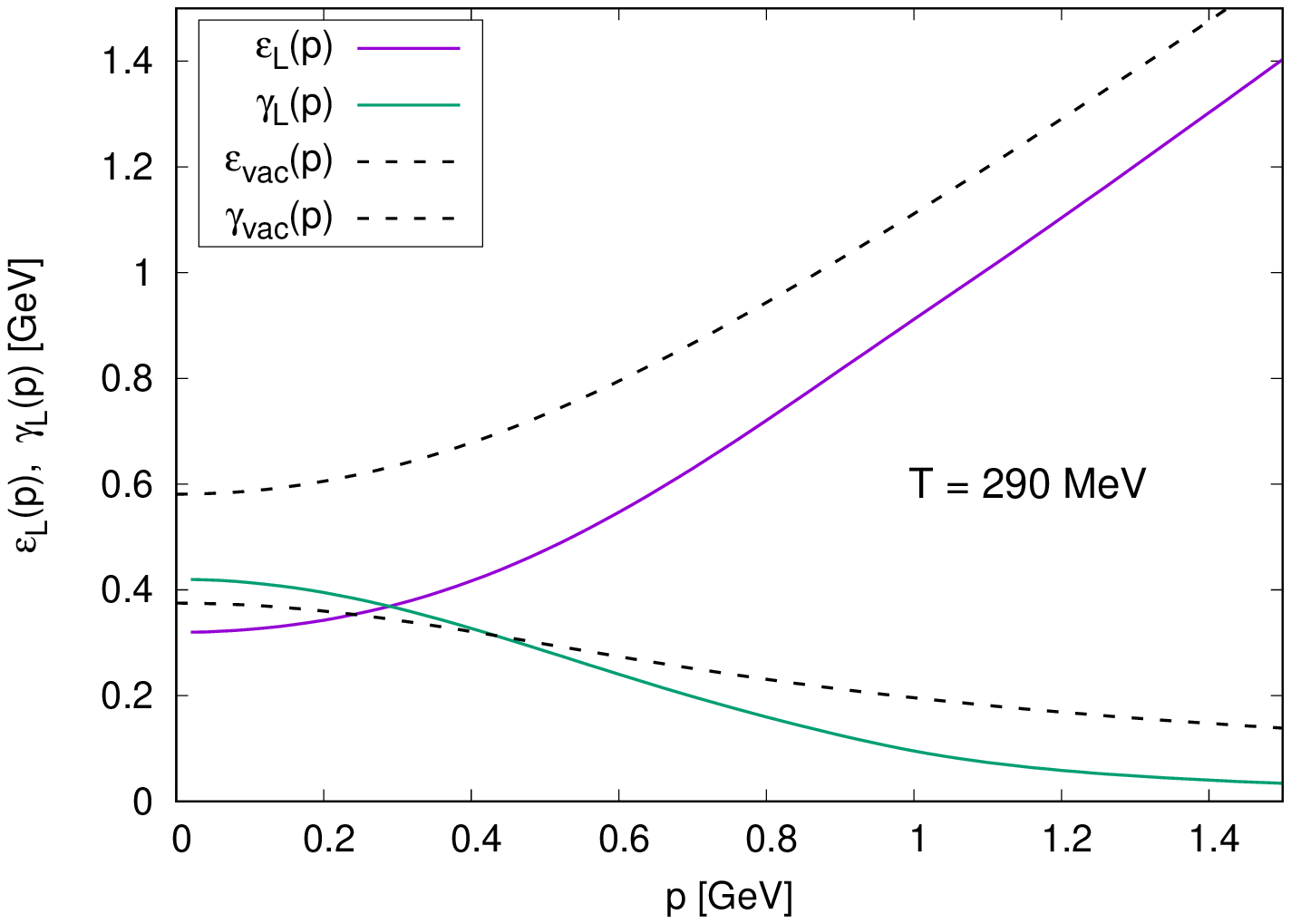}\\
 \includegraphics[width=0.30\textwidth,angle=-90]{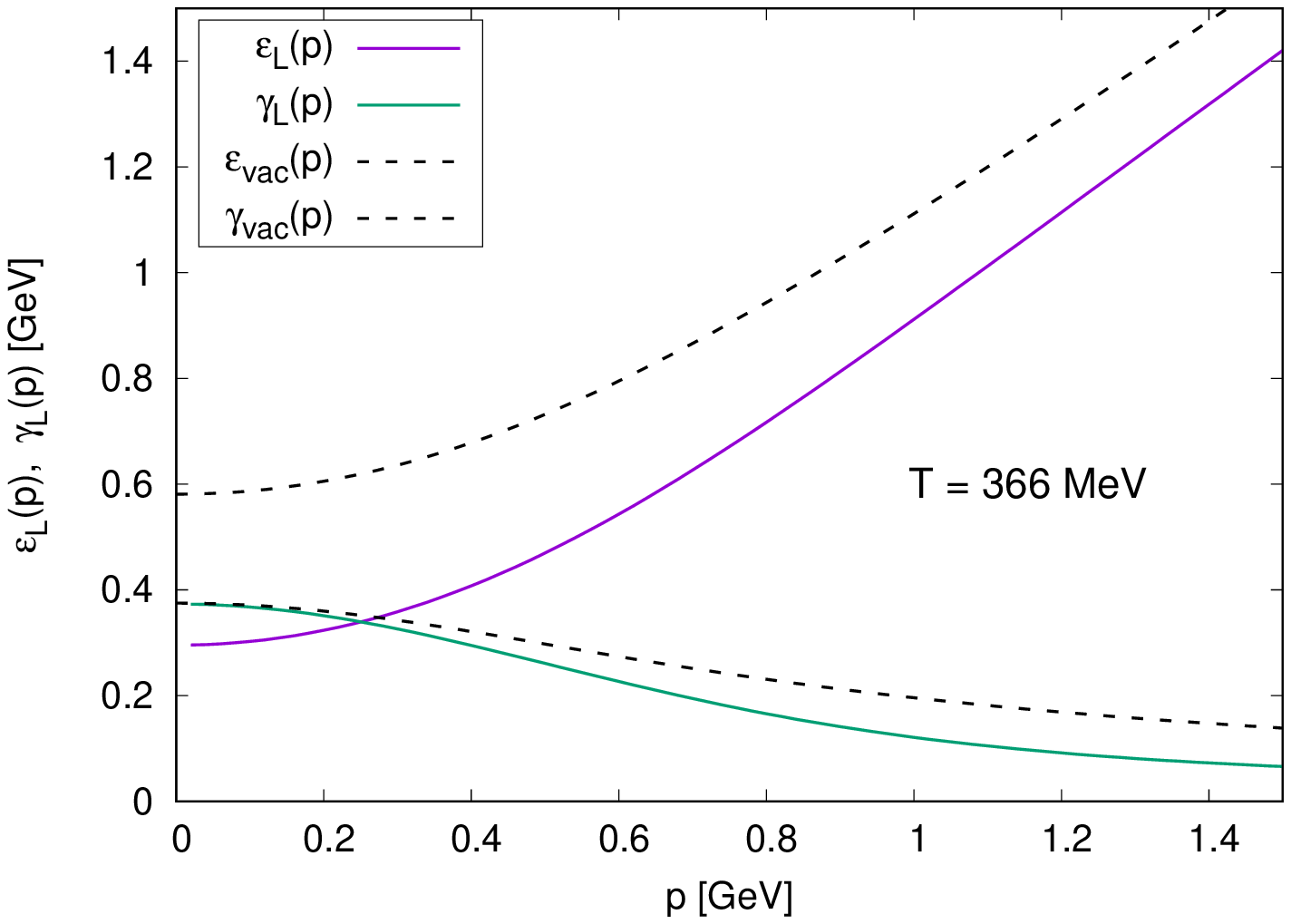}
\caption{Longitudinal dispersion relations for the gluon quasi-particles. The broken lines are the vacuum dispersion relations, common to both projections and given by Eq.~~\eqref{vacuumdisp}. The gluon mass parameters $m(T)$ and renormalization constants $\pi_{0}(T)$ used for the plots are reported in Tab.~II. Except for vanishingly small temperatures, these dispersion relations are not expected to be reliable below $|{\bf p}|\approx 500-700$~MeV.}
\end{figure*}

In the limit ${\bf p}\to 0$ and for any non-zero $\omega$, the longitudinal and the transverse projection of the gluon propagator are known to collapse to a single temperature-dependent function; as a consequence, the corresponding branches of the dispersion relations share the same zero-momentum limit. The ${\bf p}=0$ poles of the gluon propagator are located at $-i(\varepsilon_{0}(T)-i\,\gamma_{0}(T))$, where
\BE
\varepsilon_{0}(T)=\lim_{|{\bf p}|\to 0}\ \varepsilon_{T,L}({\bf p},T)\ ,\quad \gamma_{0}(T)=\lim_{|{\bf p}|\to 0}\ \gamma_{T,L}({\bf p},T)
\EE
are, respectively, the mass and the (zero-momentum) damping rate of the gluon quasi-particles. With regards to such a constraint, the optimized framework of Sec.~IIIB is inconsistent: using different mass parameters for the longitudinal and the transverse projections of the propagator causes the two branches of the dispersion relations to have unequal ${\bf p}\to 0$ limits. All the same, as previously discussed, the low-momentum limit of the longitudinal gluon propagator was found to be quantitatively unreliable at temperatures which are not vanishingly small. It follows that the ${\bf p}\to 0$ limit of the longitudinal dispersion relations cannot be trusted regardless of the inconsistency. Since only the screened expansion's transverse propagator, with the parameters in Tab.~II, was found to reproduce the lattice data at low momenta, in what follows we will make use of the transverse dispersion relations to study the behavior of $\varepsilon_{0}(T)$ and $\gamma_{0}(T)$. From first principles, it is understood that a good description of the long-wavelength longitudinal gluon excitations must yield the same results.

In Fig.~~10 we display the mass and the zero-momentum damping rate of the gluon quasi-particles as functions of the temperature. Across the critical temperature, both of them show a characteristic behavior, decreasing below $T_{c}$ and increasing again in a linear fashion above $T_{c}$. The mass decreases from $\varepsilon_{0}(0)=\varepsilon_{\text{vac}}({\bf 0})=581$~MeV to $\varepsilon_{0}(T_{c})\approx 450$~MeV, whereas the zero-momentum damping rate slightly decreases from $\gamma_{0}(0)=\gamma_{\text{vac}}({\bf 0})=375$~MeV to about $350$~MeV around $T_{c}$. The increase in the damping rate actually seems to start somewhat below the critical temperature (see the data point $T=260$~MeV in Fig.~~10); we could not determine whether this is a physically meaningful behavior or an artifact due to uncertainties in the parameters of Tab.~II.

The behavior of the gluon mass in Fig.~~10 confirms the picture of a confined gluon -- whose mass is dynamically generated through the strong interactions themselves like in the $T\to 0$ limit -- which becomes deconfined above the critical temperature $T_{c}\approx 270$~MeV. In the deconfined phase, the mass of the gluon is thermal in nature and increases linearly with the temperature. The same qualitative behavior was observed in~\cite{damp}, where the gluon mass and zero-momentum damping rate were studied in the screened expansion at finite $T$ using the same scheme of Sec.~IIIA, i.e.
taking temperature-independent values for both the gluon mass parameter $m$ and the renormalization constant $\pi_{0}$.

\begin{figure}[t] \label{zeromomentum}
  \centering
  \includegraphics[width=0.32\textwidth,angle=-90]{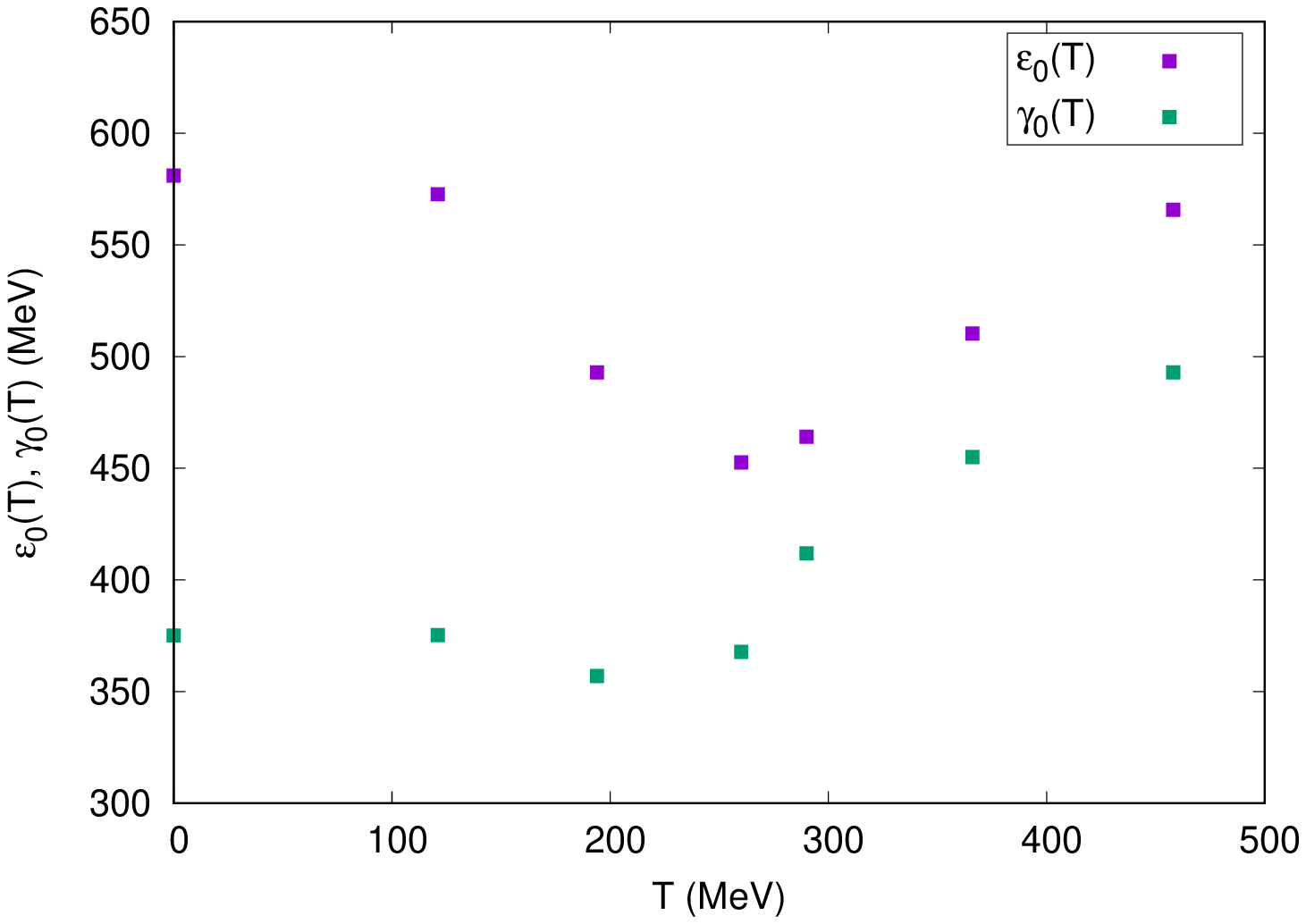}
\caption{Mass $\varepsilon_{0}(T)$ and zero-momentum damping rate $\gamma_{0}(T)$ of the gluon quasi-particles, as functions of the temperature. The parameters used for the plot are reported in Tab.~II under the transverse denomination. See text for further details.}
\end{figure}

\section{Discussion}

The comparison with the available lattice data showed that, overall, the screened expansion gives a correct
qualitative description of the gluon propagator at finite $T$. The agreement
improves if the renormalization constants are tuned at each value of the temperature. At high temperatures and deep in the IR, 
the failure to reproduce the longitudinal projection might arise from the combined effect of several issues like the need of some HTL resummation, a poor optimization and the inadequacy of the single-mass splitting of the action at a finite temperature.
Indeed, the lattice data seem to suggest that a two-mass scheme should be introduced from the beginning for
extending the screened expansion at a finite temperature.

With the exception of an infrared turnover in the longitudinal propagator, which has no counterpart in the lattice data, the qualitative behavior of the propagators seems to be correct and quite robust, irrespective of the optimization scheme.
The pole trajectories can be determined in the complex plane, yielding valuable predictions which cannot be
extracted from the lattice data in the Euclidean space. We have reported
in some detail the dispersion relations of the quasi-gluon for several temperatures across the deconfinement
transition. 

An important feature which emerges from our study is a crossover at the deconfinement transition. The energy of the quasi-particle
is suppressed by temperature in the confined phase. On the other hand, above the critical temperature, the
behavior is reversed and the energy increases as a function of temperature. The same effect can be observed for
the physical mass, defined as the long-wavelength limit $\varepsilon_{0}(T)$ of the pole's real part, as shown
in Fig.~~10. In the confined phase, the mass decreases like an order parameter being suppressed by 
the temperature. This behavior is consistent with that of a dynamical mass which is related to a condensate, the latter being expected
to vanish at the transition temperature. However, at finite temperature the quasi-gluon is also expected to 
acquire a thermal mass which increases linearly, like any other quasi-particle. The two effects might coexist across
the transition, yielding a crossover rather than a sharp transition. In the low-temperature limit the dynamical
nature of the mass dominates, while above the deconfinement transition the mass becomes a pure thermal mass.
Thus, we argue that in the low-temperature phase the mass suppression might be a signature of the dynamical
nature of the gluon mass. On the other hand, as discussed in Ref.~~\cite{damp}, the existence of an intrinsic 
damping rate, which saturates at a finite value at $T=0$, is a confirmation of the quasi-gluon scenario laid out by Stingl~\cite{stingl}. The massive gluon also has a very short finite lifetime and is canceled
from the asymptotic states~\cite{damp}, suggesting that the gluon quasi-particles of the interacting vacuum can only travel the short
distance of about a Fermi and can only exist as intermediate states at the origin of a gluon-jet event.

The issue of the gauge invariance of the poles at $T\neq0$ within the framework of the screened expansion remains, to date, unexplored. One possible development of our study at finite temperature would be to apply the guiding principles and methods of Ref.~\cite{xigauge} in order to monitor whether the Nielsen identities can be satisfied in a general covariant gauge, while fixing the values of the free parameters of the formalism from first principles. A thorough analysis of the matter would presumably require the explicit implementation of specific resummation schemes, as is already the case within the framework of ordinary thermal perturbation theory. Nonetheless, the success of the screened expansion in reproducing the lattice data -- albeit subject to a fit to the data themselves and with the limitations discussed in the previous sections -- leads us to believe that a two-mass shift of the expansion point of the thermal perturbative series may prove to be a robust enough alternative scheme already at one loop. Such a reformulation of the screened expansion requires a full recalculation of both the thermal and vacuum integrals involved in the definition of the propagators, and will be left to future studies.

\acknowledgments
The authors are in debt to Orlando Oliveira for sharing the lattice data of Ref.~~\cite{silva}.

This research was supported in part by "Piano per
la Ricerca di Ateneo 2017/2020 - Linea di intervento 2" of the University of Catania.

\newpage
\appendix

\section{one-loop graphs}

\subsection{Notation}
The Euclidean four-vector $p^\mu$ is defined as
\BE
p^\mu=({\bf p}, \omega).
\EE
where $\omega=p_4=-ip_0$ and
the Lorentz four-vector is $(p_0, {\bf p})$.

In the finite temperature formalism, $\omega=\omega_n=2\pi n T$ and the
Euclidean integral is replaced by a sum over $n$ and by a three-dimensional
integration
\BE
\int \ppp \to \int_p=T\sum_n\int\frac{{\rm d}^3{\bf p}} {(2\pi)^3}.
\EE
The generic (massive) propagator $G_m(p)$ is
\BE
G_m(p)=\frac{1}{p^2+m^2}=\frac{1}{\omega_n^2+{\bf p}^2+m^2}.
\EE

At finite temperature, it is useful to introduce the following orthogonal projectors
\begin{align}
P^T_{\mu\nu}(p)&=(1-\delta_{\mu,4})(1-\delta_{\nu,4})\left(\delta_{\mu\nu}-\frac{p_\mu p_\nu}{{\bf p}^2}\right),\nn\\
P^L_{\mu\nu}(p)&=t_{\mu\nu}(p)-P^T_{\mu\nu}(p)
\label{TL}
\end{align}
beside the Lorentz projectors
\begin{align}
t_{\mu\nu}(p)&=\delta_{\mu\nu}-\frac{p_\mu p_\nu}{p^2},\nn\\
\ell_{\mu\nu}(p)&=\frac{p_\mu p_\nu}{p^2}.
\label{tlE}
\end{align}
The trace of the projectors is
\BE 
P^T_{\mu\mu}=2,\qquad P^L_{\mu\mu}=1.
\EE

The dressed Euclidean propagator of the gluon can be written as
$\Delta^{ab}_{\mu\nu}(p)=\delta_{ab}\Delta_{\mu\nu}(p)$ where
\BE
\Delta^{-1}_{\mu\nu}(p)=G_m(p)^{-1}\>t_{\mu\nu}(p)-Ng^2\Pi_{\mu\nu}(p)+\frac{p^2}{\xi}\>\ell_{\mu\nu}(p)
\EE
and the gluon polarization is $\Pi^{ab}_{\mu\nu}(p)=Ng^2\delta_{ab}\Pi_{\mu\nu}(p)$.
Since $\Pi_{\mu\nu}(p)$ is transverse, i.e. $p^\mu\Pi_{\mu\nu}(p)=0$, in the Landau gauge
($\xi\to 0$) the dressed propagator is also transverse.
We introduce the projected polarizations
\begin{align}
\Pi_T(p)&=\frac{1}{2} P^T_{\mu\nu}(p) \Pi_{\mu\nu}(p),\nn\\
\Pi_L(p)&=P^L_{\mu\nu}(p) \Pi_{\mu\nu}(p),
\end{align}
so that the total polarization reads
\BE
\Pi_{\mu\nu}(p)= \Pi_L(p)\>P^L_{\mu\nu}(p)+\Pi_T(p)\>P^T_{\mu\nu}(p)
\EE
and the dressed propagator can be written as
\BE
\Delta_{\mu\nu}(p)=\Delta^L(p)\>P^L_{\mu\nu}(p)+\Delta^T(p)\>P^T_{\mu\nu}(p)+\frac{\xi}{p^2}\>\ell_{\mu\nu}(p),
\EE
where the projected parts are
\begin{align}
\Delta^{-1}_T(p)&=G_m(p)^{-1}-Ng^2\Pi_T(p),\nn\\
\Delta^{-1}_L(p)&=G_m(p)^{-1}-Ng^2\Pi_L(p).
\end{align}
In the Landau gauge, $\xi\to 0$, the propagator is transverse and its components are determined by the
projected polarizations $\Pi_T(p)$ and $\Pi_L(p)$.
The graphs are evaluated in the Landau gauge, using the (transverse) massive free propagator $\left[ G_m(p)\>t_{\mu\nu}(p)\right]$
in the internal gluon lines. 

The dressed Euclidean propagator of the ghost can be written as
${\cal G}_{ab}(p)=\delta_{ab}\,{\cal G}(p)$, where
\BE
{\cal G}^{-1}(p)=-G_0^{-1}(p)-Ng^2\,\Sigma(p)
\EE
and the ghost self energy is $\Sigma_{ab}(p)=\delta_{ab}\Sigma(p)$. In the graphs, the massless free propagator $-G_0(p)$ is
used in the internal ghost lines.

All the uncrossed one-loop graphs can be decoupled by the method of Ref.~~\cite{serreau} and written in terms of the set of integrals
\begin{align}
J_\alpha&=\int_k G_\alpha (k),\nn\\
I^{\alpha\beta}_{\mu\nu}(p)&=\int_k G_\alpha (k)G_\beta (p-k)k_\mu k_\nu,\nn\\
I^{\alpha\beta}(p)&=\int_k G_\alpha(k)\,G_\beta(p-k),
\label{IJ}
\end{align}
together with their projections 
\begin{align}
I^{\alpha\beta}_T(p)&=\frac{1}{2} P^T_{\mu\nu}(p)\>I^{\alpha\beta}_{\mu\nu}(p),\nn\\
I^{\alpha\beta}_L(p)&=P^L_{\mu\nu}(p)\>I^{\alpha\beta}_{\mu\nu}(p),\nn\\
I^{\alpha\beta}_{T\, p}&=\frac{1}{2} P^T_{\mu\nu}(p)\>I^{\alpha\beta}_{\mu\nu}(0),\nn\\
I^{\alpha\beta}_{L\, p}&=P^L_{\mu\nu}(p)\>I^{\alpha\beta}_{\mu\nu}(0).
\label{projI}
\end{align}
Explicit expressions are reported in Appendix B.

By exchanging $k_\mu$ and $p_\mu-k_\mu$ in the integrals, it is easy to show that 
$I^{\alpha\beta}(p)=I^{\beta\alpha}(p)$, while in general
$I^{\alpha\beta}_{\mu\nu}(p)\not=I^{\beta\alpha}_{\mu\nu}(p)$.
However, since  $p^\mu P^{L,T}_{\mu\nu}(p)=0$,
\BE
(p^\mu-k^\mu)\, P^{L,T}_{\mu\nu}(p)\, (p^\nu-k^\nu)=k^\mu k^\nu\,  P^{L,T}_{\mu\nu}(p)
\label{symmkq}
\EE
and the projected integrals turn out to be symmetric, $I^{\alpha\beta}_{L,T}= I^{\beta\alpha}_{L,T}$.

We note that $I^{\alpha\beta}_{L\, p}$ and $I^{\alpha\beta}_{T\, p}$ might depend on $p$ 
because of the explicit dependence in the projectors. For instance, let us consider any constant
integral
\BE
I_{\mu\nu}=\int_k k_\mu k_\nu f(k)=\delta_{\mu\nu}I_{\mu\mu}
\EE
which does not depend on the external momentum $p$. 
Let us denote by $I_{L,0}$, $I_{T,0}$, the non-zero components that can be written,
taking $k_\mu=(\bk,\omega_n)$, as
\begin{align}
I_{44}&=I_{L,0}=\int_k \omega_n^2 f({\bf k},\omega_n),\nn\\
I_{ii}&=I_{T,0}=\frac{1}{3}\int_k {\bf k}^2 f({\bf k},\omega_n),\qquad i=1,2,3.
\label{compon}
\end{align}
In fact, the explicit projections $I_{L,p}$,  $I_{T,p}$ can be defined and evaluated as in Eqs.~(\ref{projI}):
\begin{align}
I_{T,p}&=\frac{1}{2} P^T_{\mu\nu}(p)\>I_{\mu\nu}=I_{T,0},\nn\\
I_{L,p}&=P^L_{\mu\nu}(p)\>I_{\mu\nu}=(I_{L,0}-I_{T,0})\ \frac{{\bf p}^2}{{\bf p}^2+\omega^2}+I_{T,0}.
\end{align}
While $I_{T,p}=I_{T,0}$ and does not depend on $p$, the longitudinal projection depends on $p$ and has the different limits
\BE
\lim_{{\bf p}\to 0} I_{L,p}=I_{T,0},\quad \lim_{{\omega}\to 0} I_{L,p}=I_{L,0}.
\EE
More generally, for the integral $I^{\alpha\beta}_{\mu\nu}(p)$, which has an explicit dependence on $p$, the projections have
the following limits:
\begin{align}
I_{L,0}^{\alpha\beta}&=\lim_{{\bf p}\to 0}\left[\lim_{{\omega}\to 0} I^{\alpha\beta}_L(p)\right],
\quad I_{T,0}^{\alpha\beta}=\lim_{\omega\to 0}\left[\lim_{{\bf p}\to 0} I^{\alpha\beta}_L(p)\right],\nn\\
I_{T,0}^{\alpha\beta}&=\lim_{{\bf p}\to 0}\left[\lim_{{\omega}\to 0} I^{\alpha\beta}_T(p)\right]
=\lim_{\omega\to 0}\left[\lim_{{\bf p}\to 0} I^{\alpha\beta}_T(p)\right];
\label{limits}
\end{align}
they are related to the projections of the limit $I^{\alpha\beta}_{\mu\nu}(0)$, as defined
in Eqs.~(\ref{projI}),
\begin{align}
I^{\alpha\beta}_{T,p}&=I^{\alpha\beta}_{T,0},\nn\\
I^{\alpha\beta}_{L,p}&=(I^{\alpha\beta}_{L,0}-I^{\alpha\beta}_{T,0})\frac{{\bf p}^2}{{\bf p}^2+\omega^2}+I_{T,0}^{\alpha\beta},
\label{ILTp}
\end{align}
where
\BE
I^{\alpha\beta}_{L,0}=I^{\alpha\beta}_{44}(0),\quad I^{\alpha\beta}_{T,0}=I^{\alpha\beta}_{ii}(0).
\EE
The limits in Eq.~(\ref{limits}) agree with the physical requirement that transverse and longitudinal projections must
coincide for any $\omega$ in the limit ${\bf p}\to 0$, while they are different for any finite ${\bf p}$ in the limit $\omega\to0$.

Each crossed graph $\Pi^{\times}$, containing one insertion of the mass counterterm, can be obtained by the corresponding
uncrossed graph $\Pi$ by a simple derivative
\BE
\Pi^{\times}=-m^2\frac{\partial}{\partial m^2} \Pi.
\EE
Their explicit calculation requires the definition
of a new set of integrals $\partial I^{\alpha\beta}$, $\partial I^{\alpha\beta}_{L,T}$, $\partial J_m$, $\partial^2 J_m$:
\begin{align}
\partial I^{\alpha\beta} (p)&=\frac{\partial}{\partial \alpha^2} I^{\alpha\beta}(p),\nn\\
\partial I^{\alpha\beta}_{L,T} (p)&=\frac{\partial}{\partial \alpha^2} I^{\alpha\beta}_{L,T}(p),\nn\\
\partial J_m &=\frac{\partial}{\partial m^2} J_m,\nn\\
\partial^2 J_m &=\frac{\partial^2}{\partial (m^2)^2} J_m.\nn\\
\label{dJdef}
\end{align}

We note that the second argument ($\beta$) is kept fixed in the derivative, so that
$\partial I^{\alpha\beta}\not=\partial I^{\beta\alpha}$.
When $\alpha=\beta$ the derivative must be taken twice, so that for instance
\BE
\frac{\partial}{\partial m^2} I^{mm}=\left[\frac{\partial}{\partial \alpha^2} I^{\alpha\beta}+
\frac{\partial}{\partial \beta^2} I^{\alpha\beta}\right]_{\alpha=\beta=m}=2\partial I^{mm}.
\EE

Not all the integrals are independent. For instance, it can be easily shown that
\begin{align}
I^{\alpha\beta}(0)&=\frac{1}{\beta^2-\alpha^2}\left[J_\alpha-J_\beta\right],\nn\\
\partial J_m&=-I^{mm}(0),\nn\\
\partial I^{\alpha\beta}(0)&=\frac{1}{\beta^2-\alpha^2}\left[I^{\alpha\beta}(0)-I^{\alpha\alpha}(0)\right].
\label{identI}
\end{align}
It is useful to introduce the integrals $J_m^{L,T}$ which follow by setting $f(k)=G_m(k)$ in Eq.~(\ref{compon}),
\begin{align}
J^{L}_m&=\int_k \omega_n^2 G_m({\bf k},\omega_n),\nn\\
J^{T}_m&=\frac{1}{3}\int_k {\bf k}^2 G_m({\bf k},\omega_n),
\label{JLT}
\end{align}
so that Eqs.~(\ref{identI}) can be extended to the projected integrals,
\begin{align}
I^{\alpha\beta}_{L,T\>0}&=\frac{1}{\beta^2-\alpha^2}\left[J^{L,T}_\alpha-J^{L,T}_\beta\right],\nn\\
\partial J^{L,T}_m&=-I^{mm}_{L,T\>0},\nn\\
\partial I^{\alpha\beta}_{L,T\>0}&=\frac{1}{\beta^2-\alpha^2}\left[I^{\alpha\beta}_{L,T\>0}-I^{\alpha\alpha}_{L,T\>0}\right],
\label{identILT}
\end{align}
and, by Eq.~(\ref{ILTp}), the projections $I^{\alpha\beta}_{L,T\>p}$ can be expressed in terms of the constant integrals $J_m^{L,T}$.

\subsection{Graph 1b - (tadpole)}

Setting $d=4$, Eq.~(31) of Ref.~~\cite{genself} reads
\BE
\Pi^{(1b)}_{\mu\nu}(p)=-\int_k\left[ 3\delta_{\mu\nu}-t_{\mu\nu}(k)\right] G_m(k),
\EE
yielding
\BE
\Pi^{(1b)}_{\mu\nu}=-\left[2\delta_{\mu\nu}\>J_m+I^{m0}_{\mu\nu}(0)\right],
\label{p1btot}
\EE
where the integrals $J_\alpha$, $I^{\alpha\beta}_{\mu\nu}(p)$ were
defined in Eqs.~(\ref{IJ})
and their explicit expressions are reported in Appendix B.
The projected polarization of graph $(1b)$ is
\begin{align}
\Pi^{(1b)}_T(p)&=-\left[2J_m+I^{m0}_{T\, p}\right],\nn\\
\Pi^{(1b)}_L(p)&=-\left[2J_m+I^{m0}_{L\, p}\right].
\label{1bLT}
\end{align}
The vacuum contribution can be extracted by evaluating the integrals in the limit $T\to 0$
where $I^{m0}_{\mu\nu}(0)\to\frac{1}{4}\delta_{\mu\nu} J_m$ so that
\BE
\Pi^{(1b)}_{\mu\nu}(T=0)=-\frac{9}{4}\delta_{\mu\nu}\>J_m
\EE
in agreement with the general result of Ref.~~\cite{ptqcd2} for $d=4$.

\subsection{Graph 2b - (gluon loop)}

The general explicit expression for the graph $(2b)$ has been reported in Ref.~~\cite{genself}, for
a generic dimension $d$ and a generic free-particle propagator. In the Landau gauge, the explicit expression for $d=4$
can be written as (see also Ref.~~\cite{serreau})
\BE
\Pi^{\mu\nu}(p)=\sum_{i=1}^{4}\> \Pi^{\mu\nu}_{i}(p)
\EE
where, denoting $q=p-k$,
\newpage
\begin{widetext}
\begin{align}
\Pi_{1}^{\mu\nu}(p)&=\frac{1}{2}\ \int_{k}\ (q-k)^{\mu}\,(q-k)^{\nu}\> \left[t^{\lambda\rho}(q)\, t_{\rho\lambda}(k)\right]
\> G_{m}(k)\,G_{m}(q),\nn\\
\Pi^{\mu\nu}_{2}(p)&=\int_{k}\ t^{\mu\nu}(k)\ \left[(p+k)_\lambda\, t^{\lambda\rho} (q)\, (p+k)_\rho\right]\ G_{m}(k)\,G_{m}(q),\nn\\
\Pi^{\mu\nu}_{3}(p)&=-\int_{k}\ \left[t^{\mu\lambda}(k)\, (p+q)_{\lambda}\right]\ 
\left[t^{\nu\rho}(q)\,(p+k)_{\rho}\right]\ G_{m}(k)\,G_{m}(q),\nn\\
\Pi_{4}^{\mu\nu}(p)&=\int_{k}\ (q-k)^{\mu}\,\left[t^{\nu\lambda}(k)\, t_{\lambda\rho}(q)\,(p+k)^\rho\right]\ G_{m}(k)\,G_{m}(q)
\quad+\quad\mu\leftrightarrow\nu.
\label{Pi}
\end{align}
All integrals can be evaluated by the method of Ref.~~\cite{serreau} and written in terms of the integrals in Eq.~(\ref{IJ}). 

In some detail, 
\BE
\Pi_{1}^{\mu\nu}(p)=\frac{1}{2}\ \int_{k}\ (q-k)^{\mu}\,(q-k)^{\nu}\ \left[2+\frac{(k\cdot q)^{2}}{k^{2}q^{2}}\right]\ G_{m}(k)\,G_{m}(q),
\label{p1}
\EE
and making use of the identities
\begin{align}
2k \cdot q=(k+q)^2-k^2-q^2&=p^2-G_\alpha(k)^{-1}-G_\beta(q)^{-1}+\alpha^2+\beta^2,\nn\\
\frac{G_m(k)}{k^2}=G_0(k)G_m(k)&=\frac{1}{m^2}\left[G_0(k)-G_m(k)\right],\nn\\
(q-k)_{\mu}\,(q-k)_{\nu}&=2\,(q_{\mu}q_{\nu}+k_{\mu}k_{\nu})-p_{\mu}p_{\nu}\nn\\
\label{ident}
\end{align}
we can write
\begin{align}
\frac{(k\cdot q)^{2}}{k^{2}q^{2}}\ G_{m}(k)\,G_{m}(q)&=\frac{1}{4m^{4}}\ \Big[\big(p^{2}+2m^{2}\big)^{2}\ G_{m}(k)\,G_{m}(q)+p^{4}\ G_{0}(k)\,G_{0}(q)+\nn\\
&-\big(p^{2}+m^{2}\big)^{2}\ \Big(G_{0}(k)\,G_{m}(q)+G_{m}(k)\,G_{0}(q)\Big)\Big]+\frac{1}{4}\Big(G_{m}(k)G_{0}(k)+G_{m}(q)G_{0}(q)\Big),
\end{align}
\begin{align}
&\int_{k}\ (q-k)_{\mu}\,(q-k)_{\nu} \frac{(k\cdot q)^{2}}{k^{2}q^{2}}\ G_{m}(k)\,G_{m}(q)=
\frac{1}{4 m^{4}}\ \Big[\big(p^{2}+2m^{2}\big)^{2}\ \big(4\,I_{\mu\nu}^{mm}(p)-p_{\mu}p_{\nu}\, I^{mm}(p)\big)+\nn\\
&\qquad\qquad+p^{4}\, \big(4\,I_{\mu\nu}^{00}(p)-p_{\mu}p_{\nu}\, I^{00}(p)\big)
-2\big(p^{2}+m^{2}\big)^{2}\ \Big(2\,[I^{m0}_{\mu\nu}(p)+I^{0m}_{\mu\nu}(p)]-\,p_{\mu}p_{\nu} I^{m0}(p)\Big)\Big]+\nn\\
&\qquad\qquad+2I_{\mu\nu}^{m0}(0)+\frac{1}{2}\ p_{\mu}p_{\nu}\ I^{m0}(0),
\end{align}
\BE
\int_{k}\ (q-k)_{\mu}\,(q-k)_{\nu} \ G_{m}(k)\,G_{m}(q)=4\,I_{\mu\nu}^{mm}(p)-p_{\mu}p_{\nu}\, I^{mm}(p)
\EE
so that Eq.~(\ref{p1}) reads
\begin{align}
{\Pi_{1}}_{\mu\nu}(p)&=\frac{p^{4}}{2m^{4}}\ I_{\mu\nu}^{00}(p)+\bigg[4+\frac{(p^{2}+2m^{2})^{2}}{2m^{4}}\bigg]\ I^{mm}_{\mu\nu}(p)
-\frac{(p^{2}+m^{2})^{2}}{2m^{4}}\ \Big(I^{m0}_{\mu\nu}(p)+I^{0m}_{\mu\nu}(p)\Big)+I^{m0}_{\mu\nu}(0)+\nn\\
&-p_{\mu}p_{\nu}\ 
\bigg[\frac{p^{4}}{8m^{4}}\ I^{00}(p)+
\bigg(1+\frac{(p^{2}+2m^{2})^{2}}{8m^{4}}\bigg)\ I^{mm}(p)-\frac{(p^{2}+m^{2})^{2}}{4m^{4}}\ 
I^{m0}(p)-\frac{1}{4}\ I^{m0}(0)\bigg].
\label{p1tot}
\end{align}
\vskip 15pt
\centerline{* \quad * \quad * \quad * \quad  * \quad *  \quad *}
\vskip 15pt

The second polarization term in Eq.~(\ref{Pi}) reads
\begin{align}
{\Pi_{2}}_{\mu\nu}(p)&=4\int_{k}\  \left[\delta_{\mu\nu}-\frac{k_\mu k_\nu}{k^2}\right] \
\bigg(p^{2}-\frac{(p\cdot q)^{2}}{q^{2}}\bigg)\ G_{m}(k)\,G_{m}(q)=\nn\\
=&\ 4\int_{k}\ \left[\delta_{\mu\nu}p^2 -p^2 {k_\mu k_\nu}G_0(k)-\delta_{\mu\nu}(p\cdot q)^2 G_0(q)
+(p\cdot q)^{2} k_\mu k_\nu G_0(k) G_0(q)\right]\ G_{m}(k)\,G_{m}(q)=\nn\\
=&4\ \bigg[\delta_{\mu\nu}\,p^{2}\,I^{mm}(p)-\frac{p^{2}}{m^{2}}\ \Big(I^{0m}_{\mu\nu}(p)-I^{mm}_{\mu\nu}(p)\Big)\bigg]+
{\Pi_{2a}}_{\mu\nu}(p)+{\Pi_{2b}}_{\mu\nu}(p),
\label{p2}
\end{align}
where
\begin{align}
{\Pi_{2a}}_{\mu\nu}(p)&=-4\,\delta_{\mu\nu}\ \int_{k}\ G_{m}(q)\,G_{m}(k)\, G_{0}(k)\ (p\cdot k)^{2}
=-\frac{4}{m^{2}}\ \delta_{\mu\nu}\ \int_{k}\ G_{m}(q)\,(G_{0}(k)-G_{m}(k))\ (p\cdot k)^{2},\nn\\
{\Pi_{2b}}_{\mu\nu}(p)&=4 \int_{k}\ G_{m}(k)\,G_{m}(q)\, G_{0}(k)\,G_{0}(q)\ (p\cdot q)^{2}\ k_{\mu}k_{\nu}.
\end{align}
Using the identities
\begin{align}
2\,(p\cdot k)&=p^{2}+\beta^2-\alpha^2+G_{\alpha}^{-1}(k)-G_{\beta}^{-1}(q),\nn\\
2\,(p\cdot k)\ G_{\beta}(q)\,G_{\alpha}(k)&=(p^{2}+\beta^2-\alpha^2)\ G_{\beta}(q)\,G_{\alpha}(k)+G_{\beta}(q)-G_{\alpha}(k),\nn\\
4\,(p\cdot k)^{2}\ G_{\beta}(q)\,G_{\alpha}(k)&=(p^{2}+\beta^{2}-\alpha^2)^{2}\ G_{\beta}(q)\,G_{\alpha}(k)
+(p^{2}+\beta^{2}-\alpha^2)(G_{\beta}(q)-G_{\alpha}(k))
+2(p\cdot k)(G_{\beta}(q)-G_{\alpha}(k)),
\label{ident2}
\end{align}
which hold for any pair $k$, $q$ satisfying $k+q=p$,  we obtain for $(\alpha,\beta)=(0,m)$ and for
$(\alpha,\beta)=(m,m)$, respectively,
\begin{align}
4\,(p\cdot k)^{2}\ G_{m}(q)\,G_{0}(k)&=(p^{2}+m^{2})^{2}\ G_{m}(q)\,G_{0}(k)+(p^{2}+m^{2})(G_{m}(q)-G_{0}(k))
+2(p\cdot k)(G_{m}(q)-G_{0}(k)),\nn\\
4\,(p\cdot k)^{2}\ G_{m}(q)\,G_{m}(k)&=p^{4}\ G_{m}(q)\,G_{m}(k)+p^{2}(G_{m}(q)-G_{m}(k))+2(p\cdot k)(G_{m}(q)-G_{m}(k)),
\end{align}
so that the term $\Pi_{2a}$ can be written as
\BE
{\Pi_{2a}}_{\mu\nu}=-\frac{\delta_{\mu\nu}}{m^{2}}\ \Big[(p^{2}+m^{2})^{2}\ I^{0m}(p)-p^{4}\ I^{mm}(p)-m^{2}\ \big(p^{2}+m^{2}\big)\ I^{0m}(0)\Big].
\label{p2a}
\EE
Using the second of Eqs.~(\ref{ident}), the term $\Pi_{2b}$ can be written as
\BE
{\Pi_{2b}}_{\mu\nu}(p)=\frac{4}{m^{4}} \int_{k}\ (p\cdot q)^{2}\ k_{\mu}k_{\nu}\ 
\Big(G_{0}(k)\,G_{0}(q)-G_{0}(k)\,G_{m}(q)-G_{m}(k)\,G_{0}(q)+G_{m}(k)\,G_{m}(q)\Big),
\EE
while reversing $k$ and $q$ in Eq.~(\ref{ident2}) we obtain, for $\alpha$ and $\beta$ that take the values $0$ and $m$,
\begin{align}
4\,(p\cdot q)^{2}\ G_{0}(k)\,G_{0}(q)&=p^{4}\,G_{0}(k)\,G_{0}(q)+3\,p^{2}\,(G_{0}(k)-G_{0}(q))-2\,p\cdot k\ (G_{0}(k)-G_{0}(q)),\nn\\
4\,(p\cdot q)^{2}\ G_{0}(k)\,G_{m}(q)&=(p^{2}-m^{2})^{2}\ G_{0}(k)\,G_{m}(q)+(3\,p^{2}-m^{2})\,(G_{0}(k)-G_{m}(q))-2\,p\cdot k\ (G_{0}(k)-G_{m}(q)),\nn\\
4\,(p\cdot q)^{2}\ G_{m}(k)\,G_{0}(q)&=(p^{2}+m^{2})^{2}\ G_{m}(k)\,G_{0}(q)+(3\,p^{2}+m^{2})\,(G_{m}(k)-G_{0}(q))-2\,p\cdot k\ (G_{m}(k)-G_{0}(q)),\nn\\
4\,(p\cdot q)^{2}\ G_{m}(k)\,G_{m}(q)&=p^{4}\,G_{m}(k)\,G_{m}(q)+3\,p^{2}\,(G_{m}(k)-G_{m}(q))-2\,p\cdot k\ (G_{m}(k)-G_{m}(q)),
\end{align}
yielding for ${\Pi_{2b}}$
\begin{align}
{\Pi_{2b}}_{\mu\nu}(p)&=\frac{1}{m^{4}}\ \Big[p^{4}\ \Big(I^{00}_{\mu\nu}(p)+I^{mm}_{\mu\nu}(p)\Big)
-(p^{2}-m^{2})^{2}\ I^{0m}_{\mu\nu}(p)-(p^{2}+m^{2})^{2}\ I^{m0}_{\mu\nu}(p)\Big]+2\ I_{\mu\nu}^{0m}(0)+p_{\mu}p_{\nu}\ I^{0m}(0).
\label{p2b}
\end{align}
Adding Eq.~(\ref{p2a}) and Eq.~(\ref{p2b}) in Eq.~(\ref{p2}), the second polarization term in Eq.~(\ref{Pi}) is
\begin{align}
{\Pi_{2}}_{\mu\nu}(p)&=\delta_{\mu\nu}\ \bigg[\frac{p^{2}}{m^{2}}\ (p^{2}+4m^{2})\ I^{mm}(p)-\frac{(p^{2}+m^{2})^{2}}{m^{2}}\ I^{0m}(p)
+(p^{2}+m^{2})\ I^{0m}(0)\bigg]+p_{\mu}\,p_{\nu}\ I^{0m}(0)+\nn\\
&\qquad+\frac{p^{4}}{m^{4}}\ I_{\mu\nu}^{00}(p)+\frac{p^{2}}{m^{4}}\ (p^{2}+4m^{2})\ I^{mm}_{\mu\nu}(p)
-\frac{(p^{2}+m^{2})^{2}}{m^{4}}\ \Big(I^{m0}_{\mu\nu}(p)+I^{0m}_{\mu\nu}(p)\Big)+ 2\ I^{0m}_{\mu\nu}(0).
\label{p2tot}
\end{align}
\vskip 15pt
\centerline{* \quad * \quad * \quad * \quad  * \quad *  \quad *}
\vskip 15pt

The third polarization term in Eq.~(\ref{Pi}) can be decomposed by observing that 
\BE
\left[t^{\mu\lambda}(k)\, (p+q)_{\lambda}\right]=2\ \bigg(p^\mu-\frac{p\cdot k}{k^{2}}\ k^\mu\bigg),
\qquad \left[t^{\nu\rho}(q)\, (p+k)_{\rho}\right]=2\ \bigg(p^\nu-\frac{p\cdot q}{q^{2}}\ q^\nu\bigg),
\EE
so that, changing the integration variable from $k$ to $q$ in the $q_\nu p_\mu$ term, $\Pi_3$ reads
\begin{align}
{\Pi_3}_{\mu\nu}(p)&=-4\ \int_{k}\ \bigg(p_{\mu}p_{\nu}-\frac{p\cdot k}{k^{2}}\ (k_{\mu} p_{\nu}+k_{\nu}p_{\mu})
+\frac{(p\cdot k)(p\cdot q)}{k^{2}q^{2}}\ k_{\mu}q_{\nu}\bigg)\ G_{m}(k)\,G_{m}(q)=\nn\\
&=-4\ p_{\mu}p_{\nu}\ I^{mm}(p)+{\Pi_{3a}}_{\mu\nu}(p)+{\Pi_{3b}}_{\mu\nu}(p),
\label{p3}
\end{align}
where
\begin{align}
{\Pi_{3a}}_{\mu\nu}(p)=4\ \int_{k}\ (p\cdot k)\ (k_{\mu} p_{\nu}+k_{\nu}p_{\mu})\ G_{0}(k)\, G_{m}(k)\,G_{m}(q),\nn\\
{\Pi_{3b}}_{\mu\nu}(p)=-4\ \int_{k}\ (p\cdot k)(p\cdot q)\ k_{\mu}q_{\nu}\ G_{0}(k)\,G_{0}(q)\,G_{m}(k)\,G_{m}(q).
\end{align}
The first integral can be decomposed by using the identity
\BE
(k^{\mu} p^{\nu}+k^{\nu}p^{\mu})=p^{\mu}p^{\nu}+k^{\mu}k^{\nu}-q^{\mu}q^{\nu}
\label{identpk}
\EE
and observing that, by the second of Eqs.~(\ref{ident}) and the second of Eqs.~(\ref{ident2}),
\begin{align}
4\,(p\cdot k)\ G_{0}(k)\, G_{m}(k)\,G_{m}(q)&=\frac{4\,(p\cdot k)}{m^2}\left[G_{0}(k)G_{m}(q)-G_{m}(k)G_{m}(q)\right]=\nn\\
&=\frac{2}{m^2}\left[ (p^{2}+m^{2})\ G_{0}(k)\,G_{m}(q)-p^{2}\,G_{m}(k)\,G_{m}(q)-m^{2}\,G_{0}(k)\,G_{m}(k)\right],
\end{align}
yielding
\BE
{\Pi_{3a}}_{\mu\nu}(p)=\frac{2}{m^{2}}\ p_{\mu}p_{\nu}\ \Big[(p^{2}+m^{2})\ I^{0m}(p)-p^{2}\ I^{mm}(p)\Big]
+\frac{2}{m^{2}}\ (p^{2}+m^{2})\ \big(I^{0m}_{\mu\nu}(p)-I^{m0}_{\mu\nu}(p)\big).
\EE
\label{p3a}
In the second integral $\Pi_{3b}$  we can use the identity
\BE
k_\mu q_\nu=\left[\frac{1}{2}p_\mu p_\nu-k_\mu k_\nu\right]+\frac{1}{2}\left(k_\mu-q_\mu\right) p_\nu,
\EE
where the last term can be dropped because it is antisymmetric in the exchange of $k$ and $q$ and its
contribution to the integral is zero. Taking the second of Eqs.~(\ref{ident2}) with $\alpha=\beta=m$ and the
same equation with $\alpha=\beta=0$ and $k$,$q$ interchanged, their product can be written as
\begin{align}
4\ (p\cdot k)(p\cdot q)\ &G_{m}(k)\,G_{m}(q)\,G_{0}(k)\,G_{0}(q)=\frac{p^{4}}{m^{4}}\ \Big[G_{0}(k)\,G_{0}(q)+G_{m}(k)\,G_{m}(q)\Big]+\nn\\
&\quad+\bigg(1-\frac{p^{4}}{m^{4}}\bigg)\ \Big[G_{0}(k)\,G_{m}(q)+G_{m}(k)\,G_{0}(q)\Big]-G_{0}(q)\,G_{m}(q)-G_{0}(k)\,G_{m}(k),
\end{align}
where the second of Eqs.~(\ref{ident}) has been used for decomposing the products of more than two $G$ functions.
Then, the integral can be written
\begin{align}
{\Pi_{3b}}_{\mu\nu}(p)&=\frac{p^{4}}{m^{4}}\ \Big(I^{00}_{\mu\nu}(p)+I^{mm}_{\mu\nu}(p)\Big)+\bigg(1-\frac{p^{4}}{m^{4}}\bigg)\ 
\big(I^{0m}_{\mu\nu}(p)+I^{m0}_{\mu\nu}(p)\big)+\nn\\
&\qquad-p_{\mu}p_{\nu}\ \bigg[\frac{p^{4}}{2m^{4}}\ \big(I^{00}(p)+I^{mm}(p)\big)-\bigg(\frac{p^{4}}{m^{4}}-1\bigg)\ I^{0m}(p)\bigg]
-2\ I^{0m}_{\mu\nu}(0).
\label{p3b}
\end{align}
Adding Eq.~(\ref{p3a}) and Eq.~(\ref{p3b}) in Eq.~(\ref{p3}), the third polarization term in Eq.~(\ref{Pi}) is
\begin{align}
{\Pi_{3}}_{\mu\nu}(p)
&=\frac{p^{4}}{m^{4}}\ \Big(I^{00}_{\mu\nu}(p)+I^{mm}_{\mu\nu}(p)\Big)+\frac{3m^{4}+2m^{2}p^{2}-p^{4}}{m^{4}}\ I^{0m}_{\mu\nu}(p)
-\frac{(p^{2}+m^{2})^{2}}{m^{4}}\ I^{m0}_{\mu\nu}(p)-2\,I^{0m}_{\mu\nu}(0)+\nn\\
&\quad-p_{\mu}p_{\nu}\ \bigg[\frac{p^{4}}{2m^{4}}\ I^{00}(p)+\frac{p^{4}+4m^{2}p^{2}+8m^{4}}{2m^{4}}\ I^{mm}(p)
-\frac{(p^{2}+m^{2})^{2}}{m^{4}}\ I^{0m}(p)\bigg].
\label{p3tot}
\end{align}
\vskip 15pt
\centerline{* \quad * \quad * \quad * \quad  * \quad *  \quad *}
\vskip 15pt

The last polarization term in Eq.~(\ref{Pi}) can be decomposed by observing that
\BE
\left[t^{\nu\lambda}(k)\, t_{\lambda\rho}(q)\,(p+k)^\rho\right]=2\ \frac{(k\cdot q)}{q^{2}}\ \bigg[\frac{(k\cdot p)}{k^{2}}\ k^{\nu}-p^{\nu}\bigg].
\EE
Then, recalling that $q_\mu=p_\mu-k_\mu$, the integral reads
\BE
{\Pi_{4}}_{\mu\nu}(p)=2\ \int_{k}\bigg[\frac{(k\cdot p)}{k^{2}}\ \Big(p_\mu k_\nu- 2k_\mu k_\nu\Big)
+\ \Big(2k_\mu p_\nu-p_\mu p_\nu\Big)\bigg]
(k\cdot q)\ G_{0}(q)\,G_{m}(k)\,G_{m}(q)\>+\>\mu\leftrightarrow\nu.
\EE
Using the identity
\BE
p_{\mu}k_{\nu}+p_{\nu}k_{\mu}=p_{\mu}p_{\nu}+k_{\mu}k_{\nu}-q_{\mu}q_{\nu}
\EE
the two pieces can be added together yielding
\BE
{\Pi_{4}}_{\mu\nu}(p)=2\ \int_{k}\bigg[\frac{(k\cdot p)}{k^{2}}\ \Big(p_{\mu}p_{\nu}-3\ k_{\mu}k_{\nu}-q_{\mu}q_{\nu}\Big)
+2\ \Big(k_{\mu}k_{\nu}-q_{\mu}q_{\nu}\Big)\bigg]\ (k\cdot q)\ G_{0}(q)\,G_{m}(k)\,G_{m}(q).
\EE
The product of three $G$ functions can be decomposed by the second of Eqs.~(\ref{ident}) and the two arising terms can be written
by the first of Eqs.~(\ref{ident}), with $(\alpha,\beta)=(m,0)$ and $(\alpha,\beta)=(m,m)$, respectively,
\BE
2\ (k\cdot q)\left[ G_{0}(q)\, G_{m}(q)\right]G_{m}(k)=\frac{p^{2}+m^{2}}{m^{2}}\ G_{m}(k)\,G_{0}(q)-\frac{(p^{2}+2m^{2})}{m^{2}}\ G_{m}(k)\,G_{m}(q)
-\frac{1}{m^{2}}\ \big(G_{0}(q)-G_{m}(q)\big).
\EE
The integral then reads
\BE
{\Pi_{4}}_{\mu\nu}(p)=2\ \frac{p^{2}+m^{2}}{m^{2}}\ \big(I^{m0}_{\mu\nu}(p)-I^{0m}_{\mu\nu}(p)\big)-2\ p_{\mu}p_{\nu}\ I^{0m}(0)+
{\Pi_{4a}}_{\mu\nu}(p),
\label{p4}
\EE
where
\begin{align}
{\Pi_{4a}}_{\mu\nu}(p)&=\frac{1}{2}\ \int_{k}\ 2\ (k\cdot p)\ G_{0}(k)\ \Big(p_{\mu}p_{\nu}-3\ k_{\mu}k_{\nu}-q_{\mu}q_{\nu}\Big)\times\nn\\
&\qquad\times \bigg[\frac{p^{2}+m^{2}}{m^{2}}\ G_{m}(k)\,G_{0}(q)-\frac{(p^{2}+2m^{2})}{m^{2}}\ G_{m}(k)\,G_{m}(q)
-\frac{1}{m^{2}}\ \big(G_{0}(q)-G_{m}(q)\big)\bigg].
\end{align}
Using the first of Eqs.~(\ref{ident2}) with $\alpha=0$ and $\beta=m,0$,
\BE
2\,(k\cdot p)\ G_0(k) = (p^{2}+m^2)\ G_0(k)+1-G_{m}^{-1}(q)\, G_0(k)=p^{2}\ G_0(k)+1-G_{0}^{-1}(q)\, G_0(k),
\EE
and decoupling the product $G_m(k)G_0(k)$ by the second of Eqs.~(\ref{ident}), the term $\Pi_{4a}$ can be written as
\begin{align}
{\Pi_{4a}}_{\mu\nu}(p)&=\frac{1}{2m^{2}}\ \int_{k}\ \ \Big(p_{\mu}p_{\nu}-3\ k_{\mu}k_{\nu}-q_{\mu}q_{\nu}\Big)\ 
\bigg[\frac{p^{4}}{m^{2}}\ G_{0}(k)\,G_{0}(q)
+\frac{p^{2}(p^{2}+2m^{2})}{m^{2}}\ G_{m}(k)\,G_{m}(q)+\nn\\
&\qquad-\frac{(p^{2}+m^{2})^{2}}{m^{2}}\ G_{0}(k)\,G_{m}(q)
-\frac{p^{4}-m^{4}}{m^{2}}\ G_{m}(k)\,G_{0}(q)+m^{2}\ \bigg(G_{0}(k)\,G_{m}(k)-G_{0}(q)\,G_{m}(q)\bigg)\bigg],
\end{align}
so that the integral reads
\begin{align}
{\Pi_{4a}}_{\mu\nu}(p)&=p_{\mu}p_{\nu}\ \bigg[\frac{p^{4}}{2m^{4}}\ I^{00}(p)+\frac{p^{2}(p^{2}+2m^{2})}{2m^{4}}\ I^{mm}(p)
-\frac{p^{2}(p^{2}+m^{2})}{m^{4}}\ I^{m0}(p)+I^{0m}(0)\bigg]+\nn\\
&-2\ \frac{p^{4}}{m^{4}}\ I^{00}_{\mu\nu}(p)-2\ \frac{p^{2}(p^{2}+2m^{2})}{m^{4}}\ I^{mm}_{\mu\nu}(p)
+\frac{2p^{4}+3m^{2}p^{2}+m^{4}}{m^{4}}\ I^{0m}_{\mu\nu}(p)+\frac{2p^{4}+m^{2}p^{2}-m^{4}}{m^{4}}\ I^{m0}_{\mu\nu}(p).
\end{align}
Inserting the result in Eq.~(\ref{p4}) the fourth polarization term in Eq.~(\ref{Pi}) is
\begin{align}
{\Pi_{4}}_{\mu\nu}(p)&=p_{\mu}p_{\nu}\ \bigg[\frac{p^{4}}{2m^{4}}\ I^{00}(p)+\frac{p^{2}(p^{2}+2m^{2})}{2m^{4}}\ I^{mm}(p)
-\frac{p^{2}(p^{2}+m^{2})}{m^{4}}\ I^{m0}(p)-I^{0m}(0)\bigg]+\nn\\
&-2\ \frac{p^{4}}{m^{4}}\ I^{00}_{\mu\nu}(p)-2\ \frac{p^{2}(p^{2}+2m^{2})}{m^{4}}\ I^{mm}_{\mu\nu}(p)
+\frac{2p^{4}+3m^{2}p^{2}+m^{4}}{m^{4}}\ I^{m0}_{\mu\nu}(p)+\frac{2p^{4}+m^{2}p^{2}-m^{4}}{m^{4}}\ I^{0m}_{\mu\nu}(p).
\label{p4tot}
\end{align}
\vskip 15pt
\centerline{* \quad * \quad * \quad * \quad  * \quad *  \quad *}
\vskip 15pt

Finally, adding up the four polarization terms in Eqs.~(\ref{p1tot}),(\ref{p2tot}),(\ref{p3tot}),(\ref{p4tot}),
the total graph $(2b)$ reads
\begin{align}
&{\Pi^{(2b)}_{\mu\nu}}(p)=
\frac{p^{4}}{2m^{4}}\ I_{\mu\nu}^{00}(p)+\bigg[4+\frac{(p^{2}+2m^{2})^{2}}{2m^{4}}\bigg]\ I^{mm}_{\mu\nu}(p)
-\frac{(p^{2}+m^{2})^{2}}{2m^{4}}\ \Big(I^{m0}_{\mu\nu}(p)+I^{0m}_{\mu\nu}(p)\Big)+I^{m0}_{\mu\nu}(0)+\nn\\
&\quad+\frac{(p^{2}+m^{2})}{m^{2}}\ \Big(I^{0m}_{\mu\nu}(p)-I^{m0}_{\mu\nu}(p)\Big)+
\delta_{\mu\nu}\ \bigg[\frac{p^{2}}{m^{2}}\ (p^{2}+4m^{2})\ I^{mm}(p)-\frac{(p^{2}+m^{2})^{2}}{m^{2}}\ I^{0m}(p)
+(p^{2}+m^{2})\ I^{0m}(0)\bigg]+\nn\\
&\qquad-p_{\mu}p_{\nu}\ 
\bigg[\frac{p^{4}}{8m^{4}}\ I^{00}(p)+
\frac{(p^{4}+12m^2 p^2+44m^{4})}{8m^{4}}\ I^{mm}(p)
-\frac{(p^{2}+m^{2})(p^{2}+5m^{2})}{4m^{4}}\ I^{m0}(p)
-\frac{1}{4}\ I^{m0}(0)\bigg].
\label{p2btot}
\end{align}

\end{widetext}

The transverse projections of the graph follow by the projected integrals in Eqs.~(\ref{projI}).
We observed that by Eq.~(\ref{symmkq})
the projected integrals turn out to be symmetric, $I^{\alpha\beta}_{L,T}= I^{\beta\alpha}_{L,T}$.
Thus, the projection of the graph follows by dropping the longitudinal  and
the antisymmetric terms, and by replacing the integrals by the projected ones according to
\begin{align}
p_\mu p_\nu\ &\to\ 0,\nn\\
\left(I^{0m}_{\mu\nu}(p)-I^{m0}_{\mu\nu}(p)\right)&\to\ 0,\nn\\
\delta_{\mu\nu} \ &\to\ 1\nn\\
I^{\alpha\beta}_{\mu\nu}(p)\ &\to\ I^{\alpha\beta}_{L,T}(p)=I^{\beta\alpha}_{L,T}(p),\nn\\
I^{\alpha\beta}_{\mu\nu}(0)\ &\to\ I^{\alpha\beta}_{L,T\, p}=I^{\beta\alpha}_{L,T\, p}.
\end{align}

\subsection{Graph 2a - (ghost loop)}
In the Landau gauge, setting $d=4$ and using a free-particle propagator, the general expression of
the ghost loop (see e.g. Ref.~~\cite{genself}) reads
\BE
\Pi^{(2a)}_{\mu\nu}(p)=\int_k (p_\mu-k_\mu)k_\nu\, G_0(k)\, G_0(p-k).
\EE

By exchanging $k^\mu$ and $p^\mu-k^\mu$ the integral shows the symmetry $\Pi_{\mu\nu}=\Pi_{\nu\mu}$ so
that, using Eq.~(\ref{identpk}), we can replace
\BE
p_\mu k_\nu\ \to \ \frac{1}{2}(p_\mu k_\nu+k_\mu p_\nu)=\frac{1}{2}(k_\mu k_\nu-q_\mu q_\nu+p_\mu p_\nu).
\EE
The first two terms on the right hand side cancel in the integration yielding
\BE
\Pi^{(2a)}_{\mu\nu}(p)=\frac{1}{2} p_\mu p_\nu\ I^{00}(p)-I^{00}_{\mu\nu}(p).
\label{p2atot}
\EE
The projected ghost loop is just
\BE
\Pi^{(2a)}_{L,T}(p)=-I^{00}_{L,T}(p).
\label{p2aproj}
\EE

\subsection{Total (uncrossed) one-loop polarization}

Adding up the uncrossed one-loop graphs $(1b)$, $(2b)$ and $(2a)$, the standard  (uncrossed)  projected one-loop polarization
of Ref.~~\cite{serreau} is recovered by the sum of Eqs.~(\ref{p1btot}), (\ref{p2btot}) and (\ref{p2atot}):
\begin{widetext}
\begin{align}
\Pi_{L,T}^{1-loop}(p)=&
\bigg[\frac{p^{4}}{2m^{4}}-1\bigg]\ I_{L,T}^{00}(p)+\bigg[4+\frac{(p^{2}+2m^{2})^{2}}{2m^{4}}\bigg]\ I^{mm}_{L,T}(p)
-\frac{(p^{2}+m^{2})^{2}}{m^{4}}\ I^{m0}_{L, T}(p)+\nn\\
&\quad+\frac{p^{2}(p^{2}+4m^{2})}{m^{2}}\ I^{mm}(p)-\frac{(p^{2}+m^{2})^{2}}{m^{2}}\ I^{0m}(p)+(p^{2}+m^{2})\ I^{0m}(0)-2J_m.
\label{tot1L}
\end{align}

\subsection{Ghost self-energy}

In this work, the total one-loop ghost self energy is the sum of the standard one-loop graph and the
crossed one, which contains the insertion of a mass counterterm,
\BE
\Sigma^{tot}(p)=\left(1-m^2\frac{\partial}{\partial m^2}\right) \Sigma(p)
\label{sigtot}
\EE
where $\Sigma(p)$ is the standard one loop integral~\cite{genself,ptqcd2} in the Landau gauge,
\BE
\Sigma(p)=-\int_k \frac{k^2 p^2- (k\cdot p)^2}{k^2 (k-p)^2 (k^2+m^2)}
=-\int_k \left[p^2\, G_m(k)G_0(q)-(k\cdot p)^2\, G_0(k)G_m(k)G_0(q) \right].
\label{sig}
\EE
Using the last of Eqs.~(\ref{ident2}) with $\alpha=m$ and $\beta=0$, and decoupling the product
$G_m(k)G_0(k)$ by the second of Eqs.~(\ref{ident}), we can write
\begin{align}
4(k\cdot p)^2\, G_0(q)G_m(k)G_0(k)&=\frac{(p^2-m^2)^2}{m^2}\left[G_0(k)G_0(q)-G_m(k)G_0(q)\right]+\nn\\
&+(p^2-m^2)\left[G_0(k)G_0(q)-G_m(k)G_0(k)\right]
+2(p\cdot k)\left[G_0(q)G_0(k)-G_m(k)G_0(k)\right].
\end{align}
Then using the second of Eqs.~(\ref{ident2}) with $\alpha=\beta=0$ and dropping the vanishing
integrals
\BE
\int_k\left[G_0(q)-G_0(k)\right]=0,\qquad \int_k (p\cdot k)\, G_m(k)G_0(k)=0,
\EE
the second term of Eq.~(\ref{sig}) reads
\begin{align}
\int_k (k\cdot p)^2\, G_0(k)G_m(k)G_0(q)&=
-\frac{(p^2-m^2)^2}{4m^2}\int_k G_m(k)G_0(q)+\frac{p^4}{4m^2}\int_k G_0(k)G_0(q)
-\frac{(p^2-m^2)}{4}\int_k G_m(k)G_0(k)
\end{align}
and the (uncrossed) one-loop self energy can be written as
\BE
\Sigma(p)=-\frac{(p^2+m^2)^2}{4m^2}\, I^{m0}(p) +\frac{p^4}{4m^2}\, I^{00}(p) 
+\frac{(p^2-m^2)}{4m^2}\,(J_m-J_0) ,
\label{sig1L}
\EE
as derived in Ref.~~\cite{serreau} by the same method.
\end{widetext}

\subsection{Crossed graphs and total polarization}

The crossed graphs $(1c)$, $(2c)$, $(1d)$ and the crossed one-loop ghost self energy can be obtained by simple
derivatives. The sum of all graphs gives a total one-loop polarization that can be written
as
\BE
\Pi_{L,T}^{tot} (p)=\Pi_{L,T}^{(a-c)}(p)+\Pi_{L, T}^{(1d)}(p),
\label{PiLTtot}
\EE
where $\Pi_{L,T}^{(a-c)}(p)$ is the sum of graphs $(2a)$, $(1b)$, $(2b)$, $(1c)$, $(2c)$ 
and can be evaluated as
\BE
\Pi_{L,T}^{(a-c)}(p)=\left(1-m^2\frac{\partial}{\partial m^2}\right) \Pi_{L,T}^{1-loop} (p).
\EE
Here $\Pi_{L,T}^{1-loop}(p)$ is the projected one-loop polarization of Eq.~(\ref{tot1L}) and $\Pi_{L, T}^{(1d)}$
is the doubly crossed tadpole, with two counterterm insertions.  

The derivative acts on the coefficients of the integrals according to 
\begin{align}
\left(-m^2\frac{\partial}{\partial m^2}\right)\left[ m^2\right]&= -m^2,\nn\\
\left(-m^2\frac{\partial}{\partial m^2}\right)\left[ \frac{1}{m^2}\right]&= \frac{1}{m^2},\nn\\
\left(-m^2\frac{\partial}{\partial m^2}\right)\left[ \frac{1}{m^4}\right]&= \frac{2}{m^4}.
\label{dm}
\end{align}
The function $\Pi_{L,T}^{(a-c)}$ then reads
\begin{widetext}
\BE
\Pi_{L,T}^{(a-c)}(p)=\Pi_{L,T}^{1-loop} (p)+\left(-m^2\frac{\partial}{\partial m^2}\ \Pi_{L,T}^{1-loop}(p)\right)_{I}
-m^2 \left(\Pi_{L,T}^{1-loop}(p)\right)_{I\to \partial I},
\label{sum}
\EE
where the derivative of the coefficients is taken in the second term while the derivative of the integrals is considered
in the third term. Using Eq.~(\ref{dm}) and Eq.~(\ref{tot1L}),
\begin{align}
\left(-m^2\frac{\partial}{\partial m^2}\ \Pi_{L,T}^{1-loop}(p)\right)_{I}&=
\bigg[\frac{p^{4}}{m^{4}}\bigg]\ I_{L,T}^{00}(p)
+\bigg[\frac{p^{4}+2m^{2}p^{2}}{m^{4}}\bigg]\ I^{mm}_{L,T}(p)
-\bigg[\frac{2p^{4}+2m^{2}p^{2}}{m^{4}}\bigg]\ I^{m0}_{L, T}(p)+\nn\\
&\quad+\bigg[\frac{p^{4}}{m^{2}}\bigg]\ I^{mm}(p)
-\bigg[\frac{p^{4}-m^{4}}{m^{2}}\bigg]\ I^{0m}(p)-\bigg[m^{2}\bigg]\ I^{0m}(0),
\label{Icost}
\end{align}
while replacing the integrals $I$ by their derivatives $\partial I$, Eq.~(\ref{tot1L}) reads
\begin{align}
-m^2 \left(\Pi_{L,T}^{1-loop}(p)\right)_{I\to \partial I}=&
-\bigg[8m^2+\frac{(p^{2}+2m^{2})^{2}}{m^{2}}\bigg]\ \partial I^{mm}_{L,T}(p)
+\frac{(p^{2}+m^{2})^{2}}{m^{2}}\ \partial I^{m0}_{L, T}(p)-2p^{2}(p^{2}+4m^{2})\ \partial I^{mm}(p)+\nn\\
&+(p^{2}+m^{2})^{2}\ \partial I^{m0}(p)
-m^2(p^{2}+m^{2})\ \partial I^{m0}(0)+2m^2\ \partial J_m.
\label{Ider}
\end{align}
Summing up the contributions of Eq.~(\ref{tot1L}), (\ref{Icost}) and (\ref{Ider}) in Eq.~(\ref{sum}) and using Eq.~(\ref{identI}) we obtain 
\begin{align}
\Pi_{L,T}^{(a-c)}(p)&=
\bigg[\frac{3p^{4}}{2m^{4}}-1\bigg]\ I_{L,T}^{00}(p)+\bigg[4+\frac{3p^{4}+8m^{2}p^2+4m^4}{2m^{4}}\bigg]\ I^{mm}_{L,T}(p)
-\bigg[\frac{3p^{4}+4m^{2}p^2+m^4}{m^{4}}\bigg]\ I^{m0}_{L, T}(p)+\nn\\
&\quad+\frac{2p^{2}(p^{2}+2m^{2})}{m^{2}}\ I^{mm}(p)-\frac{2p^2(p^{2}+m^{2})}{m^{2}}\ I^{0m}(p)
-\bigg[\frac{2p^2+3m^2}{m^2}\bigg]\ J_m+\bigg[\frac{2p^2+m^2}{m^2}\bigg]\ J_0+\nn\\
&-\bigg[8m^2+\frac{(p^{2}+2m^{2})^{2}}{m^{2}}\bigg]\ \partial I^{mm}_{L,T}(p)
+\frac{(p^{2}+m^{2})^{2}}{m^{2}}\ \partial I^{m0}_{L, T}(p)-2p^{2}(p^{2}+4m^{2})\ \partial I^{mm}(p)+\nn\\
&+(p^{2}+m^{2})^{2}\ \partial I^{m0}(p)+(p^{2}+3m^{2})\  \partial J_m.
\end{align}
\end{widetext}

Finally, the doubly crossed tadpole $(1d)$ in Eq.~(\ref{PiLTtot}) can be written as~\cite{genself,ptqcd2}
\BE
\Pi_{L, T}^{(1d)}(p)=\frac{m^4}{2}\frac{\partial^2}{\partial (m^2)^2}\Pi_{L, T}^{(1b)}(p),
\EE
and using Eq.~(\ref{1bLT})
\BE
\Pi_{L, T}^{(1d)}(p)=-m^4\ \partial^2 J_m-\frac{m^4}{2}\ \partial^2 I^{m0}_{L,T\, p}\>.
\EE
By Eqs.~(\ref{identILT}) the derivative $\partial^2 I^{m0}_{L,T\, p}$ can be expressed in terms of the integrals 
$J^{L, T}_m$ and their derivatives $\partial J^{L, T}_m$, yielding 
\begin{align}
\Pi_{L, T}^{(1d)}(p)&=-m^4\ \partial^2 J_m
+\frac{1}{m^2}(J^{L,T}_{m,\, p}-J^{L,T}_{0,\, p})+\nn\\
&\qquad -\partial J^{L,T}_{m,\, p}+\frac{m^2}{2}\ \partial^2 J^{L,T}_{m,\, p},
\end{align}
where
\begin{align}
J^{T}_{m,\, p}&=J^{T}_m,\nn\\
J^{L}_{m,\, p}&=(J^{L}_m-J^{T}_m)\>\frac{{\bf p}^2}{{\bf p}^2+\omega^2}+J^{T}_m.
\end{align}

\subsection{Crossed graphs and total ghost self energy}

The total ghost self-energy $\Sigma^{tot}(p)$
can be derived by the same method, as shown in Eq.~(\ref{sigtot}),
\BE
\Sigma^{tot}(p)=\Sigma(p)+\left(-m^2\frac{\partial}{\partial m^2}\ \Sigma(p)\right)_{I}
-m^2 \left[\Sigma (p)\right]_{I\to \partial I},
\label{sigsum}
\EE
where the derivative of the coefficients is taken in the second term, while the derivative of the integrals is considered
in the third term.

Replacing the integrals $I$ by their derivatives $\partial I$, Eq.~(\ref{sig1L}) gives
\BE
-m^2 \left[\Sigma (p)\right]_{I\to \partial I}=
\frac{(p^2+m^2)^2}{4}\, \partial I^{m0}(p) 
-\frac{(p^2-m^2)}{4}\, \partial J_m ,
\label{sig1der}
\EE
while, using Eq.~(\ref{dm}), the derivative of the coefficients in Eq.~(\ref{sig1L}) gives
\begin{widetext}
\BE
\left(-m^2\frac{\partial}{\partial m^2}\ \Sigma(p)\right)_{I}=
\frac{(m^4-p^4)}{4m^2}\, I^{m0}(p) 
+\frac{p^4}{4m^2}\, I^{00}(p) 
+\frac{p^2}{4m^2}\,(J_m-J_0). 
\label{sigIcost}
\EE
The total ghost self energy then follows, adding up the contributions of Eq.~(\ref{sig1L}), (\ref{sigIcost}) and (\ref{sig1der}) in
Eq.~(\ref{sigsum})
\BE
\Sigma^{tot}(p)=
-\frac{p^2(p^2+m^2)}{2m^2}\, I^{m0}(p) 
+\frac{p^4}{2m^2}\, I^{00}(p) 
+\frac{(2p^2-m^2)}{4m^2}\,(J_m-J_0)
+\frac{(p^2+m^2)^2}{4}\, \partial I^{m0}(p) 
-\frac{(p^2-m^2)}{4}\, \partial J_m.  
\label{sigTOT}
\EE
\end{widetext}

\section{Thermal integrals}

By general arguments, the thermal integral $I(T)$ of a function $f(k)=f({\bf k},k_4)$ can be written as
\BE
I(T)=\int_k f(k)=T\sum_n\int\frac{{\rm d}^3{\bf k}} {(2\pi)^3} f({\bf k},\omega_n)=I_V+I_{Th}(T)
\label{Igen}
\EE
where, setting $k_4=\omega_n=-i k_0$,
\BE
I_V=\frac{1}{2\pi i}\int_{-i\infty}^{+i\infty} {\rm d} k_0\int \frac{{\rm d}^3{\bf k}} {(2\pi)^3} f({\bf k},-ik_0)
=\int \kkk f(k)
\EE
is the Euclidean integral at $T=0$, denoted {\it vacuum} part $I_V=I(0)$, while the
{\it thermal} part $I_{Th}(T)$ is
\BE
I_{Th}(T)=-\int \frac{{\rm d}^3{\bf k}} {(2\pi)^3} 
\sum_{\rm Resid.}\left[\frac{2 \,\Re f({\bf k},ik_0)}{e^{\beta k_0}-1}\right]_{{\rm Re}k_0>0}
\label{Ith}
\EE
where the sum is over the residues in the right complex plane of $k_0$ and the symbol $\Re f$ 
is defined as
\BE
\Re f({\bf k},ik_0)=\frac{f({\bf k},ik_0)+f({\bf k},-ik_0)}{2}.
\EE
We observe that if $f(k)$ is a complex function, then $\Re f(k)$ is not the true real part ${\rm Re} f(k)$.
The thermal part vanishes in the limit $T\to 0$.

Many of the thermal integrals were evaluated in great detail in Ref.~~\cite{serreau}. In the next sections we
collect the same results and, by the same method, we add the explicit evaluation of all the remaining integrals that
are required in the present work.

\subsection{Vacuum integrals}

The vacuum parts of all the one-loop graphs were evaluated in Ref.~~\cite{ptqcd2}. 
They can be made finite by wave function renormalization. After subtraction,
the sum of all the gluon polarization graphs in Eq.~(\ref{PiLTtot}) and of all ghost self-energy graphs
in Eq.~(\ref{sigTOT}) give the following vacuum terms at $T=0$:
\begin{align}
\Pi^{tot}_V(s)&=-\frac{3m^2\,s}{(4\pi)^2}\left[\pi_1(s)+\pi_0\right],\nn\\
\Sigma^{tot}_V(s)&=\>\frac{3m^2\,s}{(4\pi)^2}\left[\sigma_1(s)+\sigma_0\right],
\end{align}
where $s=p^2/m^2$, the constants $\pi_0$, $\sigma_0$ are arbitrary renormalization constants, depending on the subtraction point, 
and $\pi_1(s)$, $\sigma_1(s)$ are the explicit analytical functions
\begin{align}
\pi_1(x)&=\frac{5}{8x}+\frac{1}{72}\left[L_a+L_b+L_c+R_a+R_b+R_c\right],\nn\\
\sigma_1(x)&=\frac{1}{12}\left[L_g+R_g\right],
\label{FGx}
\end{align}
written in terms of the logarithmic functions $L_x$ 
\begin{align}
L_a(x)&=\frac{3x^3-34x^2-28x-24}{x}\>\times\nn\\
&\times\sqrt{\frac{4+x}{x}}
\log\left(\frac{\sqrt{4+x}-\sqrt{x}}{\sqrt{4+x}+\sqrt{x}}\right),\nn\\
L_b(x)&=\frac{2(1+x)^2}{x^3}(3x^3-20x^2+11x-2)\log(1+x),\nn\\
L_c(x)&=(2-3x^2)\log(x),\nn\\
L_g(x)&=\frac{(1+x)^2(2x-1)}{x^2}\log(1+x)-2x\log(x)
\label{logsA}
\end{align}
and of the rational parts $R_x$ 
\begin{align}
R_a(x)&=-\frac{4+x}{x}(x^2-20x+12),\nn\\
R_b(x)&=\frac{2(1+x)^2}{x^2}(x^2-10x+1),\nn\\
R_c(x)&=\frac{2}{x^2}+2-x^2,\nn\\
R_g(x)&=\frac{1}{x}+2.
\label{rational}
\end{align}

\subsection{Thermal part of $\boldsymbol{J_m}$ and $\boldsymbol{J_m^{L,T}}$}

The integral $J_m$ is defined in Eq.~(\ref{IJ}) and
has the general form of Eq.~(\ref{Igen}) with
\BE
f({\bf k},ik_0)=G_m({\bf k},ik_0)=\frac{1}{\omkm^2-k_0^2},
\label{fJ}
\EE
having denoted by $\omkm$ the positive square root
\BE
\omkm=\sqrt{\bk^2+m^2}.
\EE
The thermal part, Eq(\ref{Ith}), takes a contribution at the pole
$k_0=\omkm$, yielding
\BE
(J_m)_{Th}=-\int \frac{{\rm d}^3{\bf k}} {(2\pi)^3} 
\left[ \left(\frac{-2}{\omkm+k_0}\right)\left(\frac{1}{e^{\beta k_0}-1}\right) \right]_{k_0=\omkm},
\EE
and denoting by $n(\epsilon)$ the Bose distribution,
\BE
n(\epsilon)=\left[{e^{\beta \epsilon}-1}\right]^{-1},
\EE
we obtain
\BE
(J_m)_{Th}=\int \frac{{\rm d}^3{\bf k}} {(2\pi)^3} 
\,\frac{n(\omkm)}{\omkm}
=\int_0^\infty \frac{x^2{\rm d} x} {2\pi^2} 
\,\frac{n(\omxm)}{\omxm},
\label{Jth}
\EE
with the obvious notation $\omxm=\sqrt{x^2+m^2}$. 

In the special case $m=0$,
\BE
(J_0)_{Th}=\int_0^\infty \frac{x{\rm d} x} {2\pi^2} 
\,{n(x)}.
\EE

The thermal parts of the integrals $J_m^{L}$, $J_m^{T}$, as
defined in Eq.~(\ref{JLT}), follow immediately by replacing
$f(k)\to-k_0^2 f(k)$ and $f(k)\to \frac{1}{3}\bk^2 f(k)$,
respectively, in Eq.~(\ref{fJ}). Following the same steps as before,
the thermal parts read
\begin{align}
(J^L_m)_{Th}&=-\int_0^\infty \frac{x^2{\rm d} x} {2\pi^2} \,{\omxm}\,n(\omxm),\nn\\
(J^T_m)_{Th}&=\int_0^\infty \frac{x^4{\rm d} x} {6\pi^2} \,\frac{n(\omxm)}{\omxm}.
\label{JLTth}
\end{align}

\subsection{Thermal part of $\boldsymbol{I^{\alpha\beta}(p)}$}

The integral $I^{\alpha\beta}(p)$ is also defined in Eq.~(\ref{IJ}) and
has the general form of Eq.~(\ref{Igen}) with

\begin{align}
f(k)&=G_\alpha(k)G_\beta(p-k)=\nn\\
&\quad=\frac{1}{(\omka^2-k_0^2)\left[\ompkb^2-(p_0-k_0)^2\right]},
\label{fk}
\end{align}
where $\ompkb=\sqrt{(\bp-\bk)^2+\beta^2}$ and $-i p_0=p_4$ is the external frequency. 
The poles are at $k_0=\pm \omka$ and $k_0=p_0\pm \ompkb$. The residues are readily
evaluated:
\begin{align}
R_{\alpha}^{\pm}&=\mp\frac{1}{2\omka}G_\beta(\bp-\bk,ip_0\mp i\omka),\nn\\
R_{\beta}^{\pm}&=\mp\frac{1}{2\ompkb}G_\alpha(\bk,ip_0\pm i\ompkb),
\end{align}
and we can write
\begin{widetext}

\begin{align}
f(\bk,ik_0)&=\sum_{\pm}\frac{R^\pm_\alpha}{k_0\mp\omka}+\sum_{\pm}\frac{R^\pm_\beta}{k_0-p_0\mp\ompkb}
=A_{\alpha\beta}(\bk,\bp-\bk;ik_0,ip_0)+A_{\beta\alpha}(\bp-\bk,\bk;ip_0-ik_0,ip_0),
\label{fk2}
\end{align}
where
\BE
A_{\alpha\beta}(\bk,\bp-\bk;ik_0,ip_0)=\frac{1}{2\omka}\left[\frac{G_\beta(\bp-\bk,ip_0+i\omka)}{k_0+\omka}
-\frac{G_\beta(\bp-\bk,ip_0-i\omka)}{k_0-\omka}\right].
\label{A}
\EE
It can be easily shown that for any external frequency $\omega_n^\prime=-i p_0=2\pi T  n^\prime$ and momentum $\bp$, the
integral over $\bk$ and the sum over $\omega_n=-i k_0= 2\pi T  n$ have the property
\BE
T\sum_n\int\frac{{\rm d}^3{\bf k}} {(2\pi)^3}\ A_{\alpha\beta}(\bk,\bp-\bk;ik_0,ip_0)=
T\sum_n\int\frac{{\rm d}^3{\bf k}} {(2\pi)^3}\ A_{\alpha\beta}(\bp-\bk,\bk;ip_0-ik_0,ip_0),
\EE
which follows by replacing $\bk\to \bp-\bk$  and $k_0\to p_0-k_0$ in the integral and in the sum.
Thus, we can replace in Eq.~(\ref{Ith})
\BE
\Re f(\bk,ik_0)=\left\{\ \Re{\left[A_{\alpha\beta}(\bk,\bp-\bk;ik_0,ip_0)\right]}\quad +\quad \alpha\leftrightarrow\beta\ \right\}.
\label{delf}
\EE
Moreover, since $G_m(\bp,ip_0)=G_m(\bp,-ip_0)$, by inspection of Eq.~(\ref{A}), we observe that
$A_{\alpha\beta}(\bk,\bp-\bk;-ik_0,ip_0)=A_{\alpha\beta}(\bk,\bp-\bk;ik_0,-ip_0)$, so that 
\BE
\Re \left[ A_{\alpha\beta}(\bk,\bp-\bk;ik_0,ip_0)\right]=\frac{1}{2}\left[A_{\alpha\beta}(\bk,\bp-\bk;ik_0,ip_0)
+A_{\alpha\beta}(\bk,\bp-\bk;ik_0,-ip_0)\right].
\EE
Hereafter, the last equation is taken as the definition of the symbol $\Re$ for any generic function of $ip_0$.

In Eq.~(\ref{Ith}), the poles at $k_0=\omka$, $\omkb$ have the residues $[-n(\omka)/\omka]\, \Re\, G_\beta(\bp-\bk, ip_0-i\omka)$ and 
$[-n(\omkb)/\omkb]\,\Re\, G_\alpha(\bp-\bk, ip_0-i\omkb)$, respectively, yielding in terms of the external frequency $\omega=p_4=-ip_0$
\BE
\left[I^{\alpha\beta}(\bf p,\omega)\right]_{Th}=\int \frac{{\rm d}^3{\bf k}} {(2\pi)^3}\ 
\left\{\frac{n(\omka)}{\omka}\ \Re\, G_\beta(\bp-\bk, \omega+i\omka)\quad +\quad \alpha\leftrightarrow\beta\ \right\}.
\label{Iab}
\EE
Finally, we observe that since $G_m(\bp,ip_0)=G_m(\bp,-ip_0)$, then
\BE
\Re\, G_\beta(\bp-\bk, ip_0-i\omka)=\frac{1}{2}\left[G_\beta(\bp-\bk, \omega+i\omka)+G_\beta(\bp-\bk, \omega-i\omka)\right].
\EE
\end{widetext}
The angular integral in Eq.~(\ref{Iab}) can be evaluated exactly by writing
\BE
G_\alpha(\bp-\bk, z)=\frac{1}{g_\alpha(z,\bp^2,\bk^2)-2\bp\cdot\bk},
\EE
where, denoting $x=\sqrt{\bk^2}$ and $y=\sqrt{\bp^2}$, the function $g_\alpha(z; x^2,y^2)$ is given by
\BE
g_\alpha(z; y^2, x^2)=z^2+\alpha^2+x^2+y^2
\label{defg}
\EE
and does not depend on the angles. Moreover, we observe that
\BE
g_\alpha(z; y^2, x^2)\pm 2xy=z^2+\epsilon^2_{y\pm x, \alpha},
\label{geps}
\EE
where
\BE
\epsilon_{y\pm x, \alpha}=\sqrt{(y\pm x)^2+\alpha^2},
\EE
so that the integral over the angles can be written in terms of the function
\BE
L_\alpha(z; y,x)=\log\frac{z^2+\opa^2}{z^2+\oma^2}
\label{logalpha}
\EE
and an elementary integration gives
\begin{widetext}
\BE
\left[I^{\alpha\beta}(y,\omega)\right]_{Th}=\int_0^\infty \frac{ x{\rm d}x} {8\pi^2 y}
\left\{\frac{n(\omxa)}{\omxa}\ \Re\, L_\beta(\omega+i\omxa;\, y,x)+\quad \alpha\leftrightarrow\beta\ \right\}.
\label{Iabth}
\EE

It might be useful to evaluate the leading behavior in the long wavelength limit $\bp\to 0$ (i.e. $y\to 0$):
\BE
z^2+\epsilon^2_{y\pm x, \beta}=(z^2+\omxb^2)\left[1\pm\frac{2xy}{z^2+\omxb^2}+\frac{y^2}{z^2+\omxb^2}\right],
\EE
\BE
L_\beta(z; y,x)=\frac{4xy}{z^2+\omxb^2}-\frac{4xy^3}{(z^2+\omxb^2)^2}+\frac{16x^3y^3}{3(z^2+\omxb^2)^3}+{\cal O}(y^5),
\label{Lx0}
\EE
\BE
\left[I^{\alpha\beta}(y\to 0,\omega)\right]_{Th}\approx\int_0^\infty \frac{ x^2{\rm d}x} {2\pi^2}
\left\{\frac{n(\omxa)}{\omxa}\ \Re\,\frac{1}{(\omega+i\omxa)^2+\omxb^2}+\quad \alpha\leftrightarrow\beta\ \right\}.
\EE
Moreover, in the limit $\omega\to 0$, using Eq.~(\ref{Jth}),
\BE
\lim_{\omega\to 0}\lim_{y\to 0} \left[I^{\alpha\beta}(y,\omega)\right]_{Th}=\int_0^\infty \frac{ x^2{\rm d}x} {2\pi^2}
\left\{\frac{n(\omxa)}{\omxa}\ \frac{1}{\beta^2-\alpha^2}+\alpha\leftrightarrow\beta\right\}
=\frac{(J_\alpha)_{Th}-(J_\beta)_{Th}}{\beta^2-\alpha^2},
\EE
\end{widetext}
in agreement with the first of Eqs.~(\ref{identI}). The same limit is obtained by setting $\omega=0$ from the beginning and
exploring the leading behavior when $y\to 0$.

\subsection{Thermal part of $\boldsymbol{I^{\alpha\beta}_{L,T}(p)}$}

The projected integrals $I^{\alpha\beta}_{L,T}(p)$ were defined in Eq.~(\ref{projI}) and
have the general form of Eq.~(\ref{Igen}) with

\begin{align}
f(k)&=G_\alpha(k)G_\beta(p-k)\, \frac{k_\mu k_\nu}{c_{L,T}}\, P_{\mu\nu}^{L,T}(p),
\end{align}
where $c_L=1$ and $c_T=2$. The function $f(k)$ is the same found for the integral $I^{\alpha\beta}(p)$ 
in Eq.~(\ref{fk}), multiplied by a factor
\BE
f(k)\to f(k)\ \left[ \frac{k_\mu k_\nu}{c_{L,T}}\, P_{\mu\nu}^{L,T}(p)\right].
\label{fact}
\EE
The new factor has no poles in the complex $k_0$ plane and does not depend on the masses $\alpha$, $\beta$. 
Thus, $f(k)$ has the same pole structure of Eq.~(\ref{fk2})
with residues multiplied by the same factor. Moreover, we observe that because of Eq.~(\ref{symmkq}),
we can still exchange $k$ and $p-k$ in the integral without affecting the multiplied factor. 
Then, Eq.~(\ref{delf}) still holds with the function $A_{\alpha\beta}$
just multiplied by the same factor of Eq.~(\ref{fact}), which by an explicit calculation reads
\BE
\left[ {k_\mu k_\nu}\, P_{\mu\nu}^{L}(p)\right]
=\frac{\left[  (\bk\cdot \bp)\omega+ik_0\bp^2 \right]^2}{(\bp^2+\omega^2)\bp^2}
\EE
and
\BE
\left[ \frac{k_\mu k_\nu}{2}\, P_{\mu\nu}^{T}(p)\right]=\frac{1}{2}\left[\bk^2-\frac{(\bk\cdot\bp)^2}{\bp^2}\right],
\EE
to be evaluated at the poles $k_0=\omka$ and $k_0=\omkb$, yielding
\begin{widetext}
\begin{align}
\left[I^{\alpha\beta}_L (\bf p,\omega)\right]_{Th}&=\int \frac{{\rm d}^3{\bf k}} {(2\pi)^3}\ 
\left\{\frac{n(\omka)}{\omka}\ 
\Re\,\left[ \frac{ \left(\, (\bk\cdot \bp)\omega+i\omka\bp^2 \right)^2}{(\bp^2+\omega^2)\bp^2}\
G_\beta(\bp-\bk, \omega+i\omka)\right]\quad +\quad \alpha\leftrightarrow\beta\ \right\},\nn\\
\left[I^{\alpha\beta}_T (\bf p,\omega)\right]_{Th}&=\frac{1}{2}\int \frac{{\rm d}^3{\bf k}} {(2\pi)^3}\ 
\left[\bk^2-\frac{(\bk\cdot\bp)^2}{\bp^2}\right]\
\left\{\frac{n(\omka)}{\omka}\ \Re\, G_\beta(\bp-\bk, \omega+i\omka)\quad +\quad \alpha\leftrightarrow\beta\ \right\},
\label{ILT}
\end{align}
where the symbol $\Re$ denotes an average over $\pm\omega$ or, equivalently, an average over $\pm i\omka$.

The angular integrals can be evaluated exactly~\cite{serreau}. In the transverse projection, we can write
\begin{align}
\int \frac{{\rm d}^3{\bf k}} {(2\pi)^3}\ \left[\bk^2-\frac{(\bk\cdot\bp)^2}{\bp^2}\right]\ G_\alpha(\bp-\bk, z)
&=\int_0^\infty \frac{x^4{\rm d} x} {4\pi^2}\int_{-1}^{1} {\rm d}\cos\theta\
\frac{1-\cos^2\theta}{g_\alpha(z; y^2,x^2)-2xy\cos\theta}=\nn\\
&=\int_0^\infty \frac{x^2{\rm d} x} {8\pi^2\, y^2}\ 
\left[ g_\alpha(z; y^2,x^2)-\frac{\left(\,[g_\alpha(z; y^2,x^2)]^2-4x^2y^2\right)}{4xy}\, L_\alpha(z; y,x)\right].
\end{align}
Then, denoting by $L^T_\alpha$ the transverse logarithmic function
\BE
L^T_\alpha(z; y,x)=(z^2+\opa^2)(z^2+\oma^2)L_\alpha(z; y,x)
\EE
and using Eq.~(\ref{geps}), we can write
\BE
\left[I^{\alpha\beta}_T (y,\omega)\right]_{Th}=-\int_0^\infty \frac{ x\,{\rm d}x} {64\pi^2 y^3}
\left\{\frac{n(\omxa)}{\omxa}\ \Big[\Re\, L^T_\beta(\omega+i\omxa;\, y,x)-4xy(\omega^2+y^2+\beta^2-\alpha^2)\Big]
+\quad \alpha\leftrightarrow\beta\ \right\}.
\label{ITth}
\EE

In the longitudinal projection, the angular integration reads
\begin{align}
&\int \frac{{\rm d}^3{\bf k}} {(2\pi)^3}\ 
\left[ \frac{ \left(\, (\bk\cdot \bp)\omega+(z-\omega)\bp^2 \right)^2}{\bp^2}\right]\
\ G_\alpha(\bp-\bk, z)
=\int_0^\infty \frac{x^2{\rm d} x} {4\pi^2}\int_{-1}^{1} {\rm d}\cos\theta\
\frac{\left[x\,\omega\cos\theta+y(z-\omega)     \right]^2}{g_\alpha(z; y^2,x^2)-2xy\cos\theta}=\nn\\
&\quad=\int_0^\infty \frac{\omega^2\,x\,{\rm d} x} {32\pi^2\, y^3}\ 
\left\{
\left[ g_\alpha(z; y^2,x^2)+2y^2\left(\frac{z}{\omega}-1\right)\right]^2\ L_\alpha(z; y,x)
-4xy\, g_\alpha(z; y^2,x^2)-\frac{16xy^3(z-\omega)}{\omega}\right\}.
\end{align}
Denoting by $L^L_\alpha$ the longitudinal logarithmic function
\BE
L^L_\alpha(z; y,x)=\left[z^2+\omxa^2+y^2\left(\frac{2z}{\omega}-1\right)\right]^2\ L_\alpha(z; y,x),
\EE
using Eq.~(\ref{defg}) and observing that $\Re(z-\omega)$ vanishes when evaluated at $z=\omega\pm i\omxa$,
we can write
\BE
\left[I^{\alpha\beta}_L (y,\omega)\right]_{Th}=\frac{\omega^2} {(y^2+\omega^2)}
\int_0^\infty\, \frac{x\,{\rm d} x}{32\pi^2\,y^3}\,
\left\{\frac{n(\omxa)}{\omxa}\ \Big[\Re\, L^L_\beta(\omega+i\omxa;\, y,x)
-4xy(\omega^2+y^2+\beta^2-\alpha^2)\Big]
+\quad \alpha\leftrightarrow\beta\ \right\}.
\label{ILth}
\EE
\end{widetext}

\subsection{Thermal part of $\boldsymbol{\partial J_m}$, $\boldsymbol{\partial J_m^{L,T}}$, $\boldsymbol{\partial^2 J_m}$ and $\boldsymbol{\partial^2 J_m^{L,T}}$}

The thermal parts of $\partial J_m$ and $\partial J_m^{L,T}$ can be obtained by
a simple derivative of the thermal parts of $J_m$ and $J_m^{L,T}$, respectively,
according to the definition of the integrals in Eq.~(\ref{dJdef}).
For a function of $\omxm$ 
\BE
\frac{\partial}{\partial m^2}=\frac{1}{2\omxm}\,\frac{\partial}{\partial\omxm}=\frac{1}{2x}\,\frac{\partial}{\partial x},
\EE
so that it might be useful to integrate by parts, using Eq.~(\ref{Jth}):
\begin{align}
&(\partial J_m)_{Th}=\frac{\partial}{\partial m^2} \int_0^\infty \frac{x^2{\rm d} x} {2\pi^2} 
\,\frac{n(\omxm)}{\omxm}=\nn\\
&\quad=\int_0^\infty \frac{x\,{\rm d} x} {4\pi^2}\frac{\partial}{\partial x}\left[\frac{n(\omxm)}{\omxm}\right]=
-\int_0^\infty \frac{{\rm d} x} {4\pi^2}\,\frac{n(\omxm)}{\omxm}.
\label{dJth}
\end{align}
A plain further derivative gives
\BE
(\partial^2 J_m)_{Th}=\int_0^\infty \frac{{\rm d} x} {8\pi^2}\,\frac{n(\omxm)}{\omxm^3}-\frac{1}{T}\, J_m^{\,nn/\epsilon\epsilon},
\EE
where
\BE
J_m^{\,nn/\epsilon\epsilon}=\int_0^\infty \frac{{\rm d} x} {8\pi^2}\,\left[\frac{n(\omxm)\,n(-\omxm)}{(\omxm)^2}\right].
\EE
By the same method, using Eq.~(\ref{JLTth}),
\begin{align}
(\partial J^L_m)_{Th}&=\int_0^\infty \frac{{\rm d} x} {4\pi^2} \,{\omxm}\,n(\omxm),\nn\\
(\partial J^T_m)_{Th}&=-\int_0^\infty \frac{x^2\,{\rm d} x} {4\pi^2} \,\frac{n(\omxm)}{\omxm}=-\frac{1}{2}\, (J_m)_{Th},
\label{dJLTth}
\end{align}
and by a plain further derivative
\begin{align}
(\partial^2 J^L_m)_{Th}&=-\frac{1}{2}(\partial J_m)_{Th}+\frac{1}{T}\,J_m^{\,nn},\nn\\
(\partial^2 J^T_m)_{Th}&=-\frac{1}{2}\,(\partial J_m)_{Th},
\end{align}
where
\BE
J_m^{\,nn}=\int_0^\infty \frac{{\rm d} x} {8\pi^2}\,n(\omxm)\,n(-\omxm).
\EE

\subsection{Thermal part of $\boldsymbol{\partial I^{\alpha\beta}(p)}$}

The thermal part of $\partial I^{\alpha\beta}(p)$  can be obtained by
a derivative of the thermal part of $I^{\alpha\beta}(p)$, using
the explicit expression of Eq.~(\ref{Iabth}) 
\BE
\left[\partial I^{\alpha\beta}(y,\omega)\right]_{Th}=
\int_0^\infty \frac{ x{\rm d}x} {8\pi^2 y}
\left\{\frac{\partial}{\partial \alpha^2}{\cal A}+\frac{\partial}{\partial \alpha^2}{\cal B}\right\},
\label{dIab1}
\EE
where
\begin{align}
{\cal A}&=\frac{n(\omxa)}{\omxa}\ \Re\, L_\beta(\omega+i\omxa;\, y,x),\nn\\
{\cal B}&=\frac{n(\omxb)}{\omxb}\ \Re\, L_\alpha(\omega+i\omxb;\, y,x).
\end{align}
Using $\omxa$ as independent variable, with $\omxa{\rm d}\omxa=x\,{\rm d}x$, 
we can write $x=\sqrt{\omxa^2-\alpha^2}$ and eliminate the explicit dependence on $x$ in the function ${\cal A}$.
The total derivative of ${\cal A}$ reads
\BE
\frac{{\rm d}{\cal A}}{{\rm d}\,\omxa}=\left(\frac{\partial {\cal A}}{\partial \omxa}\right)_{\displaystyle{x}}
+\left(\frac{\partial {\cal A}}{\partial x}\right)_{\displaystyle{\omxa}} \left(\frac{{\rm d} x} {{\rm d} \omxa}\right),
\EE
and observing that
\BE
\left(\frac{\partial {\cal A}}{\partial \omxa}\right)_{\displaystyle{x}}=2\omxa \left(\frac{\partial {\cal A}}{\partial \alpha^2}\right),
\quad \left(\frac{{\rm d} x} {{\rm d} \omxa}\right)=\frac{\omxa}{x},
\EE
it can be written as
\BE
\frac{{\rm d}{\cal A}}{{\rm d}\,\omxa}=2\omxa\, \left(\frac{\partial {\cal A}}{\partial \alpha^2}\right)
+\frac{\omxa}{x}\, \left(\frac{\partial {\cal A}}{\partial x}\right)_{\displaystyle{\omxa}},
\EE
so that the first derivative in Eq.~(\ref{dIab1}) follows as
\BE
\frac{\partial {\cal A}}{\partial \alpha^2}=\frac{1}{2\omxa}\,\frac{{\rm d}{\cal A}}{{\rm d}\,\omxa}
-\frac{1}{2x}\,\left(\frac{\partial {\cal A}}{\partial x}\right)_{\displaystyle{\omxa}}.
\label{dA}
\EE
Moreover, observing that
\begin{widetext}
\BE
\left(\frac{\partial {\cal A}}{\partial x}\right)_{\displaystyle{\omxa}}=
2x\,\left(\frac{\partial {\cal A}}{\partial \opb^2}+\frac{\partial {\cal A}}{\partial \omb^2}\right)
+2y\,\left(\frac{\partial {\cal A}}{\partial \opb^2}-\frac{\partial {\cal A}}{\partial \omb^2}\right),
\EE
we find, explicitly,
\begin{align}
\frac{1}{2x}\,\left(\frac{\partial {\cal A}}{\partial x}\right)_{\displaystyle{\omxa}}&=
\frac{n(\omxa)}{\omxa}\ \Re\left[\frac{1}{z_\alpha^2+\opb^2}-\frac{1}{z_\alpha^2+\omb^2}\right]+
\left(\frac{y}{x}\right)\frac{n(\omxa)}{\omxa}\ \Re\left[\frac{1}{z_\alpha^2+\opb^2}+\frac{1}{z_\alpha^2+\omb^2}\right]
\end{align}
where $z_\alpha=\omega+i\omxa$. On the other hand, a simple derivative gives
\BE
\frac{\partial{\cal B}}{\partial \alpha^2}=\frac{n(\omxb)}{\omxb}\ \Re\left[\frac{1}{z_\beta^2+\opa^2}-\frac{1}{z_\beta^2+\oma^2}\right],
\label{dB}
\EE
where $z_\beta=\omega+i\omxb$.
Finally, inserting Eq.~(\ref{dA}) in Eq.~(\ref{dIab1}) and changing the integration variable $x\, {\rm d}x=\omxa{\rm d}\omxa$ in the first
term, the integral of the total derivative gives a vanishing contribution at $x=\infty$ and $x=0$, since $L_\beta\to 0$. 
Collecting the other terms, we find
\begin{align}
\left[\partial I^{\alpha\beta}(y,\omega)\right]_{Th}&=
-\int_0^\infty \frac{{\rm d}x} {8\pi^2}\,
\frac{n(\omxa)}{\omxa}\ \Re\left[\frac{1}{(\omega+i\omxa)^2+\opb^2}+\frac{1}{(\omega+i\omxa)^2+\omb^2}\right]+\nn\\
&\quad +\int_0^\infty \frac{x\,{\rm d}x} {8\pi^2 y}\left\{
\frac{n(\omxb)}{\omxb}\ \Re\left[\frac{1}{(\omega+i\omxb)^2+\opa^2}-\frac{1}{(\omega+i\omxb)^2+\oma^2}\right]
\quad - \quad \left(\alpha\leftrightarrow\beta\right) 
\right\},
\label{dIab2}
\end{align}
where the second integral is zero if $\alpha=\beta$.

\subsection{Thermal part of $\boldsymbol{\partial I^{\alpha\beta}_{L,T}(p)}$}

The thermal part of the projected integrals $\partial I^{\alpha\beta}_{L,T}(p)$  can be obtained by
a derivative of the thermal part of $I^{\alpha\beta}_{L,T}(p)$, using
the explicit expressions of Eqs.~(\ref{ILth}),(\ref{ITth}):
\begin{align}
\left[\partial I^{\alpha\beta}_L (y,\omega)\right]_{Th}&=\frac{\omega^2} {(y^2+\omega^2)}
\int_0^\infty\, \frac{x\,{\rm d} x}{32\pi^2\,y^3}\,
\left\{\frac{\partial}{\partial \alpha^2}{\cal A}_L+\frac{\partial}{\partial \alpha^2}{\cal B}_L\right\},\nn\\
\left[\partial I^{\alpha\beta}_T (y,\omega)\right]_{Th}&=-\int_0^\infty \frac{ x\,{\rm d}x} {64\pi^2 y^3}\,
\left\{\frac{\partial}{\partial \alpha^2}{\cal A}_T+\frac{\partial}{\partial \alpha^2}{\cal B}_T\right\},
\label{dILT1}
\end{align}
where
\begin{align}
{\cal A}_{L,T}&=\frac{n(\omxa)}{\omxa}\ \Big[\Re\, L^{L,T}_\beta(\omega+i\omxa;\, y,x)
-4xy(\omega^2+y^2+\beta^2-\alpha^2)\Big],\nn\\
{\cal B}_{L,T}&=\frac{n(\omxb)}{\omxb}\ \Big[\Re\, L^{L,T}_\alpha(\omega+i\omxb;\, y,x)
-4xy(\omega^2+y^2+\alpha^2-\beta^2)\Big].
\end{align}
Because of the explicit dependence on $\alpha$, Eq.~(\ref{dA}) is modified as
\BE
\frac{\partial {\cal A}_{L,T}}{\partial \alpha^2}=4xy\,\frac{n(\omxa)}{\omxa}+
\frac{1}{2\omxa}\,\frac{{\rm d}{\cal A}_{L,T}}{{\rm d}\,\omxa}
-\frac{1}{2x}\,\left(\frac{\partial {\cal A}_{L,T}}{\partial x}\right)_{\displaystyle{\omxa}}
\label{dALT},
\EE
while Eq.~(\ref{dB}) becomes
\begin{align}
\frac{\partial{\cal B}_{L}}{\partial \alpha^2}&=\frac{n(\omxb)}{\omxb}\,\Bigg\{
\Re\Bigg[\left(\frac{1}{z_\beta^2+\opa^2}-\frac{1}{z_\beta^2+\oma^2}\right)
\left(z_\beta^2+\omxa^2+y^2\Big(\frac{2z_\beta}{\omega}-1\Big)\right)^2+\nn\\
&\qquad\qquad\qquad\qquad \qquad\qquad\qquad
+2\left(z_\beta^2+\omxa^2+y^2\Big(\frac{2z_\beta}{\omega}-1\Big)\right)\,L_\alpha(z_\beta; y,x)\Bigg]
-4xy\Bigg\},\nn\\
\frac{\partial{\cal B}_{T}}{\partial \alpha^2}&=\frac{n(\omxb)}{\omxb}\,\Bigg\{
\Re\Bigg[\left(2z_\beta^2+\oma^2+\opa^2\right)\,L_\alpha(z_\beta; y,x)\Bigg]-8xy\Bigg\},
\label{dBLT}
\end{align}
where $z_\beta=\omega+i\omxb$. Moreover, an explicit calculation gives
\begin{align}
-\frac{1}{2x}\,\left(\frac{\partial {\cal A}_{L}}{\partial x}\right)_{\displaystyle{\omxa}}&=
-\frac{1}{2x}\,\frac{n(\omxa)}{\omxa}\,\Bigg\{\Re\Bigg[
4x\left(z_\alpha^2+\omxb^2+y^2\Big(\frac{2z_\alpha}{\omega}-1\Big)\right)\,L_\beta(z_\alpha; y,x)\,+\nn\\
&+2\left(z_\alpha^2+\omxb^2+y^2\Big(\frac{2z_\alpha}{\omega}-1\Big)\right)^2
\left(\frac{x+y}{z_\alpha^2+\opb^2}-\frac{x-y}{z_\alpha^2+\omb^2}\right)
\Bigg] -4y(\omega^2+y^2+\beta^2-\alpha^2)\Bigg\},\nn\\
-\frac{1}{2x}\,\left(\frac{\partial {\cal A}_{T}}{\partial x}\right)_{\displaystyle{\omxa}}&=
-\frac{n(\omxa)}{\omxa}\,\Bigg\{\Re\Bigg[
\left(2z_\alpha^2+\omb^2+\opb^2-4y^2\right)\,L_\beta(z_\alpha; y,x)\Bigg]-4xy\Bigg\},
\label{dALTx}
\end{align}
where $z_\alpha=\omega+i\omxa$.

Inserting Eqs.~(\ref{dALT}),(\ref{dBLT}) in Eq.~(\ref{dILT1}) and dropping
the integral of the total derivative which gives a vanishing contribution, we find
\begin{align}
&\left[\partial I^{\alpha\beta}_L (y,\omega)\right]_{Th}=-\frac{\omega^2} {y^2(y^2+\omega^2)}
\int_0^\infty \frac{{\rm d} x}{32\pi^2}\,\frac{n(\omxa)}{\omxa}\,
\Re\Bigg[\bigg(\frac{1}{z_\alpha^2+\opb^2}+\frac{1}{z_\alpha^2+\omb^2}\bigg)
\bigg(z_\alpha^2+\omxb^2+y^2(\frac{2z_\alpha}{\omega}-1)\bigg)^2\Bigg]\ +\nn\\
&\qquad+\frac{\omega^2} {y^2(y^2+\omega^2)}\int_0^\infty \frac{{\rm d} x}{16\pi^2}\,\frac{n(\omxa)}{\omxa}\,
\left[\omega^2+y^2+\beta^2-\alpha^2\right]\,
+\,\frac{\omega^2} {y^2(y^2+\omega^2)}\int_0^\infty \frac{x^2{\rm d} x}{8\pi^2}\,
\left[\frac{n(\omxa)}{\omxa}-\frac{n(\omxb)}{\omxb}\right]\ +\nn\\
&\qquad+\frac{\omega^2} {y^3(y^2+\omega^2)}\Bigg\{
\int_0^\infty \frac{x{\rm d} x}{16\pi^2}\,\frac{n(\omxb)}{\omxb}\,
\Re\Bigg[ \bigg(z_\beta^2+\omxa^2+y^2(\frac{2z_\beta}{\omega}-1)\bigg) L_\alpha(z_\beta;\, y,x) \Bigg]
\quad - \quad \left(\alpha\leftrightarrow\beta\right) 
\Bigg\}\ +\nn\\
&\qquad+\frac{\omega^2} {y^3(y^2+\omega^2)}\Bigg\{
\int_0^\infty \frac{x{\rm d} x}{32\pi^2}\,\frac{n(\omxb)}{\omxb}\,
\Re\Bigg[ \bigg(z_\beta^2+\omxa^2+y^2(\frac{2z_\beta}{\omega}-1)\bigg)^2 
\bigg(\frac{1}{z_\beta^2+\opa^2}-\frac{1}{z_\beta^2+\oma^2}\bigg)\Bigg]
\>-\>\left(\alpha\leftrightarrow\beta\right) 
\Bigg\}\ ,\nn\\
&\left[\partial I^{\alpha\beta}_T (y,\omega)\right]_{Th}=-\frac{1} {y}\,
\int_0^\infty \frac{x{\rm d} x}{16\pi^2}\,\frac{n(\omxa)}{\omxa}\,
\Re\,L_\beta(z_\alpha; y,x)+\frac{1}{y^2}\,\int_0^\infty \frac{x^2{\rm d} x}{8\pi^2}\,
\left[\frac{n(\omxb)}{\omxb}-\frac{n(\omxa)}{\omxa}\right]+\nn\\
&\qquad\qquad\qquad\qquad\qquad+\frac{1}{y^3}\,\Bigg\{
\int_0^\infty \frac{x{\rm d} x}{32\pi^2}\,\frac{n(\omxa)}{\omxa}\,
\Re\bigg[
\left(z_\alpha^2+\beta^2+x^2+y^2\right)\,L_\beta(z_\alpha; y,x)\bigg]
\>-\>\left(\alpha\leftrightarrow\beta\right) \Bigg\},
\label{dILT2}
\end{align}
\end{widetext}
where, as before, $z_\alpha=\omega+i\omxa$ and $z_\beta=\omega+i\omxb$. We observe that most of these
integrals are antisymmetric in the mass arguments $\alpha,\beta$ and their contribution is zero if $\alpha=\beta=m$.

It is instructive to explore the leading behavior in the limit $p\to 0$. According to Eq.~(\ref{limits}), the
longitudinal projection $\partial I^{\alpha\beta}_L$ tends to the
value $\partial I^{\alpha\beta}_{L,0}$ if $\omega$ is set to zero first and the limit $y\to 0$
is studied afterwards. Setting $\omega\to 0$ in Eq.~(\ref{dILT2}), the only terms of $\partial I^{\alpha\beta}_L$ 
that do not vanish are those containing
the factor $(2z/\omega)^2$. Observing that $z_\alpha^2\to -\omxa^2$ and that, in the limit $y\to 0$,
\begin{align}
\frac{1}{z_\alpha^2+\opb^2}+\frac{1}{z_\alpha^2+\omb^2}&\to \frac{2}{\beta^2-\alpha^2},\nn\\
\frac{1}{z_\alpha^2+\opb^2}-\frac{1}{z_\alpha^2+\omb^2}&\to \frac{-4xy}{(\beta^2-\alpha^2)^2},
\end{align}
we obtain the leading behavior
\begin{align}
\left[\partial I^{\alpha\beta}_{L,0}\right]_{Th}&=\frac{\left(\partial J^{L}_\alpha\right)_{Th}}{\beta^2-\alpha^2}
+\frac{\big(J^{L}_\alpha\big)_{Th} - \big(J^{L}_\beta\big)_{Th} }{(\beta^2-\alpha^2)^2},
\end{align}
having made use of the explicit expressions of $\big(J^{L}_m\big)_{Th}$, $\left(\partial J^{L}_m\right)_{Th}$
as reported in Eqs.~(\ref{JLTth}),(\ref{dJLTth}). The result is in agreement with the general relations of Eq.~(\ref{identILT}).

The transverse projection, $\partial I^{\alpha\beta}_T$, tends to a different value, $\partial I^{\alpha\beta}_{T,0}$,
in the same limit. Using Eq.~(\ref{Lx0}),
\begin{align}
\left(z^2+\omxb^2+y^2\right)L_\beta(z_\alpha; y,x)&\approx{4xy}+\frac{16x^3y^3}{3(z^2+\omxb^2)^2}+{\cal O}(y^5)\nn\\
L_\beta(z_\alpha; y,x)&\approx\frac{4xy}{(z^2+\omxb^2)}+{\cal O}(y^3),
\end{align}
and inserting the expansions in $\partial I^{\alpha\beta}_T$, in Eq.~(\ref{dILT2}), the terms $y^{-2}$ cancel exactly while 
the leading term is of order $\sim y^0$, so that we can safely take the limit $y\to 0$. The leading term reads
\begin{align}
&\left[\partial I^{\alpha\beta}_T (0,\omega)\right]_{Th}=
-\int_0^\infty \frac{x^2{\rm d} x}{4\pi^2}\,\frac{n(\omxa)}{\omxa}\,
\Re\,\frac{1}{(z_\alpha^2+\omxb^2)}\,+\nn\\
&+\int_0^\infty \frac{x^4{\rm d} x}{6\pi^2}\,
\left\{ \frac{n(\omxa)}{\omxa}\, \Re\,\frac{1}{(z_\alpha^2+\omxb^2)^2}
-\left(\alpha\leftrightarrow\beta\right) \right\}.
\label{y0}
\end{align}
The expansion holds for any value of
$\omega$, even $\omega=0$, so that we can exchange the limits for the transverse projection.
Setting $\omega=0$ and $z_\alpha^2=-\omxa^2$, we can simply write $(z_\alpha^2+\omxb^2)=(\beta^2-\alpha^2)$
and the leading term reads
\begin{align}
\left[\partial I^{\alpha\beta}_{T,0}\right]_{Th}&=\frac{\left(\partial J^{T}_\alpha\right)_{Th}}{\beta^2-\alpha^2}
+\frac{\big(J^{T}_\alpha\big)_{Th} - \big(J^{T}_\beta\big)_{Th} }{(\beta^2-\alpha^2)^2},
\end{align}
having made use of the explicit expressions of $\big(J^{T}_m\big)_{Th}$, $\left(\partial J^{T}_m\right)_{Th}$
as reported in Eqs.~(\ref{JLTth}),(\ref{dJLTth}). Again, the result is in agreement with the general 
relations of Eq.~(\ref{identILT}).

On the other hand, the limits cannot be interchanged for the longitudinal projection 
$\partial I^{\alpha\beta}_L$ which tends to the same limit of the transverse projection,
$\partial I^{\alpha\beta}_{T,0}$,
if $y$ is set to zero first and the limit $\omega\to 0$ is taken afterwards. Taking $\omega$ finite,
we can write in the limit $y\to 0$
\begin{widetext}
\begin{align}
&\Re\Bigg[\bigg(\sum_{\pm}\frac{1}{z_\alpha^2+\epsilon_{y\pm x,\beta}^2}\bigg)
\bigg(z_\alpha^2+\omxb^2+y^2(\frac{2z_\alpha}{\omega}-1)\bigg)^2\Bigg]\approx
2(\omega^2+y^2+\beta^2-\alpha^2)+8x^2y^2\Re\,\frac{1}{(z_\alpha^2+\omxb^2)}+{\cal O}(y^4);
\end{align}
then, using the expansion
\begin{align}
&\bigg(\frac{1}{z_\beta^2+\opa^2}-\frac{1}{z_\beta^2+\oma^2}\bigg)\approx
-\frac{4xy}{(z_\beta^2+\omxa^2)^2}+\frac{8xy^3}{(z_\beta^2+\omxa^2)^3}
-\frac{16x^3y^3}{(z_\beta^2+\omxa^2)^4}+{\cal O}(y^5),
\end{align}
we can write
\begin{align}
&\Re\Bigg[ \bigg(z_\beta^2+\omxa^2+y^2(\frac{2z_\beta}{\omega}-1)\bigg)^2 
\bigg(\frac{1}{z_\beta^2+\opa^2}-\frac{1}{z_\beta^2+\oma^2}\bigg)\Bigg]
\approx\Re\Bigg[-4xy-\frac{16xy^3(z_\beta-\omega)}{\omega(z_\beta^2+\omxa^2)}
-\frac{16x^3y^3}{(z_\beta^2+\omxa^2)^2}+{\cal O}(y^5)\Bigg],
\end{align}
and finally, using Eq.~(\ref{Lx0}),
\begin{align}
&\Re\Bigg[ 2\bigg(z_\beta^2+\omxa^2+y^2(\frac{2z_\beta}{\omega}-1)\bigg) L_\alpha(z_\beta;\, y,x) \Bigg]
\approx \Re\Bigg[8xy+\frac{16xy^3(z_\beta-\omega)}{\omega(z_\beta^2+\omxa^2)}
+\frac{32x^3y^3}{3(z_\beta^2+\omxa^2)^2}+{\cal O}(y^5)\Bigg].
\end{align}
\end{widetext}
Inserting the expansions in $\partial I^{\alpha\beta}_L$, in Eq.~(\ref{dILT2}), again
the negative powers of $y$ cancel exactly. We can safely set $y=0$ and the same identical
expression of Eq.~(\ref{y0}) is recovered, yielding
\BE
\left[\partial I^{\alpha\beta}_L (0,\omega)\right]_{Th}=
\left[\partial I^{\alpha\beta}_T (0,\omega)\right]_{Th}
\EE
for any finite $\omega$,
as expected in the long wavelength limit where no special direction in space is defined,
in agreement with Eq.~(\ref{limits}).

\end{document}